\documentclass{ptephy_v1}
\preprintnumber{2207.05752} 
\usepackage[dvipdfmx]{hyperref}
\usepackage{amsmath}
\usepackage{graphics}
\usepackage{bm}
\usepackage[normalem]{ulem}

\begin{document}

\title{Kinetic theory of discontinuous shear thickening of a moderately dense inertial suspension of frictionless soft particles}

\author{Satoshi Takada}
\affil{Department of Mechanical Systems Engineering and Institute of Engineering,
    Tokyo University of Agriculture and Technology,
    2-24-16 Naka-cho, Koganei, Tokyo 184-8588, Japan 
    \email{takada@go.tuat.ac.jp}}

\author{Kazuhiro Hara
\thanks{Present address: 
    {\it Suzuki Motor Corporation, 
    300 Takatsukacho, Minami-ku, Hamamatsu, Shizuoka 432-8611, Japan}}}
\affil{Department of Industrial Technology and Innovation,
    Tokyo University of Agriculture and Technology,
    2--24--16 Naka-cho, Koganei, Tokyo 184--8588, Japan}

\author{Hisao Hayakawa}
\affil{Yukawa Institute for Theoretical Physics, 
    Kyoto University,
    Kitashirakawa Oiwakecho, Sakyo-ku, Kyoto 606--8502, Japan
    \email{hisao@yukawa.kyoto-u.ac.jp}}

\begin{abstract}%
We demonstrate that a discontinuous shear thickening (DST) can take place even in a moderately dense inertial suspension consisting of frictionless soft particles.
This DST can be regarded as an ignited-quenched transition in the inertial suspension. 
An approximate kinetic theory well recovers the results of the Langevin simulation in the wide range of the volume fraction without any fitting parameters.
\end{abstract}

\subjectindex{A57, J01}

\maketitle

\section{Introduction}\label{sec:intro}
When we apply a simple shear for dense suspensions, the viscosity exhibits a discontinuous jump at a certain shear rate.
This discontinuous change of the viscosity is known as the discontinuous shear thickening (DST)~\cite{Wagner09,Brown14,Ness22,Seto13}.
The normal stress difference is also discontinuously changed associated with the DST~\cite{Laun94, Cwalina14}.
The DST can be observed even in frictional dry granular materials~\cite{Otsuki11}.
Although the DST is analogous to the first-order phase transition in equilibrium, the DST takes place only in nonequilibrium situations.
The DST is closely related to the shear jamming~\cite{Bi11,Fall15,Peters16,Otsuki20,Pradipto20}.
Thus, the DST is important to study the physics of densely packed systems.

Although there are some debates~\cite{Hoffman72, Hoffman74, Hoffman98, Brady85, Brady88, Bender96, Wagner09, Melrose04, Farr97, Wang20} on the origin of the DST, 
frictional contacts between particles are believed to be the main origin of the DST~\cite{Otsuki11,Seto13,Kawasaki18,Wyart14}.
One of the natural questions is whether the DST-like process can happen even if we are interested in suspensions consisting of frictionless particles.

A DST-like phenomenon can be observed in inertial suspensions, a model of aerosols~\cite{Friedlander,Koch2001,Hu12}, in which collisions between particles play important roles.
There are several theoretical studies of inertial suspensions consisting of frictionless hard-core particles based on the kinetic theory under the influence of Stokes' drag~\cite{Hayakawa19, Tsao95, Sangani96, Koch99, Koch01, Garzo12, Saha17, Alam19, Saha20, Hayakawa17, Takada20}.
The theoretical prediction quantitatively reproduces the results of simulation in the wide range of the volume fraction ($\varphi\lesssim0.50$)~\cite{Hayakawa17,Takada20}.
It is noteworthy that the DST-like behavior, caused by an ignited-quenched transition of the kinetic temperature, can be observed only in dilute inertial suspensions of hard-core particles.
Namely, the DST becomes the continuous shear thickening (CST) if the volume fraction $\varphi$ is larger than a few percent~\cite{Sangani96,Saha17,Hayakawa17,Takada20}.
This behavior is completely different from the DST commonly observed in colloidal suspensions in which the DST can be observed only in dense suspensions. 

Sugimoto and Takada recently developed the kinetic theory of dilute inertial suspensions comprising frictionless soft particles~\cite{Sugimoto20}.
They discovered that discontinuous changes in kinetic temperature and viscosity can occur twice. Their theoretical results agreed with the simulation results without any fitting parameters. 
This is a remarkable result, although the second discontinuous change is difficult to observe in real experiments because the kinetic temperature in the ignited phase becomes approximately $10^6$ times larger than that in the quenched phase.
For later convenience, we call the phase with the shear rate larger than the second critical shear rate the exploded phase.

This paper extends the analysis of dilute suspensions discussed in Ref.~\cite{Sugimoto20} to denser situations with the aid of the Enskog theory~\cite{Hayakawa17,Takada20,Resibois,Garzo99,Lutsko05,Garzo13,Gonzalez19,Garzo}.
Detailed discussions for hydrodynamic interactions between particles based on simulations and comparing such systems with the kinetic theory without hydrodynamic interactions are presented in a paper~\cite{hydro}. 
We also note that the recent molecular dynamics (MD) simulation for a mixture of elastic molecules and granular grains recovers the results of the kinetic theory of inertial suspensions~\cite{Gonzalez23}. 

The organization of this paper is as follows.
In the next section, we introduce the Langevin equation used for the simulation of inertial suspensions.
In Sect.~\ref{simulation}, we explain the simulation protocol.
In Sect.~\ref{kinetic_theory}, we develop the kinetic theory of inertial suspensions under the influence of Stokes' drag in a simple shear flow, and derive a set of dynamic equations that describes the rheology of this system.
In Sect.~\ref{rheology}, we present the results of the steady rheology obtained from both the kinetic theory and the Langevin simulation, 
in which we verify the existence of DST-like processes in the wide range of parameters' space. 
In Sect.~\ref{sec:conclusion}, we conclude and discuss our results.
In Appendix~\ref{hydro}, we briefly discuss the effect of hydrodynamic lubrication interactions between grains discussed in Ref.~\cite{hydro}.
Appendix~\ref{frame_McLennan} explains the framework of the kinetic theory used in this paper.
In Appendix~\ref{sec:scattering}, we summarize the expressions of the scattering angle and the turning point of soft-core particles.
Appendix~\ref{sec:replacement} compares the results of simulation based on Cauchy's contact stress with an approximate expression for the soft-core systems by using the collisional contribution to the stress in an inner-hard core model.
In Appendix~\ref{sec:convergence}, we discuss the convergence of the expression of the stress tensor in terms of a series expansion of the dimensionless shear rate.
In Appendix~\ref{sec:dilute}, we evaluate the collision moment $\overleftrightarrow{\Lambda}$ in dilute soft-core systems.

\section{Langevin model}\label{Langevin_model}
\begin{figure}[htbp]
    \centering
    \includegraphics[width=0.6\linewidth]{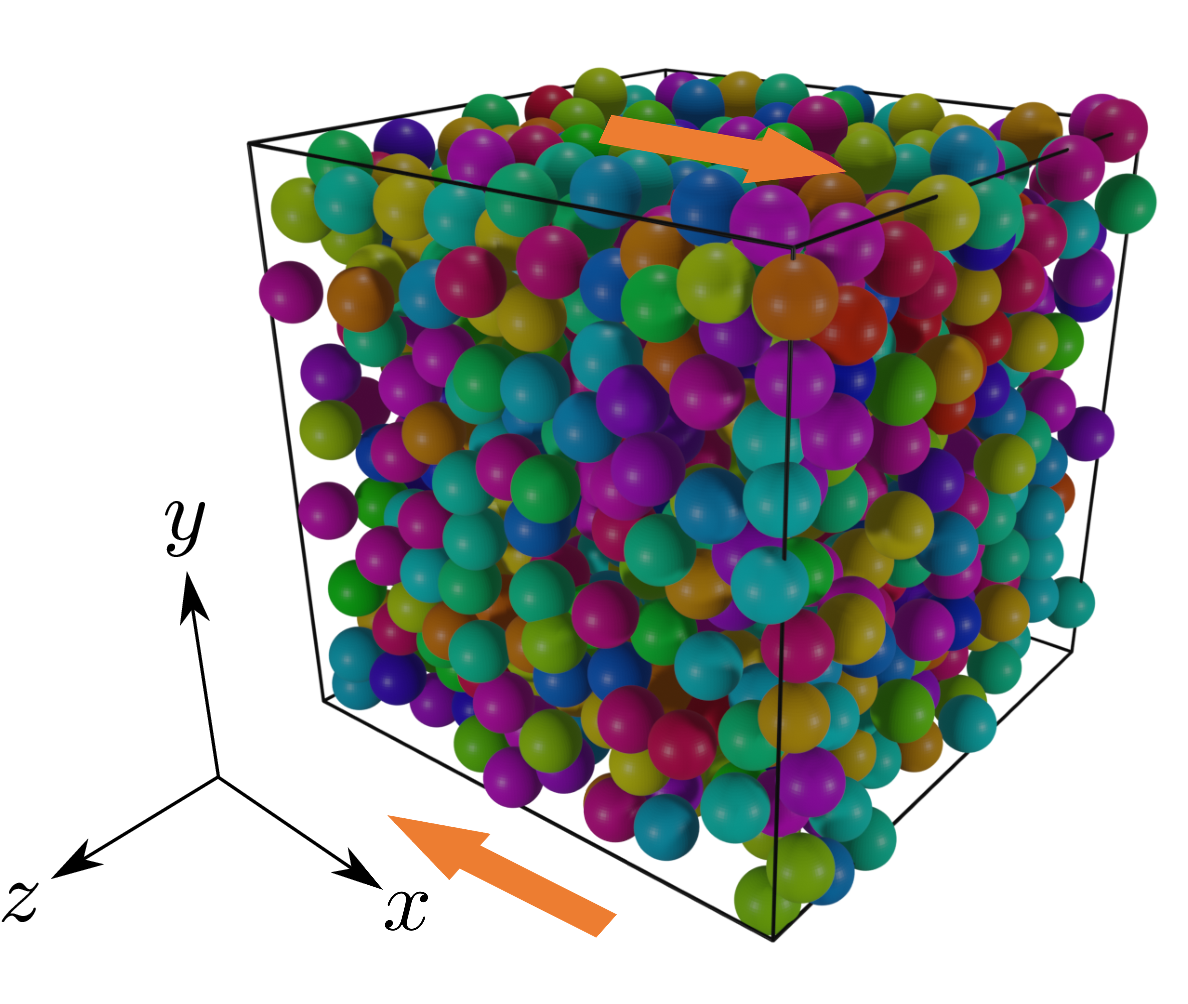}
    \caption{A snapshot of our system.
    Particles are initially distributed at random.
    The arrows indicate the shear direction.}
    \label{fig:setup}
\end{figure}

We consider $N$ monodisperse frictionless soft particles (mass $m$ and diameter $d$ of each particle), which are suspended in a solvent (the viscosity $\eta_0$) and are confined in a three-dimensional cubic box with the linear size $L$ as shown in Fig.~\ref{fig:setup}.
We assume that the contact force between particles is described by the harmonic potential
\begin{equation}
    U(r)=\frac{\varepsilon}{2}\left(1-\frac{r}{d}\right)^2\Theta\left(1-\frac{r}{d}\right),
    \label{eq:harmonic_potential}
\end{equation}
where $r$ is the inter-particle distance and $\varepsilon$ is the energy scale to characterize the repulsive interaction, and $\Theta(x)$ is the step function satisfying $\Theta(x)=1$ ($x\ge 0$) and $0$ ($x<0$).
Although clustering effects caused by attractive interactions between particles cannot be ignored in realistic situations,
such effects are suppressed if particles are charged~\cite{Derjaguin41,Verwey,Israelachvili} or if the temperature is high enough~\cite{Kawasaki14,Hayakawa19,Bossis89,Mari15}.

Equation of motion of the suspended particle $i$ at time $t$ with its position $\bm{r}_i$ under a simple shear with the shear rate $\dot\gamma$ is given by
\begin{equation}
    \frac{d\bm{p}_i}{dt} 
    = \sum_{j\neq i}\bm{F}_{ij} 
    - \zeta \bm{p}_i 
    + \bm{\xi}_i,
    \label{eq:Langevin}
\end{equation}
where  
$\bm{p}_i\equiv m\bm{V}_i$ with
$\bm{V}_i\equiv \bm{v}_i-\dot\gamma y_i\hat{\bm{e}}_x$ with the unit vector $\hat{\bm{e}}_x$ parallel to $x$ direction and the velocity $\bm{v}_i$ of $i-$th particle is the peculiar momentum, 
$\bm{F}_{ij}\equiv -\partial U(r_{ij})/\partial \bm{r}_{ij}$ is the inter-particle force between $i$--th and $j$--th particles with $\bm{r}_{ij}\equiv \bm{r}_i - \bm{r}_j$ and $r_{ij}\equiv |\bm{r}_{ij}|$, and $T_\mathrm{env}$ is the environmental (solvent) temperature.
Note that the drag coefficient $\zeta$ is expressed as $\zeta=3\pi\eta_0d/m$.
The noise $\bm{\xi}_i(t)=\xi_{i,\alpha}(t) \hat{\bm{e}}_\alpha$ satisfies the fluctuation-dissipation relation:
\begin{equation}
    \langle \bm{\xi}_i(t)\rangle=\bm{0},\quad
    \langle \xi_{i,\alpha}(t) \xi_{j,\beta}(t^\prime) \rangle
    = 2m T_\mathrm{env} \zeta \delta_{ij}\delta_{\alpha\beta}\delta(t-t^\prime) ,
    \label{eq:white_Gaussian}
\end{equation}
where $\langle \cdot \rangle$ expresses the average over the noise.

The assumptions behind Eqs.~\eqref{eq:harmonic_potential}, \eqref{eq:Langevin}, and \eqref{eq:white_Gaussian} are summarized as follows. 
(i) Suspended particles are monodisperse and frictionless.
(ii) The inter-particle force is given by the harmonic potential in Eq.~\eqref{eq:harmonic_potential} and collisions between particles are elastic.
(iii) Suspended particles are agitated by the white Gaussian noise as in Eq.~\eqref{eq:white_Gaussian}.
(iv) Suspended particles also feel Stokes' drag while the hydrodynamic interaction between particles is ignored.
(v) The environmental temperature $T_\mathrm{env}$ is independent of the motion of suspended particles.
(vi) Perfect density matching between solvent and particles is assumed.
Although aerosols cannot satisfy the density matching condition, the sedimentation is negligible within the observation time for small suspended particles~\cite{Koch2001,Hu12}. 
Note that the role of hydrodynamic interactions among particles in inertial suspensions is analyzed in another paper~\cite{hydro} (
See also Appendix \ref{hydro} for the short-range contributions).

In our analysis, we also adopt the SLLOD dynamics~\cite{Evans84, Evans} to simulate the shear flow with the aid of the Lees-Edwards boundary condition~\cite{Lees72} (see Fig.~\ref{fig:setup}).
As far as we have checked, the uniform flow is stable once the system reaches a steady state.
In this paper variables are scaled by the mass $m$, the diameter $d$, and the drag coefficient $\zeta$.
We have introduced the volume fraction $\varphi$, the (dimensionless) shear rate $\dot\gamma^*$, the softness $\varepsilon^*$, and the strength of the noise $\xi_\mathrm{env}$ as the dimensionless control parameters.
These dimensionless parameters are connected to the introduced physical variables as
\begin{equation}
    \varphi \equiv \frac{\pi}{6}\frac{Nd^3}{L^3},\quad
    \dot\gamma^* \equiv \frac{\dot\gamma}{\zeta},\quad
    \varepsilon^*\equiv \frac{\varepsilon}{md^2\zeta^2},\quad
    \xi_\mathrm{env}\equiv \sqrt{\frac{T_\mathrm{env}}{m}}\frac{1}{d \zeta}.
    \label{eq:dimless_quantities}
\end{equation}
In this study, the ranges of the parameters are $0<\varphi\le 0.5$, $0.1\le \dot\gamma^*\le 100$, $10^{-1}\le \varepsilon^*\le 10^8$, and $10^{-1}\le\xi_\mathrm{env}\le10^2$.

What we are interested in are the viscosity $\eta$ and the kinetic temperature $T$ defined as
\begin{equation}\label{eq:eta&T}
    \eta \equiv \frac{\sigma_{xy}}{\dot\gamma} ,\quad
    T\equiv \frac{1}{N}\sum_i \frac{p_i^2}{3m},
\end{equation}
in the simulation.
Note that the stress tensor $\sigma_{\alpha\beta}$ consists of two parts:
\begin{equation}
    \sigma_{\alpha\beta}
    =\sigma_{\alpha\beta}^k + \sigma_{\mathrm{S},\alpha\beta}^c ,
    \label{eq:sigma_scalar}
\end{equation}
where the kinetic and contact stresses, $\sigma_{\alpha\beta}^k$ and $\sigma_{\mathrm{S},\alpha\beta}^c$ are, respectively, given by
\begin{equation}
    \sigma_{\alpha\beta}^k
    \equiv -\frac{1}{L^3}\sum_i \frac{p_{i,\alpha}p_{i,\beta}}{m},
    \quad
    \sigma_{\mathrm{S},\alpha\beta}^c
    \equiv -\frac{1}{L^3}\sum_i\sum_{j<i}r_{ij,\alpha}F_{ij,\beta} .
    \label{eq:sigma_sim}
\end{equation}
Here, the subscript S for $\sigma_{\mathrm{S},\alpha\beta}^c$ stands for the expression for soft-core systems.
We can evaluate the dimensionless viscosity $\eta^*\equiv \eta \zeta/(nT_\mathrm{env})$ and dimensionless temperature $\theta$ from the simulation as
\begin{equation}
    \eta^*
    \equiv \frac{\Pi_{xy}+\Pi_{\mathrm{S},xy}^c}{\dot\gamma^*}, \quad
    \theta\equiv \frac{T}{T_{\mathrm{env}}} 
\end{equation}
with
\begin{equation}
    \Pi_{\alpha\beta}
    \equiv \frac{\sigma_{\alpha\beta}^k}{nT_\mathrm{env}}
    +\theta \delta_{\alpha\beta},\quad
    \Pi_{\mathrm{S},\alpha\beta}^c
    \equiv \frac{\sigma_{\mathrm{S},\alpha\beta}^c}{nT_\mathrm{env}},
    \label{eq:Pi_c_S}
\end{equation}
where $n\equiv N/L^3$ is the number density of the suspended particles.

\section{Simulation protocol}\label{simulation}
To perform the simulation, we fix $N=1000$.
We adopt the time increment for the simulations $\Delta t=10^{-2}\min(d/\sqrt{2T/m}, d\sqrt{m/\varepsilon})$, where $\min(a, b)$ chooses the smaller one between $a$ and $b$.
From test simulations, we have verified the choice of $\Delta t$ is small enough to get convergent results.

We prepare the equilibrium initial conditions where the velocity distribution satisfies the Maxwellian with the (dimensionless) temperature $\theta=1$.
Then, we add the shear with the shear rate $\dot\gamma$ at $t=0$, and the system evolves by Eqs.~\eqref{eq:Langevin} and \eqref{eq:white_Gaussian} at $t>0$.
The system is considered to have reached a steady state when the strain $\dot\gamma t$ reaches $1000$, where the temporal fluctuations of observables are sufficiently small.

In the simulation, we control the four parameters: the volume fractions $\varphi$, the dimensionless shear rate $\dot\gamma^*$, the dimensionless softness $\varepsilon^*$, and the strength of the noise $\xi_\mathrm{env}$.
Sets of the values of these parameters used in the simulation are listed in Tables \ref{table:set1}--\ref{table:set3}.

\begin{table}[htbp]
    \centering
    \caption{Sets of control parameters for the flow curves with fixing $\xi_\mathrm{env}=1.0$ and $\varepsilon^*=10^4$.
    Here, the parameter sets for which simulations are performed are marked with a check $\checkmark$, while those not simulated are left blank.}
    \begin{tabular}{cc|ccccccccc}
        \hline
        & & \multicolumn{9}{c}{$\dot\gamma^*$} \\
        & & $10^0$ & $10^{0.2}$& $10^{0.4}$& $10^{0.6}$& $10^{0.8}$& $10^{1}$& $10^{1.2}$& $10^{1.4}$& $10^{1.6}$\\
        \hline
        &$0.10$ & $\checkmark$ & $\checkmark$ & $\checkmark$& $\checkmark$& $\checkmark$& $\checkmark$& $\checkmark$& $\checkmark$& $\checkmark$\\
        &$0.20$ & $\checkmark$ & $\checkmark$& $\checkmark$& $\checkmark$& $\checkmark$& $\checkmark$& $\checkmark$& $\checkmark$& $\checkmark$\\
        $\varphi$&$0.30$ & $\checkmark$ & $\checkmark$& $\checkmark$& $\checkmark$& $\checkmark$& $\checkmark$& $\checkmark$& $\checkmark$& $\checkmark$\\
        &$0.40$ & $\checkmark$ & $\checkmark$& $\checkmark$& $\checkmark$& $\checkmark$& $\checkmark$& $\checkmark$& $\checkmark$& $\checkmark$\\
        &$0.50$ & $\checkmark$ & $\checkmark$& $\checkmark$& $\checkmark$& $\checkmark$& $\checkmark$& $\checkmark$& $\checkmark$& $\checkmark$\\
        \hline
    \end{tabular}
    \label{table:set1}
\end{table}
\begin{table}[htbp]
    \centering
    \caption{Sets of control parameters for the flow curves with fixing $\varphi=0.40$ and $\xi_\mathrm{env}=1.0$.
    Here, the parameter sets for which simulations are performed are marked with a check $\checkmark$, while those not simulated are left blank.}
    \begin{tabular}{cc|ccccccccccc}
        \hline
        & & \multicolumn{11}{c}{$\dot\gamma^*$} \\
        & & $10^0$ & $10^{0.2}$& $10^{0.4}$& $10^{0.6}$& $10^{0.8}$& $10^{1}$& $10^{1.2}$& $10^{1.4}$& $10^{1.6}$& $10^{1.8}$& $10^{2}$\\
        \hline
        &$10^0$ & $\checkmark$ & $\checkmark$& $\checkmark$& $\checkmark$& $\checkmark$& $\checkmark$& $\checkmark$& $\checkmark$& $\checkmark$& & \\
        &$10^2$ & $\checkmark$ & $\checkmark$& $\checkmark$& $\checkmark$& $\checkmark$& $\checkmark$& $\checkmark$& $\checkmark$& $\checkmark$& & \\
        &$10^3$ &  & & & & & & & & $\checkmark$& & \\
        $\varepsilon^*$&$10^4$ & $\checkmark$ & $\checkmark$& $\checkmark$& $\checkmark$& $\checkmark$& $\checkmark$& $\checkmark$& $\checkmark$& $\checkmark$& & \\
        &$10^5$ & & & & & & & & & $\checkmark$& & \\
        &$10^6$ & $\checkmark$ & $\checkmark$& $\checkmark$& $\checkmark$& $\checkmark$& $\checkmark$& $\checkmark$& $\checkmark$& $\checkmark$& $\checkmark$& $\checkmark$\\
        &$10^7$ & & & & & & & & & $\checkmark$& & \\
        &$10^8$ & $\checkmark$ & $\checkmark$& $\checkmark$& $\checkmark$& $\checkmark$& $\checkmark$& $\checkmark$& $\checkmark$& $\checkmark$& $\checkmark$& $\checkmark$\\
        \hline
    \end{tabular}
    \label{table:set2}
\end{table}
\begin{table}[htbp]
    \centering
    \caption{Sets of control parameters for the flow curves with fixing $\varphi=0.30$ and $\varepsilon^*=10^4$.
    Here, the parameter sets for which simulations are performed are marked with a check $\checkmark$, while those not simulated are left blank.}
    \begin{tabular}{cl|ccccccccc}
        \hline
        & & \multicolumn{9}{c}{$\dot\gamma^*$} \\
        & & $10^0$ & $10^{0.2}$& $10^{0.4}$& $10^{0.6}$& $10^{0.8}$& $10^{1}$& $10^{1.2}$& $10^{1.4}$& $10^{1.6}$\\
        \hline
        &$10^{-1}$ & $\checkmark$ & $\checkmark$& $\checkmark$& $\checkmark$& $\checkmark$& $\checkmark$& $\checkmark$& $\checkmark$& $\checkmark$\\
        $\xi_\mathrm{env}$&$10^{0}$ & $\checkmark$ & $\checkmark$& $\checkmark$& $\checkmark$& $\checkmark$& $\checkmark$& $\checkmark$& $\checkmark$& $\checkmark$\\
        &$10^{1}$ & $\checkmark$ & $\checkmark$& $\checkmark$& $\checkmark$& $\checkmark$& $\checkmark$& $\checkmark$& $\checkmark$& $\checkmark$\\
        &$10^{2}$ & $\checkmark$ & $\checkmark$& $\checkmark$& $\checkmark$& $\checkmark$& $\checkmark$& $\checkmark$& $\checkmark$& $\checkmark$\\
        \hline
    \end{tabular}
    \label{table:set3}
\end{table}

\section{Kinetic theory of inertial suspensions}\label{kinetic_theory}

In this section, we develop the kinetic theory of inertial suspensions consisting of frictionless soft particles.
In the first subsection, we present the kinetic equation for the one-body distribution for the inertial suspension and the moment equations for the stress tensor.
Because the kinetic equation cannot be solved exactly in general situations, we adopt some assumptions to simplify the analysis in the subsequent subsections.
In the second subsection, we introduce the Enskog approximation to solve the kinetic theory approximately. 
In the third subsection, we employ Grad's approximation to obtain a set of closure equations.

\subsection{Kinetic equation for inertial suspensions}

The kinetic theory is a powerful tool for describing the behavior of inertial suspensions.
The basic assumption of the kinetic theory is that both the random noise and collisions between particles are important, 
though the collisions have been ignored in the analysis of most of the theories for colloidal suspensions so far.

\begin{figure}[htbp]
    \centering
    \includegraphics[width=0.6\linewidth]{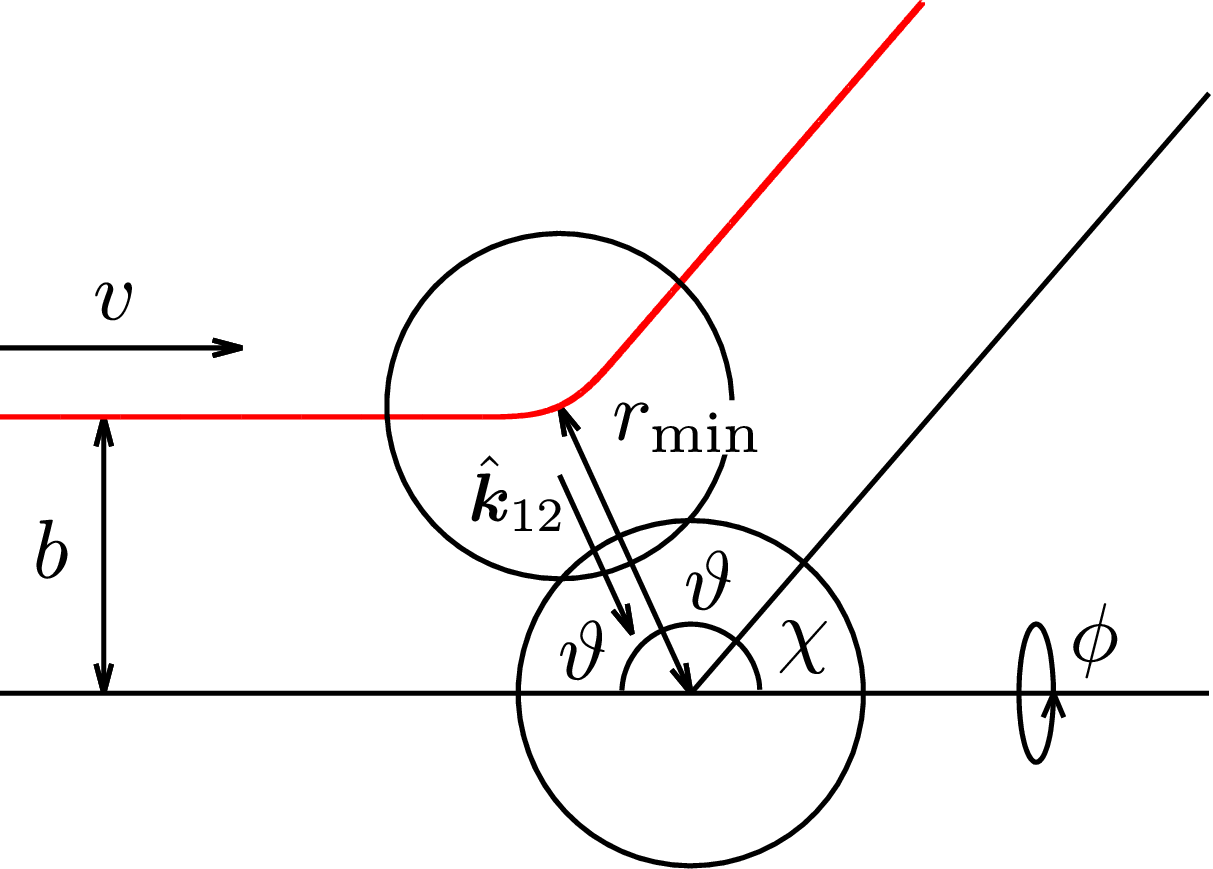}
    \caption{A schematic of a scattering process between two colliding particles with the impact parameter $b$ and the relative speed $v$, where the scattering angle $\chi$ and the closest distance $r_\mathrm{min}$ are given by Eqs.~\eqref{eq:chi_r_min}.}
    \label{fig:scattering}
\end{figure}

As shown in Appendices~\ref{sec:framework} and~\ref{McLennan}, the kinetic equation for the one-body distribution function $f(\bm{R}_1,\bm{V}_1;t)$ can be written as
\begin{equation}
    \left[\frac{\partial}{\partial t}
    -\dot\gamma V_{1,y}\frac{\partial}{\partial V_{1,x}} 
   + \bm{V}_1\cdot \frac{\partial}{\partial \bm{R}_1}
    \right]
    f(1)
    \approx
    \zeta
    \frac{\partial}{\partial \bm{V}_1}\cdot\left[\left(\bm{V}_1 + \frac{T_\mathrm{env}}{m}\frac{\partial}{\partial \bm{V}_1}\right)f(1)\right]
    +J[\bm{V}_1|f^{(2)}(1,2)] ,
    \label{eq:kinetic_eq}
\end{equation}
where $\bm{R}_1\equiv \bm{r}_1-\dot\gamma t y_1 \hat{\bm{e}}_x$,
$(1)$ and $(1,2)$ are the abbreviations of $(\bm{R}_1,\bm{V}_1;t)$ and $(\bm{R}_1,\bm{V}_1,\bm{R}_2,\bm{V}_2;t)$, respectively.
To be consistent with the simulation setup, we ignore the hydrodynamic interaction between particles.
If the interaction is short-ranged and collisions are elastic, $J[\bm{V}|f^{(2)}(1,2)]$ is given by~\cite{Resibois}
\begin{align}\label{J(f^2)}
    &J[\bm{V}_1|f^{(2)}(1,2)]
    = \int d\bm{V}_2 \int d\bm{\Omega} \mathcal{S}(\chi, V_{12})V_{12}
    \left[f^{(2)}(1^{\prime\prime},2^{\prime\prime})
    - f^{(2)}(1,2)\right],
\end{align}
where $V_{12}\equiv | \bm{V}_1-\bm{V}_2|$, $(1^{\prime\prime},2^{\prime\prime})$ expresses a set of pre-collisional positions and velocities of $(1,2)$, $\chi$ is the scattering angle, $d\bm{\Omega}=\sin\chi d\chi d\phi$, and $\mathcal{S}(\chi, V)$ is the differential collision cross section (see also Fig.~\ref{fig:scattering}).
Then, the particle trajectory is bent for $r_\mathrm{min}<r\le d$ and is reflected at $r=r_\mathrm{min}$, where $r_\mathrm{min}$ is the turning point of the trajectory given by Eq.~\eqref{eq:r_min}, which depends on the impact parameter $b$ and the relative speed $v$ between two colliding particles.
The explicit relationship between $\chi$ and the impact parameter $b$ is also presented in Appendix \ref{sec:scattering}.
It is noted that the impact parameter $b$ is related to the differential cross section as $\mathcal{S}(\chi, V)\sin\chi = b|\partial b/\partial \chi|$ \cite{Chapman, Hirschfelder}.

As in the case of hard-core collisions, the relationships between the pre-collisional velocities $(\bm{V}_1^{\prime\prime}, \bm{V}_2^{\prime\prime})$ and the post-collisional ones $(\bm{V}_1, \bm{V}_2)$ are expressed as
\begin{equation}
    \begin{cases}
    \bm{V}_1^{\prime\prime} = \bm{V}_1 - (\bm{V}_{12}\cdot \hat{\bm{k}}_{12})\hat{\bm{k}}_{12}\\
    \bm{V}_2^{\prime\prime} = \bm{V}_2 + (\bm{V}_{12}\cdot \hat{\bm{k}}_{12})\hat{\bm{k}}_{12}
    \end{cases},
    \label{eq:coll_rule}
\end{equation}
where $\hat{\bm{k}}_{12}\equiv \bm{r}_{21}/r_{21}$. 
Note that $(\bm{V}_1^{\prime\prime}, \bm{V}_2^{\prime\prime})$ and $(\bm{V}_1, \bm{V}_2)$ must be measured for $r>d$, i.e. without influence of the soft-core potential.

Now, let us assume that the system is uniform, and thus we can ignore the term $\bm{V}_1\cdot \partial f(\bm{R}_1, \bm{V}_1, t)/\partial \bm{R}_1$ in Eq.~\eqref{eq:kinetic_eq}.
\footnote{If there exists such a term, we cannot get a closed equation for the stress.}
Under this assumption, by multiplying $mV_\alpha V_\beta$ into Eq.~\eqref{eq:kinetic_eq} and integrating over $\bm{V}_1$, one obtains an approximate equation
\begin{align}
    &\frac{d}{d t} \sigma_{\alpha\beta}^k
    +\dot\gamma \left(\delta_{\alpha x}\sigma_{y\beta}^k + \delta_{\beta x}\sigma_{y\alpha}^k\right)
    \approx -2\zeta \left(\sigma_{\alpha\beta}^k +n T_\mathrm{env}\delta_{\alpha\beta}\right) 
    + \Lambda_{\alpha\beta},
    \label{eq:evol_P}
\end{align}
where $\overleftrightarrow{\sigma^k}$ is the kinetic contribution of the stress tensor is expressed as~\footnote{Eq.~\eqref{eq:sigma^k} is alternative expression of the kinetic stress in Eq.~\eqref{eq:sigma_sim} using the distribution function $f(1)$ under the spatially uniform assumption.
}
\begin{equation}\label{eq:sigma^k}
    \overleftrightarrow{\sigma^k} \equiv -m\int d\bm{V} \bm{\mathrm{V}}\bm{\mathrm{V}} f(1),
\end{equation}
and the collision moment by
\begin{equation}
    \overleftrightarrow{\Lambda}
    \equiv -m \int d\bm{V} \bm{\mathrm{V}}\bm{\mathrm{V}} J[\bm{V}|f^{(2)}].
    \label{Lambda_ab}
\end{equation}
Let us introduce the theoretical temperature $T$ and two anisotropic temperatures $\Delta T$ and $\delta T$, respectively, as
\begin{equation}\label{eq:T&DT&dT}
    T= -\frac{\sigma^k_{\alpha\alpha}}{3n},\quad
    \Delta T \equiv -\frac{\sigma^k_{xx}-\sigma^k_{yy}}{n},\quad
    \delta T \equiv -\frac{\sigma^k_{xx}-\sigma^k_{zz}}{n},
\end{equation}
where we have adopted Einstein's convention in which double Greek characters take summation over $x$, $y$, and $z$, i.e. $\sigma^k_{\alpha\alpha}\equiv \sigma^k_{xx}+\sigma^k_{yy}+\sigma^k_{zz}$ throughout this paper~\footnote{
Note that the expression of the kinetic temperature $T$ in Eq.~\eqref{eq:T&DT&dT} can be obtained from Eqs.~\eqref{eq:eta&T} and \eqref{eq:sigma_sim}.}.
Then, Eq.~\eqref{eq:evol_P} can be rewritten as
\begin{subequations}\label{eq:dynamic_eqs}
\begin{align}
    \frac{d T}{d t} 
    &= \frac{2}{3n}\dot\gamma \sigma^k_{xy}
    +2\zeta(T_\mathrm{env}-T)-\frac{1}{3n}\Lambda_{\alpha\alpha},\\
    \frac{d \Delta T}{d t}
    &= \frac{2}{n}\dot\gamma \sigma^k_{xy} -2\zeta\Delta T
    -\frac{\delta \Lambda_{xx}-\delta \Lambda_{yy}}{n},\\
    \frac{d \delta T}{d t}
    &= \frac{2}{n}\dot\gamma \sigma^k_{xy} -2\zeta\delta T
    -\frac{2\delta \Lambda_{xx}+\delta \Lambda_{yy}}{n},\\
    \frac{d \sigma^k_{xy}}{d t}
    &= -\dot\gamma \sigma^k_{yy}-2\zeta \sigma^k_{xy}
    +\Lambda_{xy},
\end{align}
\end{subequations}
where 
\begin{equation}
    \delta \Lambda_{xx} 
    \equiv \Lambda_{xx}-\frac{1}{3}\Lambda_{\alpha\alpha},\quad
    \delta \Lambda_{yy} 
    \equiv \Lambda_{yy}-\frac{1}{3}\Lambda_{\alpha\alpha}.
    \label{eq:delta_Lambda_xx_yy}
\end{equation}
For practical calculations, let us rewrite the set of Eqs.~\eqref{eq:dynamic_eqs} in the dimensionless forms.
Let us introduce the dimensionless temperatures and kinetic deviatoric stress as
\begin{equation}
    \Delta \theta \equiv \frac{\Delta T}{T_\mathrm{env}},\quad
    \delta \theta \equiv \frac{\delta T}{T_\mathrm{env}},
\end{equation}
as well as the dimensionless time $\tau\equiv \dot\gamma t$.
Using these quantities, the set of Eqs.~\eqref{eq:dynamic_eqs} is rewritten as
\begin{subequations}\label{eq:nondim_dynamic_eqs}
\begin{align}
    \frac{d \theta}{d \tau} 
    &= \frac{2}{3}\dot\gamma^* \Pi_{xy}
    +2(1-\theta)-\frac{1}{3}\Lambda_{\alpha\alpha}^*,\\
    \frac{d \Delta \theta}{d \tau}
    &= 2\dot\gamma^* \Pi_{xy} -2\Delta\theta
    -\delta \Lambda_{xx}^* + \delta \Lambda_{yy}^*,\\
    \frac{d \delta \theta}{d \tau}
    &= 2\dot\gamma^* \Pi_{xy} -2\zeta\delta \theta
    -2\delta \Lambda_{xx}^* - \delta \Lambda_{yy}^*,\\
    \frac{d \Pi_{xy}}{d \tau}
    &= -\dot\gamma^* \Pi_{yy}-2\Pi_{xy}
    +\Lambda_{xy}^*,
\end{align}
\end{subequations}
with
\begin{equation}
    \Lambda_{\alpha\beta}^*\equiv \frac{\Lambda_{\alpha\beta}}{n\zeta T_\mathrm{env}},\quad
    \delta \Lambda_{\alpha\beta}^*\equiv \frac{\delta\Lambda_{\alpha\beta}}{n\zeta T_\mathrm{env}}.
    \label{eq:nondim_Lambda_ab}
\end{equation}

So far, we have not used any assumption in the kinetic theory for spatially uniform simple shear flows.
However, these equations Eqs.~\eqref{eq:nondim_dynamic_eqs} cannot be solved since it contains the two-body distribution $f^{(2)}(1,2)$.
In the subsequent subsections, we adopt some assumptions to solve Eqs.~\eqref{eq:nondim_dynamic_eqs}.

\subsection{Enskog approximation}

In this subsection, we adopt the Enskog approximation to obtain a closure of the one-body distribution.
It should be noted that the Enskog approximation is only applicable to hard-core systems because we cannot describe continuous changes in the distribution function for soft-core systems during contact in real soft-core situations.
This problem does not appear in the Boltzmann equation for dilute gases because we do not need to consider the position dependence of the distribution function~\cite{Chapman}.  

Instead of the analysis of real soft-core systems without approximation, we approximate a soft-core collision by a hard-core collision at the turning point $r=r_\mathrm{min}$ introduced in Eq.~\eqref{eq:r_min}. 
Therefore, the collision integral $J[\bm{V}|f^{(2)}(1,2)]$ can be approximated as
\begin{align}
    J[\bm{V}|f^{(2)}(1,2)]
    &\approx \int d\bm{V}_2 \int d\bm{\Omega} \mathcal{S}(\chi, V_{12})V_{12}\nonumber\\
    &\hspace{1em}\times\left[f^{(2)}(\bm{R}_1, \bm{V}_1^{\prime\prime}, \bm{R}_1-r_\mathrm{min}\hat{\bm{k}}_{12}, \bm{V}_2^{\prime\prime};t)
    - f^{(2)}(\bm{R}_1, \bm{V}_1,\bm{R}_1+r_\mathrm{min}\hat{\bm{k}}_{12}, \bm{V}_2;t)\right].
    \label{eq:J_Enskog}
\end{align}
Within this approximation, the effect of softness is absorbed in the differential cross section $\mathcal{S}(\chi, V_{12})$.

Let us adopt the decoupling (Enskog) approximation of the two-body distribution function in which the two-body distribution function can be expressed as a product of the one-body distribution functions multiplied by the correlation function.
Such a procedure is used for the Boltzmann equation for dilute gases and the Enskog equation for moderately dense gases~\cite{Takada20,Sangani96,Koch99,Garzo12,Resibois, Garzo99,Lutsko05,Garzo13,Gonzalez19,Garzo}.  
Since the collision is characterized by the turning point $r=r_\mathrm{min}$, we may utilize the decoupling approximation for hard-core particles as
\begin{equation}
    f^{(2)}(1,2)
    \approx g_0 f(1) f(2),
    \label{eq:decoupling}
\end{equation}
where the radial distribution function at contact $g_0\equiv (1-\varphi/2)/(1-\varphi)^3$ is the empirical dimensionless geometric factor for $\varphi\le 0.49$~\cite{Carnahan69}.
Note that the volume fraction is not measured by the turning point but by the outer boundary of the acting potential force.
It is also noted that the effect of shear on $g_0$ is ignored, though $g_0$ depends on the angle as discussed in Refs.~\cite{Brady97, Suzuki19, Banetta19}.
The last assumption we use is
\begin{equation}
    f\left(\bm{R}\mp r_\mathrm{min}\hat{\bm{k}}_{12}, \bm{v}_1;t\right)
    \approx f\left(\bm{V}_1\pm \dot\gamma r_\mathrm{min}\hat{k}_{12,y}\hat{\bm{e}}_x, t\right) ,
\end{equation}
since we are interested in spatially uniform cases~\cite{Hayakawa17}.
This assumption means that only affine deformations are considered. 
The validity of the above approximations can be verified by the comparison of the theoretical results with numerical results.
Under these assumptions, the kinetic equation \eqref{eq:kinetic_eq} is converted to the Enskog equation for the inertial suspension of frictionless soft-core particles~\cite{Takada20,Sangani96,Koch99,Garzo12,Resibois,Garzo99,Lutsko05,Garzo13,Gonzalez19,Garzo} as
\begin{equation}\label{Enskog_innertial_suspension}
    \left(\frac{\partial}{\partial t}-\dot\gamma V_{1,y} \frac{\partial}{\partial V_{1,x}}\right)f(\bm{V}_1,t)
    =
    \zeta
    \frac{\partial}{\partial \bm{V}_1}\cdot\left[\left(\bm{V}_1 + \frac{T_\mathrm{env}}{m}\frac{\partial}{\partial \bm{V}_1}\right)f(1)\right]
    + J_\mathrm{E}[\bm{V}_1|f, f]
\end{equation}
with
\begin{align}\label{eq:J_E}
    J_\mathrm{E}[\bm{V}_1|f, f]
    &= g_0\int d\bm{V}_2\int d\Omega \mathcal{S}(\chi,V_{12})V_{12}\nonumber\\
    &\hspace{1em}\times
    \left[
    f\left(\bm{V}_1^{\prime\prime},t\right)
    f\left(\bm{V}_2^{\prime\prime}+\dot\gamma r_\mathrm{min} \hat{k}_{12,y}\hat{\bm{e}}_x,t\right)
    -f\left(\bm{V}_1,t\right)
    f\left(\bm{V}_2-\dot\gamma r_\mathrm{min} \hat{k}_{12,y} \hat{\bm{e}}_x,t\right)\right],
\end{align}
where the subscript $\mathrm{E}$ is attached to represent the quantity under the Enskog approximation.
Then, the collision moment~\eqref{Lambda_ab} becomes~\cite{Takada20}
\begin{equation}
    \Lambda_{\alpha\beta}^\mathrm{E}
    = \overline{\Lambda}_{\alpha\beta}^\mathrm{E}
    -\dot\gamma(\delta_{\alpha x}\sigma^c_{\mathrm{H},y\beta} + \delta_{\beta x}\sigma^c_{\mathrm{H},y\alpha}),
\end{equation}
where the quantity $\overline{\Lambda}^\mathrm{E}$ is defined by \cite{Takada20}
\begin{align}
    \overline{\Lambda}_{\alpha\beta}^\mathrm{E}
    &\equiv \frac{m}{2}g_0 \int d\bm{V}_1 \int d\bm{V}_2
    \int d\Omega \mathcal{S}(\chi,V_{12})V_{12}
    (\bm{V}_{12}\cdot \hat{\bm{k}}_{12})\nonumber\\
    &\hspace{1em}
    \times \left[V_{12,\alpha}\hat{k}_{12,\beta} 
    + V_{12,\beta}\hat{k}_{12,\alpha}
    - 2(\bm{V}_{12}\cdot \hat{\bm{k}}_{12})
    \hat{k}_{12,\alpha}\hat{k}_{12,\beta})\right]
    f\left(\bm{V}_1+\dot\gamma r_\mathrm{min} \hat{k}_{12,y} \hat{\bm{e}}_x,t\right)
    f\left(\bm{V}_2,t\right).
    \label{eq:overline_Lambda}
\end{align}
Here, we adopt the approximation that the contact stress can be approximated by the collisional contribution to the stress of a hard-core system with the core diameter $r_\mathrm{min}$ as
\begin{align}
    \sigma_{\mathrm{H},\alpha\beta}^c
    &\equiv -\frac{m}{2}g_0\int d\bm{V}_1 \int d\bm{V}_2 \int d\bm{\Omega}
    \mathcal{S}(\chi, V_{12}) V_{12}
    r_\mathrm{min}(\bm{V}_{12}\cdot \hat{\bm{k}}_{12})
    \hat{k}_{12,\alpha} \hat{k}_{12,\beta}\nonumber\\
    &\hspace{1em}\times f\left(\bm{V}_1+ \frac{1}{2}\dot\gamma r_\mathrm{min}\hat{k}_{12,y}\hat{\bm{e}}_x, t\right)
    f\left(\bm{V}_2- \frac{1}{2}\dot\gamma r_\mathrm{min}\hat{k}_{12,y}\hat{\bm{e}}_x, t\right) ,
    \label{eq:Pc}
\end{align}
where the subscript H stands for the expression for hard-core systems.
Note that this $\sigma_{\mathrm{H},\alpha\beta}^c$ can be obtained once we know the one-body distribution function $f(\bm{V},t)$.
Although the theoretical consistency between Eqs.~\eqref{eq:sigma_sim} and \eqref{eq:Pc} cannot be justified, we regard $\sigma_{\mathrm{S},\alpha\beta}^c$ as the correspondence of $\sigma_{\mathrm{H},\alpha\beta}^c$.
This replacement sufficiently reproduces the simulation results as demonstrated in Appendix~\ref{sec:replacement}.

When we analyze collision processes of soft-core particles, we need to consider the time dependence of differential cross section $\mathcal{S}(\chi, V_{12})$, 
which is too complicated to handle precisely.
As previously mentioned in the previous subsection, the relationship between $b$ and $\chi=\pi-2\vartheta$ is given by
$\mathcal{S}(\chi, V)\sin\chi = b|\partial b/\partial \chi|$.
Using $d\hat{\bm{k}}_{12}=\sin\vartheta d\vartheta d\phi$ (see also Fig.~\ref{fig:scattering}) and that the distribution function may be independent of the angle $\phi$, one can get
\begin{equation}
    \int d\Omega \mathcal{S}(\chi,V_{12})V_{12}
    = 4\int d\vartheta d\phi \sin\vartheta \mathcal{S}(\chi,V_{12})V_{12}\cos\vartheta
    = 4\int d\hat{\bm{k}}_{12} \mathcal{S}(\chi,V_{12})(V_{12}\cdot \hat{\bm{k}}_{12}).
    \label{eq:int_Omega_k12}
\end{equation}
Then, Eq.~\eqref{eq:overline_Lambda} can be rewritten as
\begin{align}
    \overline{\Lambda}_{\alpha\beta}^\mathrm{E}
    &= 4\times \frac{m}{2}g_0 \int d\bm{V}_1 \int d\bm{V}_2
    \int d\hat{\bm{k}}_{12} \mathcal{S}(\chi,V_{12})
    (\bm{V}_{12}\cdot \hat{\bm{k}}_{12})^2\nonumber\\
    &\hspace{1em}
    \times \left[V_{12,\alpha}\hat{k}_{12,\beta} 
    + V_{12,\beta}\hat{k}_{12,\alpha}
    - 2(\bm{V}_{12}\cdot \hat{\bm{k}}_{12})
    \hat{k}_{12,\alpha}\hat{k}_{12,\beta})\right]
    f\left(\bm{V}_1+\dot\gamma r_\mathrm{min} \hat{k}_{12,y} \hat{\bm{e}}_x,t\right)
    f\left(\bm{V}_2,t\right).
    \label{eq:overline_Lambda_ver2}
\end{align}
Since we ignore the continuous change of the distribution function during a contact, we can use
\begin{equation}
    \mathcal{S}(\chi,V_{12})
    \approx \frac{d^2}{4}\Omega_{2,2}^*
    \Theta(\bm{V}_{12}\cdot \hat{\bm{k}}_{12}) ,
    \label{eq:replacement}
\end{equation}
where $\Omega_{2,2}^*$ is defined as~\cite{Sugimoto20, Chapman, Hirschfelder}
\begin{equation}
    \Omega_{k,\ell}^*\left(\frac{T}{\varepsilon}\right)
    \equiv \int_0^\infty dy y^{2\ell+3} e^{-y^2}
    \int_0^1 db^* b^*
    \left[1-\sin^k \chi\left(b^*, 2y\sqrt{T^*}\right)\right]
    \label{eq:def_Omega_kl}
\end{equation}
with $b^*\equiv b/d$ and $T^*\equiv T/\varepsilon$.
Note that $\Omega_{2,2}^*$ expresses the softness. 
The expression of Eq.~\eqref{eq:replacement} contains a decoupling approximation where we ignore the $\Omega$ dependence of the distribution function.
Note that this $\Omega_{k,\ell}^*$ is independent of $\varphi$~\cite{Chapman, Hirschfelder}.

So far, the used approximations are reasonable, but it is still difficult to solve the Enskog equation because $r_\mathrm{min}$ depends on $b^*$ and $T^*$.
Then, we adopt a crucial simplification for $r_\mathrm{min}$ by the replacement with $d$ as
\begin{equation}\label{r_min_to_d}
    r_\mathrm{min}\to d
\end{equation}
for later discussion. 
Although this simplification cannot be justified, as will be shown, the theory based on this approximation leads to reasonable results.

\begin{figure}[htbp]
    \centering
    \includegraphics[width=0.6\linewidth]{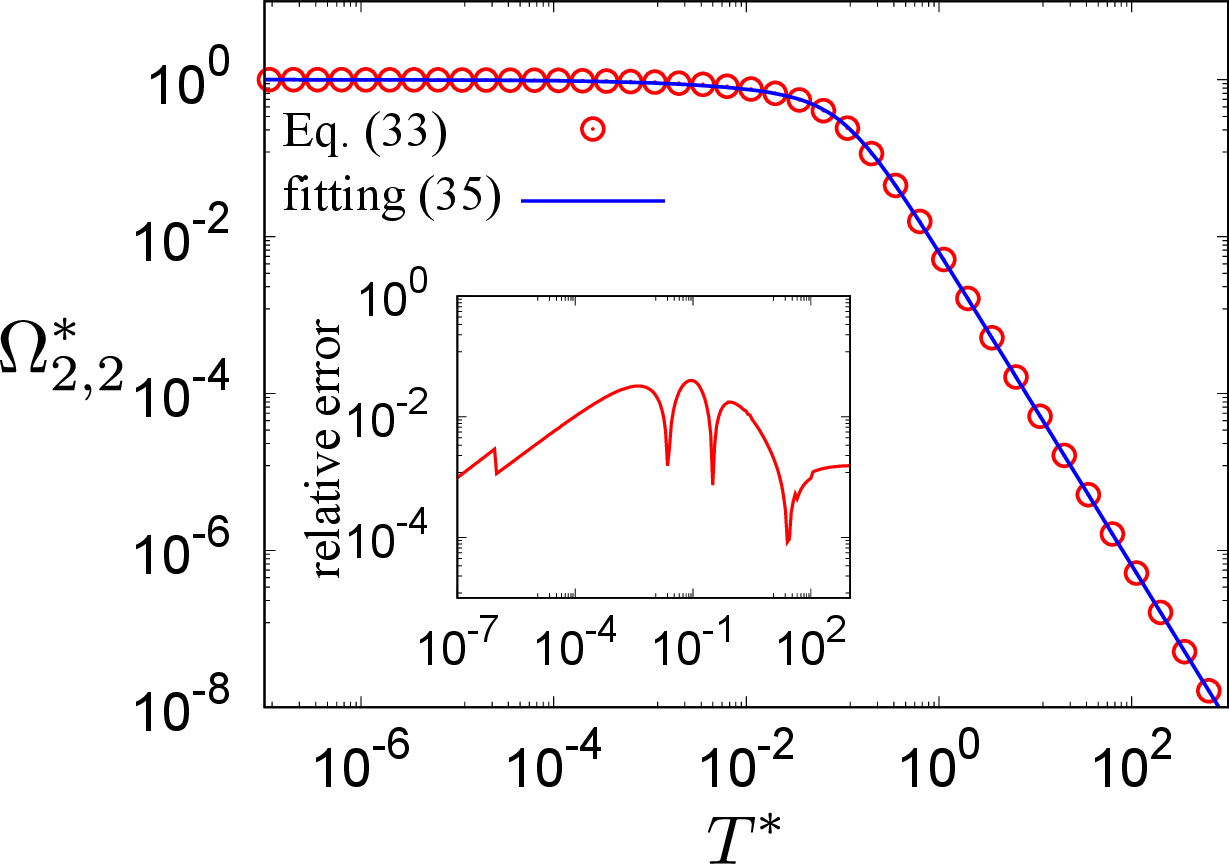}
    \caption{Temperature dependence of $\Omega_{2,2}^*$ given by Eq.~\eqref{eq:def_Omega_kl} (open circles) and its fitting curve $\Omega_{2,2}^{ \mathrm{fit}*}$ given by Eq.~\eqref{eq:Omega22_fitting} (solid line).
    The inset shows the relative error $|\Omega_{2,2}^*-\Omega_{2,2}^{\mathrm{fit}*}|/\Omega_{2,2}^*$ against the temperature.
    }
    \label{fig:Omega22}
\end{figure}

The temperature dependence of $\Omega_{2,2}^*$ is plotted in Fig.~\ref{fig:Omega22} as a theoretical result.
The softness parameter of the differential cross section $\Omega_{2,2}^*$ becomes $1$ in the low-temperature limit. 
This is because the kinetic energy is sufficiently smaller than the potential energy in this limit, which means that the trajectories of particles become almost the same as those of hard-core particles.
$\Omega_{2,2}^*$ also behaves as $\Omega_{2,2}^*\sim (T/\varepsilon)^{-2}$ and $1-\Omega_{2,2}^*\sim (T/\varepsilon)^{1/2}$ in the high and low temperature regimes, respectively~\cite{Sugimoto20}.

We also find that the temperature dependence of $\Omega_{2,2}^*$ can be well fitted by:
\begin{equation}
    \Omega_{2,2}^{\mathrm{fit}*}(T^*)=
    \frac{1}{1+a_0\sqrt{T^*}+a_1T^*+a_2 T^{*2}},
    \label{eq:Omega22_fitting}
\end{equation}
with 
\begin{equation}\label{a_0-a_2}
    a_0 \approx 2.6206,\quad
    a_1 \approx 0.39208,\quad
    a_2 \approx 154.37.
\end{equation}
The relative error $|\Omega_{2,2}^*-\Omega_{2,2}^{\mathrm{fit}*}|/\Omega_{2,2}^*$ is smaller than $2\times 10^{-2}$ as shown in the inset of Fig.~\ref{fig:Omega22}.
Thus, it is sufficient to use Eq.~\eqref{eq:Omega22_fitting} for the practical usage of $\Omega_{2,2}^*$.

After using the replacement \eqref{r_min_to_d}, Eq.~\eqref{eq:overline_Lambda_ver2} is rewritten as
\begin{align}
    \overline{\Lambda}_{\alpha\beta}^\mathrm{E}
    &= \frac{m}{2}d^3g_0\Omega_{2,2}^* \int d\bm{V}_1 \int d\bm{V}_2
    \int d\hat{\bm{k}}_{12}
    \Theta(\bm{V}_{12}\cdot \hat{\bm{k}}_{12})
    (\bm{V}_{12}\cdot \hat{\bm{k}}_{12})^2\nonumber\\
    &\hspace{1em}
    \times \left[V_{12,\alpha}\hat{k}_{12,\beta} 
    + V_{12,\beta}\hat{k}_{12,\alpha}
    - 2(\bm{V}_{12}\cdot \hat{\bm{k}}_{12})
    \hat{k}_{12,\alpha}\hat{k}_{12,\beta})\right]
    f\left(\bm{V}_1+\dot\gamma d \hat{k}_{12,y} \hat{\bm{e}}_x,t\right)
    f\left(\bm{V}_2,t\right).
    \label{eq:overline_Lambda_ver3}
\end{align}
Substituting Eqs.~\eqref{eq:replacement}--\eqref{a_0-a_2} to Eq.~\eqref{eq:Pc}, we obtain an approximate expression for the collisional contribution to the stress as
\begin{align}
    \sigma_{\mathrm{H},\alpha\beta}^c
    &\approx -\frac{m}{2}d^3 g_0\Omega_{2,2}^*
    \int d\bm{V}_1 \int d\bm{V}_2 \int d\hat{\bm{k}}_{12}
    \Theta(\bm{V}_{12}\cdot \hat{\bm{k}}_{12})
    (\bm{V}_{12}\cdot \hat{\bm{k}}_{12})^2
    \hat{k}_{12,\alpha} \hat{k}_{12,\beta}\nonumber\\
    &\hspace{1em}\times f\left(\bm{V}_1+ \frac{1}{2}\dot\gamma d\hat{k}_{12,y}\hat{\bm{e}}_x, t\right)
    f\left(\bm{V}_2- \frac{1}{2}\dot\gamma d\hat{k}_{12,y}\hat{\bm{e}}_x, t\right).
    \label{eq:Pc_2}
\end{align}

\subsection{Rheology based on Grad's approximation}

Although we have adopted several assumptions to simplify the calculation in the previous subsection, it is still difficult to solve a set of equations~\eqref{eq:dynamic_eqs} because the collision moment $\Lambda_{\alpha\beta}$ is an integral of a nonlinear function of $f(\bm{V},t)$.
It is known that Grad's approximation~\cite{Grad49,Garzo,Torrilhon16}
\begin{equation}
    f(\bm{V},t)\approx 
    n v_\mathrm{T}^{-3}\tilde{f}_\mathrm{M}(\bm{c},t)\left(1-\frac{\Pi_{\alpha\beta}}{\theta}c_\alpha c_\beta\right) 
    \label{eq:Grad}
\end{equation}
with 
\begin{align}
    \bm{c} &\equiv \frac{\bm{V}}{v_\mathrm{T}},
    \quad
    v_\mathrm{T} \equiv \displaystyle\sqrt{\frac{2T}{m}},
\end{align}
and the dimensionless Maxwell distribution
\begin{equation}
    \tilde{f}_\mathrm{M}(\bm{c},t)
    =\pi ^{-3/2}
    \exp\left(-c^2\right),
\end{equation}
yields good approximations for hard-core~\cite{Hayakawa19, Hayakawa17, Garzo13, Garzo, Garzo02, Chamorro15,Takada20,Santos04, Takada18, Lun84, Jenkins852D, Jenkins853D, Takada21_Mpemba, Takada21_binary}  
and dilute soft-core~\cite{Sugimoto20, Takada21PG} systems.
An extended Grad's approximation is also used for non-Brownian suspensions, in which $\Pi_{\alpha\beta}$ is replaced with the counterpart of the contact stress~\cite{Suzuki19}.
We adopt Grad's approximation for moderately dense inertial suspensions consisting of soft-core particles in this paper.

We follow the parallel procedure to those used in Refs.~\cite{Takada20,Santos98,Montanero99},
where the collisional moment and the contact stress are written in a series of an expansion parameter given by
\begin{equation}
    \tilde{\dot{\gamma}}\equiv \frac{\dot\gamma^*}{\xi_\mathrm{env}\sqrt{\theta}}.
\end{equation}
Thus, the corresponding terms in Eq.~\eqref{eq:nondim_Lambda_ab} are, respectively, given by 
\footnote{The expressions of the coefficients $C_1^{(i)}$, $C_2^{(i)}$, $C_3^{(i)}$, and $C_4^{(i)}$ in Eqs.~\eqref{eq:Lambda_expansion} and $C_5^{(i)}$ in Eq.~\eqref{eq:sigma_expansion} are equivalent to $\widetilde{\Lambda}_{\alpha\alpha}^{(i)*}$, $\widetilde{\Lambda}_{xy}^{(i)*}$, $\delta\widetilde{\Lambda}_{xx}^{(i)*}$, $\delta\widetilde{\Lambda}_{yy}^{(i)*}$, and $\widetilde{\Pi}_{\mathrm{H},xy}^{(i)*}$ given by Eqs.~(3.11) and listed in Tables I and II of Ref.\ \cite{Takada20}, respectively.}
\begin{subequations}\label{eq:Lambda_expansion}
\begin{align}
    \Lambda_{\alpha\alpha}^*
    &\approx \varphi g_0 \xi_\mathrm{env} \theta^{3/2}\Omega_{2,2}^*
    \sum_{i=0}^{N_\mathrm{c}}
    C_1^{(i)}\tilde{\dot\gamma}^i,&\quad
    \Lambda_{xy}^*
    &\approx \varphi g_0 \xi_\mathrm{env} \theta^{3/2}\Omega_{2,2}^*
    \sum_{i=0}^{N_\mathrm{c}}
    C_2^{(i)}\tilde{\dot\gamma}^i,\\
    \delta\Lambda_{xx}^*
    &\approx \varphi g_0 \xi_\mathrm{env} \theta^{3/2}\Omega_{2,2}^*
    \sum_{i=0}^{N_\mathrm{c}}
    C_3^{(i)}\tilde{\dot\gamma}^i,&\quad
    \delta\Lambda_{yy}^*
    &\approx \varphi g_0 \xi_\mathrm{env} \theta^{3/2}\Omega_{2,2}^*
    \sum_{i=0}^{N_\mathrm{c}}
    C_4^{(i)}\tilde{\dot\gamma}^i.
\end{align}
\end{subequations}
Strictly speaking, we should set $N_\mathrm{c}\to\infty$ in Eqs.~\eqref{eq:Lambda_expansion}.
This treatment with $N_\mathrm{c}\to \infty$ has been performed for sheared dry granular gases with hard-core particles~\cite{Iizuka}, but nobody has succeeded for our setup. 
However, it is sufficient to terminate the calculation at a finite order $N_\mathrm{c}$ for practical use.
This is because the convergence is fast enough when the collisions are elastic (see Appendix \ref{sec:convergence}).
In this paper, we adopt $N_\mathrm{c}=2$.

After putting some assumptions explained in the previous subsections, we can rewrite the set of (dimensionless) dynamic equations~\eqref{eq:nondim_dynamic_eqs}.
We note that the coefficients $C_1^{(i)}$, $C_2^{(i)}$, $C_3^{(i)}$, and $C_4^{(i)}$ in Eqs.~\eqref{eq:Lambda_expansion} can be written in terms of these dimensionless quantities \cite{Takada20,Santos98,Montanero99}.
Then, with the aid of Eqs.~\eqref{eq:Lambda_expansion}, the set of (dimensionless) dynamic equations \eqref{eq:nondim_dynamic_eqs} is rewritten as
\begin{subequations}\label{eq:dynamic_eqs_ver2}
\begin{align}
    \frac{d \theta}{d\tau}
    &= \frac{2}{3}\dot\gamma^* \Pi_{xy}
    +2(1-\theta)-\frac{1}{3}\varphi g_0 \xi_\mathrm{env}\theta^{3/2}\Omega_{2,2}^*
    \sum_{i=0}^{N_\mathrm{c}}
    C_1^{(i)}\tilde{\dot\gamma}^i,\\
    \frac{d \Delta\theta}{d\tau}
    &= 2\dot\gamma^* \Pi_{xy} -2\Delta\theta
    - \varphi g_0 \xi_\mathrm{env}\theta^{3/2}\Omega_{2,2}^*
    \sum_{i=0}^{N_\mathrm{c}}
    \left(C_3^{(i)}-C_4^{(i)}\right)\tilde{\dot\gamma}^i,\\
    \frac{d\delta\theta}{d\tau}
    &= 2\dot\gamma^* \Pi_{xy} -2\delta\theta
    - \varphi g_0 \xi_\mathrm{env}\theta^{3/2}\Omega_{2,2}^*
    \sum_{i=0}^{N_\mathrm{c}}
    \left(2C_3^{(i)}+C_4^{(i)}\right)\tilde{\dot\gamma}^i,\\
    \frac{d \Pi_{xy}}{d\tau}
    &= \dot\gamma^* \left(\theta-\frac{2}{3}\Delta\theta + \frac{1}{3}\delta \theta\right)-2\Pi_{xy}
    +\varphi g_0 \xi_\mathrm{env}\theta^{3/2}\Omega_{2,2}^*
    \sum_{i=0}^{N_\mathrm{c}}
    C_2^{(i)}\tilde{\dot\gamma}^i.
\end{align}
\end{subequations}
We can solve the set of equations \eqref{eq:dynamic_eqs_ver2} numerically to obtain $\theta$, $\Delta \theta$, $\delta\theta$, and $\Pi_{xy}$.
Although the metastable state should be a steady-state solution of Eqs.~\eqref{eq:dynamic_eqs_ver2}, the finding of the metastable solution is numerically non-trivial. 
The metastable state at a given shear rate is obtained from steady states with a slightly different shear rate by gradually changing the shear rate on the low-shear and high-shear sides with a rate of change $10^{0.02}$ when the shear rate is increased and $10^{-0.02}$ when decreased.
This process is used to draw the flow curve.
Substituting the steady-state solutions of Eqs.~\eqref{eq:dynamic_eqs_ver2} into Eq.~\eqref{eq:Grad} we can obtain the approximate velocity distribution function.
Then, inserting this distribution function into Eq.~\eqref{eq:Pc_2}, we can determine the collisional contribution to the stress as a series of $\tilde{\dot\gamma}$ as
\begin{equation}\label{eq:sigma_expansion}
    \sigma_{\mathrm{H},xy}^c
    \approx \frac{\pi}{6}n^2d^3g_0 T\Omega_{2,2}^*
    \sum_{i=0}^{N_\mathrm{c}}
    C_5^{(i)}\tilde{\dot\gamma}^i.
\end{equation}

Therefore, we obtain the approximate expression of the dimensionless viscosity as
\begin{equation}
    \eta^*
    \approx \frac{\Pi_{xy}+\Pi_{\mathrm{H},xy}^c}{\dot\gamma^*},
    \label{eq:eta_linear}
\end{equation}
with
\begin{equation}
    \Pi_{\mathrm{H},xy}^c
    \equiv \frac{\sigma_{\mathrm{H},xy}^c}{nT_\mathrm{env}}.
    \label{eq:nondim_quantities}
\end{equation}

To close this section, we list the assumptions used in the theoretical analysis:
(i) We ignore the spatial inhomogeneity in the simple shear flow.
(ii) We adopt the Enskog approximation even for soft-core systems, in which we regard the turning diameter $r_\mathrm{min}$ as the impact-speed dependent inner-hard core. Then, the contact stress for the soft-core system is approximated by that for the hard-core system in Eq.~\eqref{eq:Pc}, which is evaluated by the product of the one-body distribution function.
(iii) Using the decoupling approximation, the differential cross section is reduced to $\Omega_{2,2}^*d^2/4$ as in Eq.~\eqref{eq:replacement}.
To evaluate $\Omega_{2,2}^*$ we adopt a Pad\'{e} approximation as in Eq.~\eqref{eq:Omega22_fitting}. 
(iv) We replace $r_\mathrm{min}$ with $d$ as in Eq.~\eqref{r_min_to_d} after we extract the soft-core form factor $\Omega_{2,2}^*$.
This is the most crucial approximation.
(v) We adopt Grad's approximation Eq.~\eqref{eq:Grad}.
(vi) We have introduced the truncation in Eqs.~\eqref{eq:Lambda_expansion}.
We note that assumptions (i) and (v) are commonly used even for the dilute system \cite{Sugimoto20} and assumption (vi) is adopted for the moderately dense hard-core system~\cite{Hayakawa17, Takada20, Montanero99, Santos98}, while the other assumptions (ii)--(iv) are newly introduced for moderately dense soft-core systems in this paper.

\section{Steady rheology}\label{rheology}

In this section, we present the theoretical predictions of steady rheology based on the solutions of a set of dynamic equations~\eqref{eq:dynamic_eqs_ver2} and compare the results with those by the simulation.
As already mentioned, we adopt $N_\mathrm{c}=2$ for the analysis in Eqs.~\eqref{eq:Lambda_expansion} and \eqref{eq:sigma_expansion}.
Although the results of the linear theory with $N_\mathrm{c}=1$ are a little deviated from those of $N_\mathrm{c}\ge 2$, there are some analytic expressions in the linear theory as in Appendix \ref{app:linear_theory}.
As explained in Sect.~\ref{Langevin_model}, we study the rheology in the ranges of $0<\varphi\le 0.5$, $0.1\le \dot\gamma^*\le 100$, $10^{-1}\le \varepsilon^*\le 10^8$, and $10^{-1}\le\xi_\mathrm{env}\le10^2$.

\begin{figure}[htbp]
    \centering
    \includegraphics[width=\linewidth]{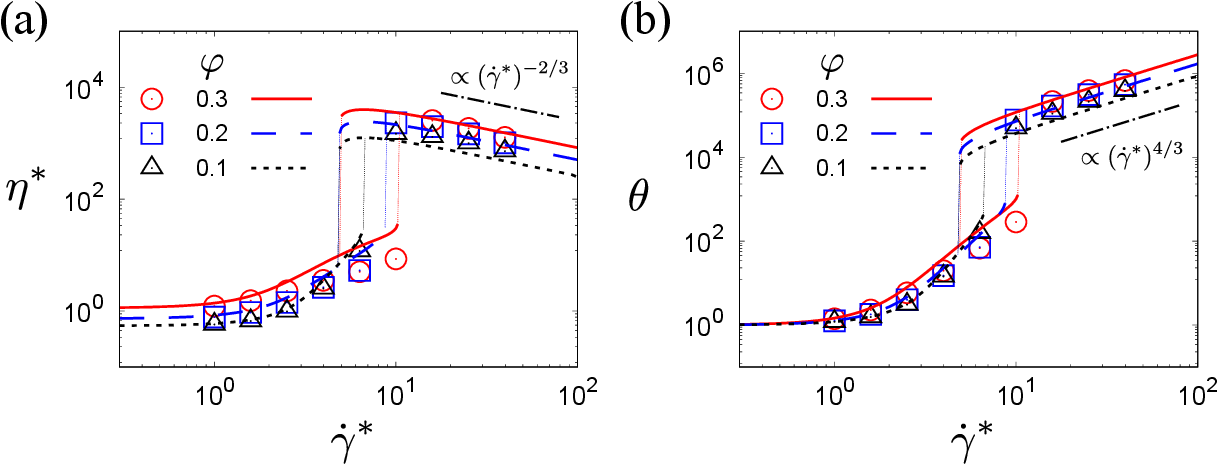}
    \caption{Plots of the theoretical (a) $\eta^*$ and (b) $\theta$ against $\dot\gamma^*$ for $\varphi=0.10$ (dotted line), $0.20$ (dashed line), and $0.30$ (solid line) with fixing $\xi_\mathrm{env}=1.0$, $\varepsilon^*=10^4$, and $N_\mathrm{c}=2$, where the theoretical curves are obtained by Eqs.~\eqref{eq:dynamic_eqs_ver2}--\eqref{eq:nondim_quantities}.
    The left and the right vertical dotted lines are discontinuous changes of $\eta^*$ and $\theta$ from the CST and the exploded phases, respectively.
    The symbols for circles ($\varphi=0.3$), squares ($\varphi=0.20$), and triangles ($\varphi=0.1$) correspond to the simulation results.
    The guide lines represent $\eta^*\propto (\dot\gamma^*)^{-2/3}$ and $\theta\propto (\dot\gamma^*)^{4/3}$, respectively.}
    \label{fig:flow_curves_dilute}
\end{figure}
\begin{figure}[htbp]
    \centering
    \includegraphics[width=\linewidth]{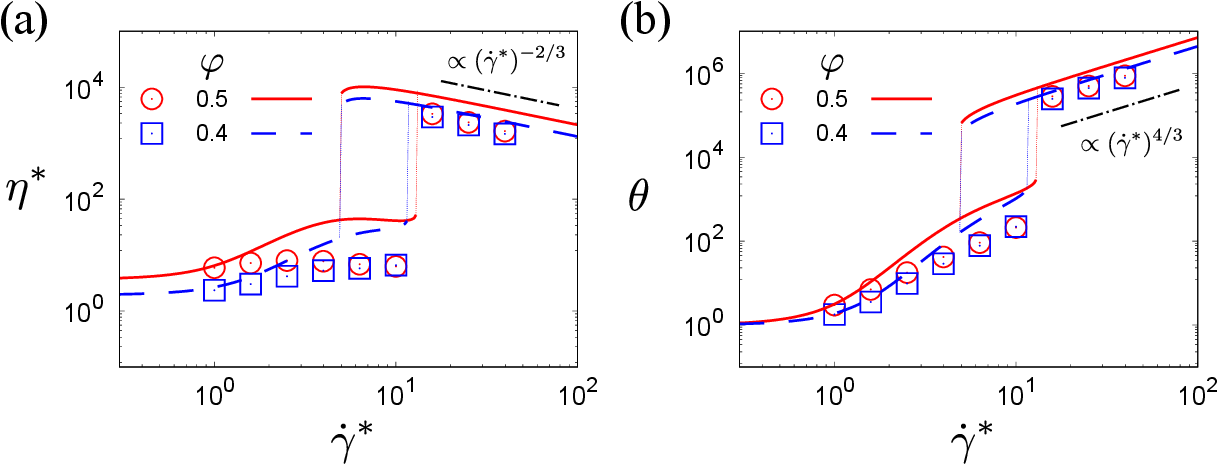}
    \caption{Plots of the theoretical (a) $\eta^*$ and (b) $\theta$ against $\dot\gamma^*$ for $\varphi=0.40$ (dotted line) and $0.50$ (dashed line) with fixing $\xi_\mathrm{env}=1.0$, $\varepsilon^*=10^4$, and $N_\mathrm{c}=2$, where the theoretical curves are obtained by Eqs.~\eqref{eq:dynamic_eqs_ver2}--\eqref{eq:nondim_quantities}.
    The left and the right vertical dotted lines are discontinuous changes of $\eta^*$ and $\theta$ from the CST and the exploded phases, respectively.
    The symbols for circles ($\varphi=0.50$) and squares ($\varphi=0.40$) correspond to the simulation results.
    The guide lines represent $\eta^*\propto (\dot\gamma^*)^{-2/3}$ and $\theta\propto (\dot\gamma^*)^{4/3}$, respectively.}
    \label{fig:flow_curves_dense}
\end{figure}

Figures \ref{fig:flow_curves_dilute} and \ref{fig:flow_curves_dense} present a comparison between the theoretical predictions for $N_\mathrm{c}=2$ regarding $\eta^*$ and $\theta$ as functions of $\dot\gamma^*$, and the simulation results for particle volume fractions $\varphi = 0.10$, $0.20$, and $0.30$ in Fig.~\ref{fig:flow_curves_dilute}, and $\varphi = 0.40$ and $0.50$ in Fig.~\ref{fig:flow_curves_dense}, with $\varepsilon^* = 10^4$ and $\xi_\mathrm{env} = 1.0$. 
Note that there is no ignited-quenched transition for these densities, i.e., we can observe only CST for hard-core systems~\cite{Saha17, Hayakawa17}.
The theoretical values of $\eta^*$ and $\theta$ are derived by solving the system of equations in Eqs.~\eqref{eq:dynamic_eqs_ver2} with the assistance of Eq.~\eqref{eq:eta_linear}. 
These equations are numerically solved starting with an initial kinetic temperature of $\theta = 1$ at $\tau = 0$. 
As shown in the figures, the theoretical results exhibit reasonable agreement with the simulation data even for $\varphi = 0.50$, although the agreement is somewhat less precise for the dense suspensions compared to the dilute case. 
It is important to note that all theoretical flow curves exhibiting DST-like transitions are associated with hysteresis.
This means that the CST phase changes discontinuously to the exploded phase at $\dot\gamma=\dot\gamma_{\mathrm{c}_{\underrightarrow{\mathrm{Ex}}}}\approx 10$ when increasing the shear rate quasi-stationary, and conversely, the discontinuous jump from the exploded to the CST phases at $\dot\gamma=\dot\gamma_{\mathrm{c}_{\underleftarrow{\mathrm{Ex}}}}\approx 6(\lesssim \dot\gamma_{\mathrm{c}_{\underrightarrow{\mathrm{Ex}}}})$ when decreasing.
For clarity, we refer to these transitions as DST or DST-like changes in subsequent discussions.

Remarkably, DST-like behavior is observed across a broad range of volume fractions, from the dilute limit~\cite{Sugimoto20} up to $\varphi = 0.50$, as illustrated in Figs.~\ref{fig:flow_curves_dilute} and \ref{fig:flow_curves_dense}. 
This finding contrasts with conventional DST, which is typically observed only in dense colloidal suspensions~\cite{Wagner09,Brown14,Ness22,Seto13,Mari15}. 
Furthermore, the discontinuous transitions of $\eta^*$ and $\theta$ observed here across such a wide range of $\varphi$ differ from those in hard-core inertial suspensions, where DST transitions only occur above a critical volume fraction, typically a few percent~\cite{Sangani96,Saha17,Hayakawa19, Hayakawa17, Takada20}. 
Our results suggest that the DST-like transitions observed in this study are driven by the softness of the particles and their inertial effects. 
Additionally, the system exhibits shear thinning behavior in the exploded phase. 
As reported in Ref.~\cite{Sugimoto20}, the quantity $\Omega_{2,2}^*$ follows the scaling law $\Omega_{2,2}^* \sim \theta^{-2}$ for $\dot\gamma^* \gg 1$. 
Consequently, the viscosity and temperature should scale as $\eta^* \sim (\dot\gamma)^{*-2/3}$ and $\theta \sim (\dot\gamma)^{*4/3}$, respectively, in the exploded phase. 
These theoretical predictions align well with the simulation results shown in Figs.~\ref{fig:flow_curves_dilute} and \ref{fig:flow_curves_dense}. 
Thus, the crucial assumption in Eq.~\eqref{r_min_to_d} does not invite significant discrepancies between the theoretical and simulation results.

\begin{figure}[htbp]
    \centering
    \includegraphics[width=\linewidth]{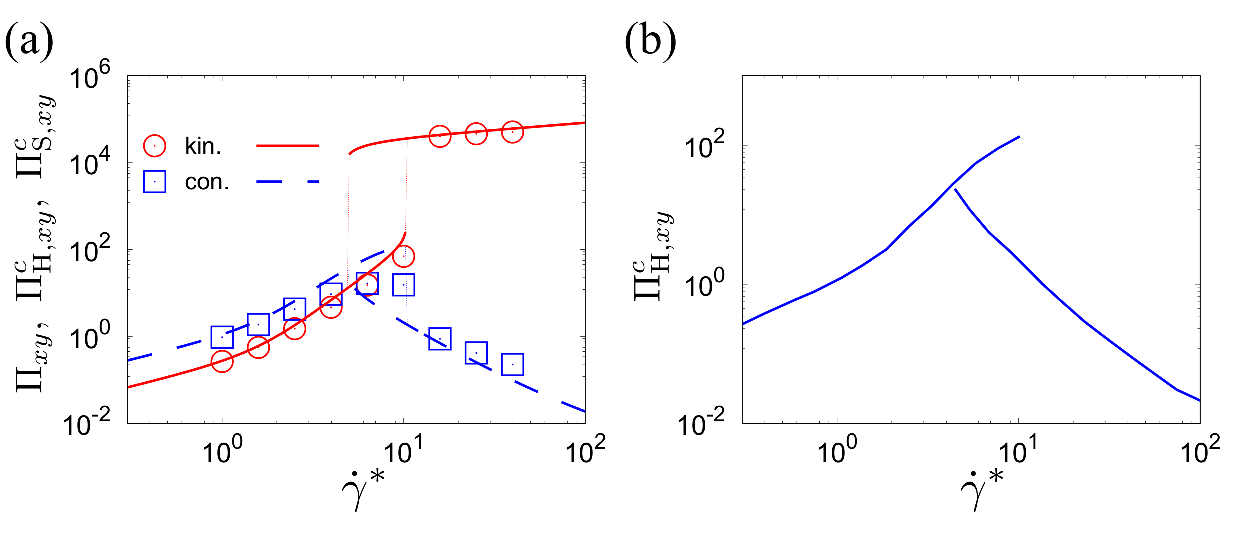}
    \caption{Plots of the theoretical kinetic shear stress $\Pi_{xy}$ (red solid line) and contact shear stress $\Pi_{\mathrm{H},xy}^{c}$ (blue dashed line) against $\dot\gamma^*$ for $\varphi=0.30$ with fixing $\xi_\mathrm{env}=1.0$, $\varepsilon^*=10^4$, and $N_\mathrm{c}=2$, where the theoretical curves are obtained by Eqs.~\eqref{eq:dynamic_eqs_ver2}--\eqref{eq:nondim_quantities}.
    The symbols (circles for the kinetic stress and squares for the contact stress $\Pi_{\mathrm{S},xy}^{c}$) correspond to the simulation results.
    The vertical dotted lines are discontinuous changes of the shear stress between the CST and the exploded phases.
    }
    \label{fig:shear_stress}
\end{figure}

Interestingly, there also exist two branches of the contact stress $\Pi_{xy}^c$ as shown in Fig.~\ref{fig:shear_stress}.
These two branches jump discontinuously at $\dot\gamma^*=\dot\gamma_{\mathrm{c}_{\underleftarrow{\mathrm{Ex}}}}^*\approx 7$ and $\dot\gamma_{\mathrm{c}_{\underrightarrow{\mathrm{Ex}}}}^*\approx 10$, where $d \eta^*/d \dot\gamma^*$ diverges.
These points correspond to where the viscosity or the temperature also jump.
When we focus on the exploded phase, the contact stress $\Pi^c_{xy}$ is decreasing against the shear rate while the kinetic stress abruptly increases in this region as shown in Fig.~\ref{fig:shear_stress}.
This trend is because the duration time becomes negligibly small when the kinetic temperature becomes sufficiently large~\cite{Sugimoto20}.
Thus, the DST can be observed only when we consider the kinetic stress.
Our result is consistent with Kawasaki {\it et al.}~\cite{Kawasaki14} which presented only the contact stress.

\begin{figure}[htbp]
    \centering
    \includegraphics[width=\linewidth]{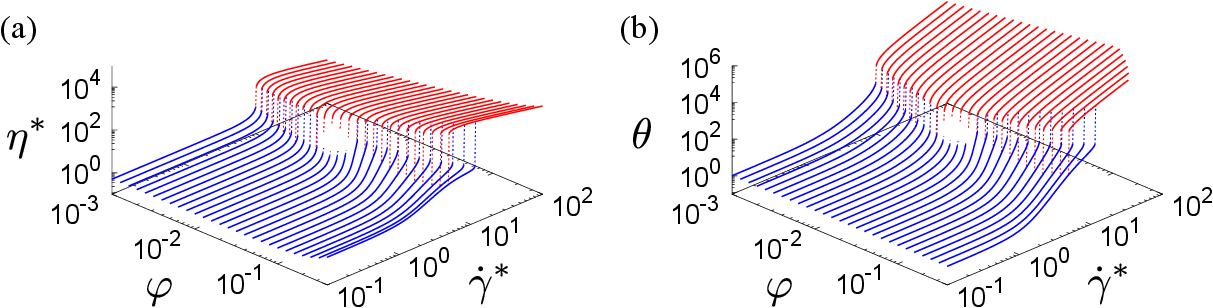}
    \caption{The theoretical flow curves for (a) $\eta^*$ and (b) $\theta$ against $\varphi$ and $\dot\gamma^*$ for $\xi_\mathrm{env}=1.0$ and $\varepsilon^*=10^4$, respectively, where the theoretical curves are obtained by Eqs.~\eqref{eq:dynamic_eqs_ver2}--\eqref{eq:nondim_quantities}.
    Here, the red and the blue solid lines express the flow curves in the exploded and quenched phases, respectively. 
    }
    \label{fig:flow_curves_3D_1e4}
\end{figure}
\begin{figure}[htbp]
    \centering
    \includegraphics[width=\linewidth]{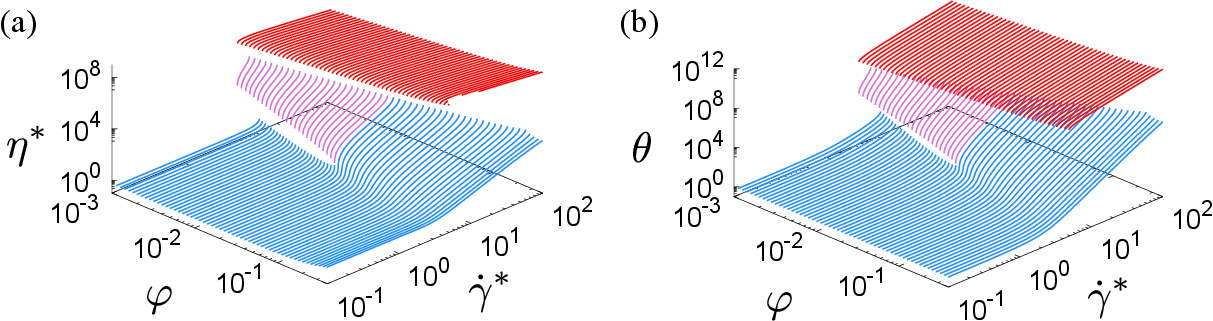}
    \caption{The theoretical flow curves for (a) $\eta^*$ and (b) $\theta$ against $\varphi$ and $\dot\gamma^*$ for $\xi_\mathrm{env}=1.0$ and $\varepsilon^*=10^8$, respectively, where the theoretical curves are obtained by Eqs.~\eqref{eq:dynamic_eqs_ver2}--\eqref{eq:nondim_quantities}.
    Here, the red and the purple lines represent the exploded and the ignited phases, respectively.
    The blue lines correspond to the quenched phase for $\varphi\lesssim 1.8\times10^{-2}$ while the quenched and the ignited phases merge and become continuous for $\varphi\gtrsim 1.8\times 10^{-2}$.}
    \label{fig:flow_curves_3D_1e8}
\end{figure}

Figures \ref{fig:flow_curves_3D_1e4} and \ref{fig:flow_curves_3D_1e8} also exhibit how the DST-like transitions of $\eta^*$ and $\theta$ depend on $\varphi$ and $\dot\gamma^*$ for $\varepsilon^*=10^4$ and $10^8$, respectively, with fixing $\xi_\mathrm{env}=1.0$ based on the kinetic theory with $N_\mathrm{c}=2$ in Eqs.~\eqref{eq:Lambda_expansion}.
As indicated in Ref.~\cite{Sugimoto20}, there are two-step DST-like changes of $\eta^*$ and $\theta$ when the softness becomes larger.
When we focus on the results for $\varepsilon^*=10^8$ in Fig.~\ref{fig:flow_curves_3D_1e8}, the CST phase separates into the quenched and the ignited phases for $\varphi\lesssim 1.8\times 10^{-2}$.
This change is discontinuous at $\dot\gamma=\dot\gamma_{\mathrm{c}_{\mathrm{Q}\leftarrow\mathrm{I}}}$ when increasing the shear rate and $\dot\gamma=\dot\gamma_{\mathrm{c}_{\mathrm{Q}\rightarrow\mathrm{I}}}(\lesssim \dot\gamma_{\mathrm{c}_{\mathrm{Q}\leftarrow\mathrm{I}}})$ when decreasing the shear rate.
This means that this change also exhibits the hysteresis.
Here, the disappearance of the first DST-like change when increasing the volume fraction $\varphi$ corresponds to the disappearance of the DST-like behavior in hard-core systems~\cite{Saha17, Hayakawa17}.
On the other hand, the second DST-like change survives in the wide range of $\varphi$ for larger $\varepsilon^*$ (see Fig.~\ref{fig:flow_curves_3D_1e8}), which is newly observed in this paper.

\begin{figure}[htbp]
    \centering
    \includegraphics[width=\linewidth]{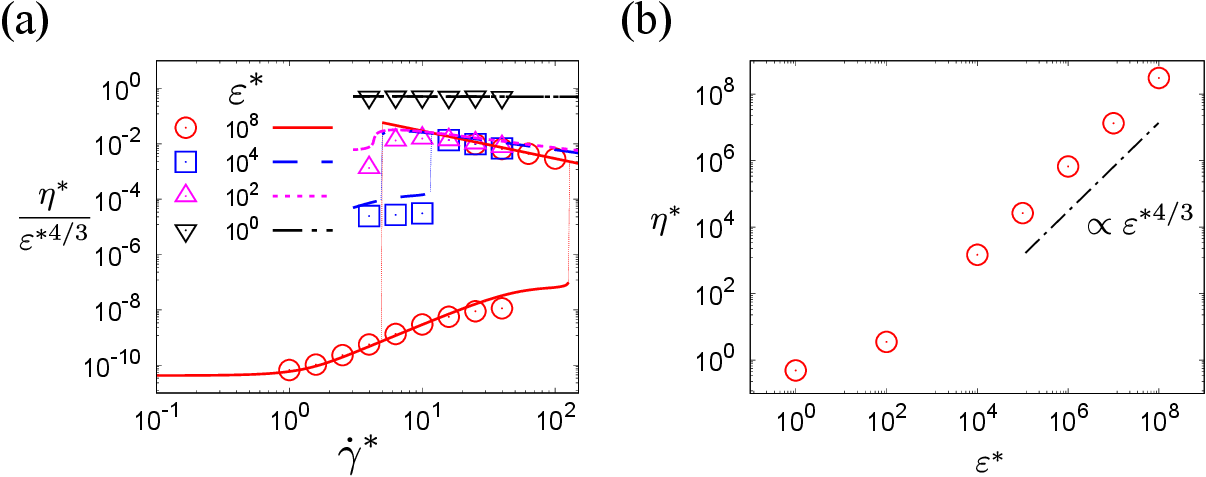}
    \caption{(a) Plots of $\eta^*/\varepsilon^{*4/3}$ against $\dot\gamma^*$ for various $\varepsilon^*$ with fixing $\varphi=0.40$ and $\xi_\mathrm{env}=1.0$.
    The theoretical results are expressed as lines (red solid lines for $\varepsilon^*=10^8$, blue dashed line for $\varepsilon^*=10^4$, purple dotted line for $\varepsilon^*=10^2$, and black chain line for $\varepsilon^*=1$), where the theoretical curves are obtained by Eqs.~\eqref{eq:dynamic_eqs_ver2}--\eqref{eq:nondim_quantities}.
    The symbols (circles for $\varepsilon^*=10^8$, squares for $\varepsilon^*=10^4$, purple upper triangles for $\varepsilon^*=10^2$ and black lower triangles for $\varepsilon^*=1$) are obtained by the simulation. 
    The vertical dotted lines are discontinuous changes of $\eta^*$ between the CST and the exploded phases.
    (b) Plot of the viscosity in the exploded phase at $\dot\gamma^*=10^{1.6}$ against the softness $\varepsilon^*$ when we fix $\varphi=0.40$ and $\xi_\mathrm{env}=1.0$.
    The guided line represents $\eta^*\propto (\varepsilon^*)^{4/3}$.
    }
    \label{fig:flow_curves_scaled}
\end{figure}

These results are counterintuitive, but as shown in Fig.~\ref{fig:flow_curves_scaled}(a), $\eta^*$ can be scaled by $\epsilon^{*4/3}$ in the vicinity of $\dot\gamma_\mathrm{c}^*$ except for extremely soft particles (such as $\varepsilon^*=1$).
The scaling $\eta^*\sim \epsilon^{*4/3}$ can be confirmed in Fig.~\ref{fig:flow_curves_scaled}(b), where the softness dependence of the viscosity at $\dot\gamma^*=10^{1.6}$ is plotted.
This means that the viscosity in the exploded phase diverges in the hard-core limit.
Thus, we can only observe the CST for moderately dense inertial suspensions consisting of hard-core particles \cite{Hayakawa17, Takada20}.
As mentioned above, the ignited phase disappears when the particles become softer as shown in Fig.~\ref{fig:flow_curves_3D_1e4}.
This is because the values of the exploded phase decreases as the softness $\varepsilon^*$ becomes smaller, then the ignited and the exploded phases merge together.

\begin{figure}[htbp]
    \centering
    \includegraphics[width=\linewidth]{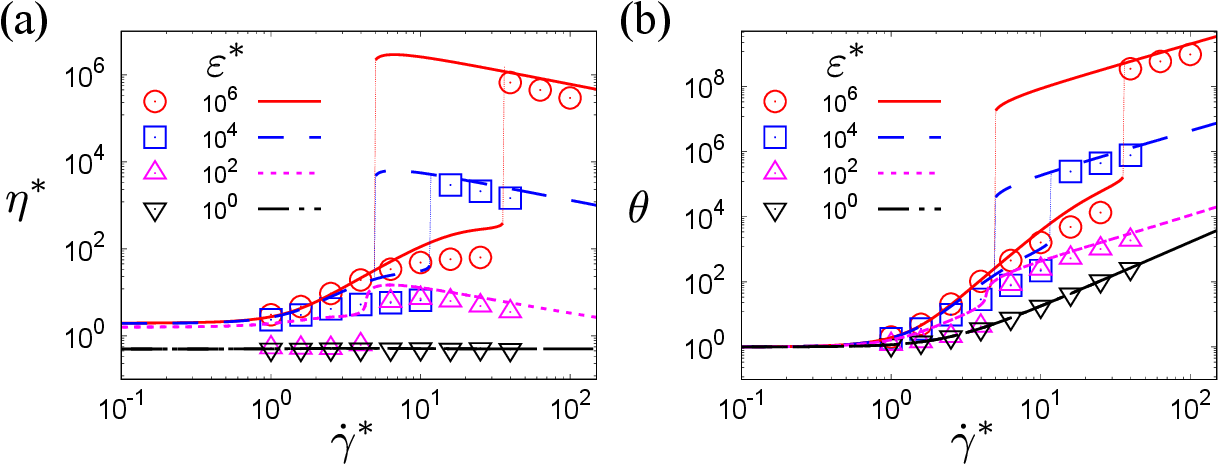}
    \caption{Plots of (a) $\eta^*$ and (b) $\theta$ against $\dot\gamma^*$ for $\varepsilon^*=1, 10^2, 10^4$, and $10^6$
    with fixed $\varphi=0.40$ and $\xi_\mathrm{env}=1.0$.
    Here, the lines express the theoretical flow curves with $N_\mathrm{c}=2$ obtained by Eqs.~\eqref{eq:dynamic_eqs_ver2}--\eqref{eq:nondim_quantities}.
    The corresponding symbols are obtained by the simulation. 
    The vertical dotted lines are discontinuous changes of $\eta^*$ and $\theta$ between the CST and the exploded phases.}
    \label{fig:flow_curves_eps}
\end{figure}

We also check how the rheology depends on the softness parameter $\varepsilon^*$.
Figure \ref{fig:flow_curves_eps} exhibits the comparison of the theoretical results with those of the simulations for $\varepsilon^*=10^0$, $10^2$, $10^4$, and $10^6$ with fixed $\varphi=0.40$ and $\xi_\mathrm{env}=1.0$.
As can be seen in Fig.~\ref{fig:flow_curves_eps}, the theory gives reasonable agreement with the results of simulations.
Figure~\ref{fig:flow_curves_eps} also indicates that the DST-like discontinuous changes of $\eta^*$ and $\theta$ of moderately dense inertial suspensions disappear, at least, for $\varepsilon^* \le 10$.

\begin{figure}[htbp]
    \centering
    \includegraphics[width=\linewidth]{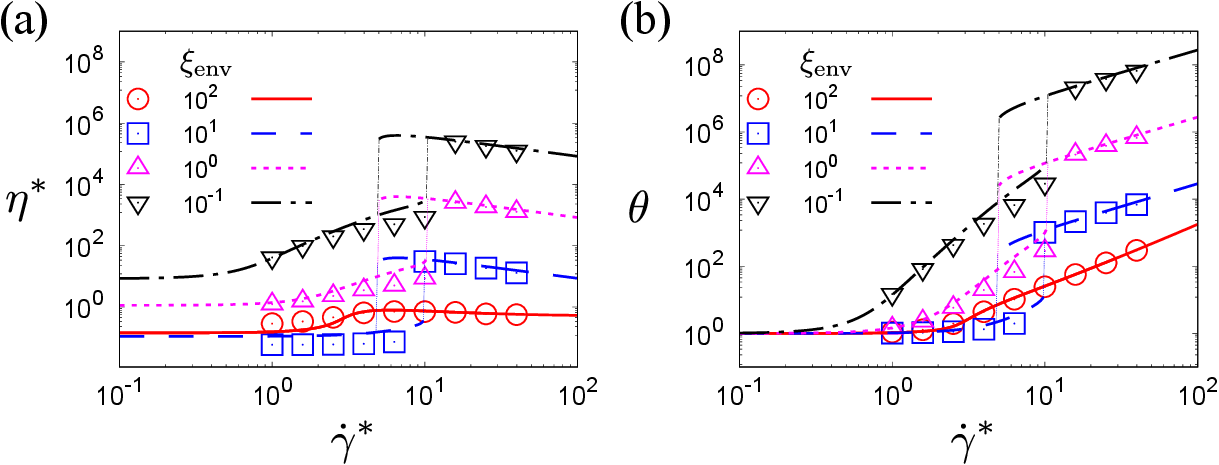}
    \caption{Plots of (a) $\eta^*$ and (b) $\theta$ against $\dot\gamma^*$ for $\xi_\mathrm{env}=10^{-1}$, $10^0$, $10^1$, and $10^2$ when we fix $\varphi=0.30$ and $\varepsilon^*=10^4$.
    Here, the lines express the theoretical expressions, where the theory is given by Eqs.~\eqref{eq:dynamic_eqs_ver2}--\eqref{eq:nondim_quantities}.
    The corresponding symbols are obtained by the simulation.
    The vertical dotted lines are discontinuous changes of $\eta^*$ and $\theta$ between the CST and the exploded phases.
    }
    \label{fig:flow_curves_XI}
\end{figure}

\begin{figure}[htbp]
    \centering
    \includegraphics[width=0.6\linewidth]{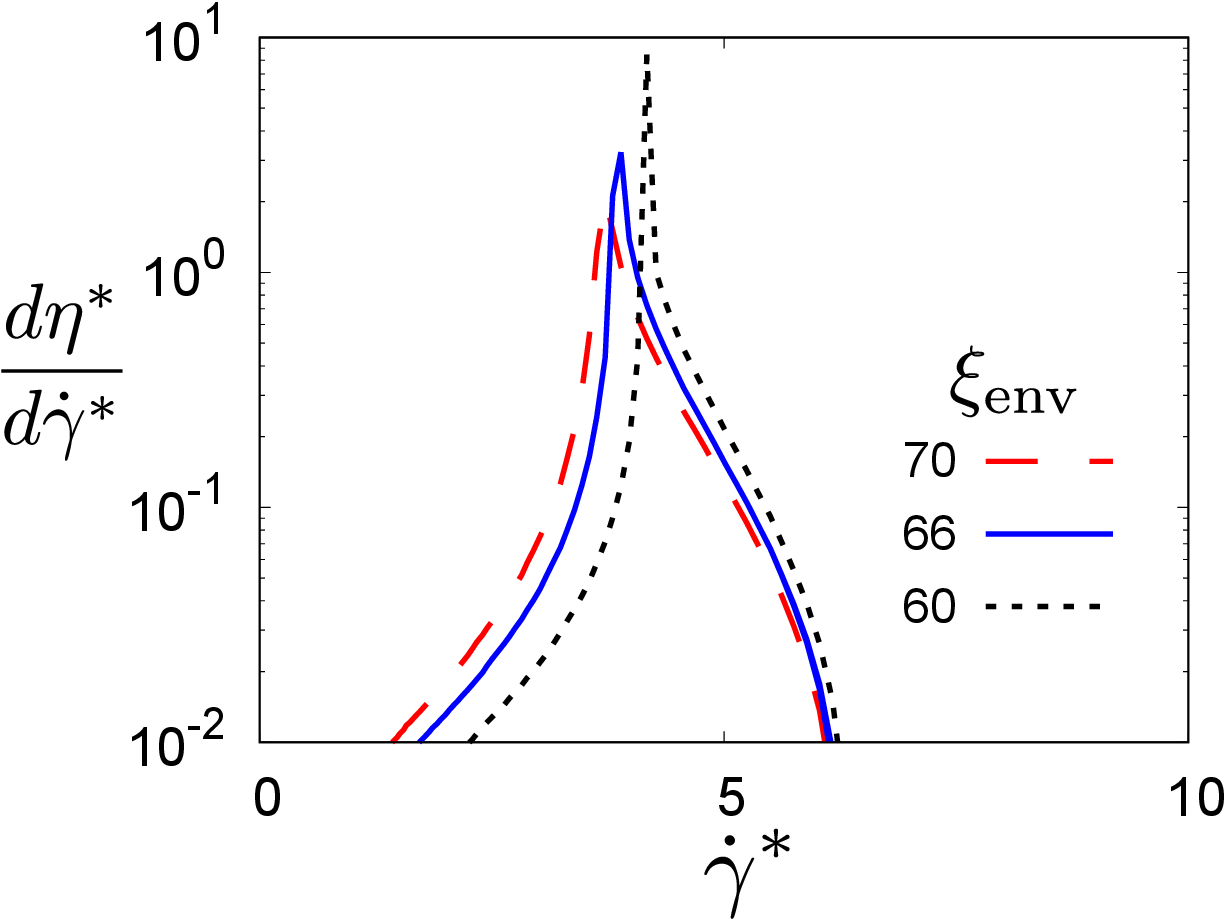}
    \caption{Plots of $d\eta^*/d\dot\gamma^*$ against $\dot\gamma^*$ obtained from the theory for $\xi_\mathrm{env}=60$, $66$, and $70$ when we fix $\varphi=0.30$ and $\varepsilon^*=10^4$.}
    \label{fig:flow_curves_critical}
\end{figure}

Next, let us also investigate $\xi_\mathrm{env}$ dependence of the flow curves.
Figure \ref{fig:flow_curves_XI} demonstrates that the theory works well in the wide range of $\xi_\mathrm{env}$ when we fix $\varphi=0.30$ and $\varepsilon^*=10^4$.
The DST-like changes of $\eta^*$ and $\theta$ disappear if $\xi_\mathrm{env}$ is large enough such as $\xi_\mathrm{env}=10^2$.
To clarify the critical $\xi_{\rm env}$, we control $\xi_{\rm env}$ in the range $60\le \xi_{\rm env}\le 70$ with fixing $\varphi=0.30$ and $\varepsilon^*=10^4$.
As shown in Fig.~\ref{fig:flow_curves_critical}, 
the divergence of the slope $d\eta^*/d\dot\gamma^*$ disappears at around $\xi_\mathrm{env}\approx 66$.
This means that there is a transition from DST to CST at this point.

\begin{figure}[htbp]
    \centering
    \includegraphics[width=\linewidth]{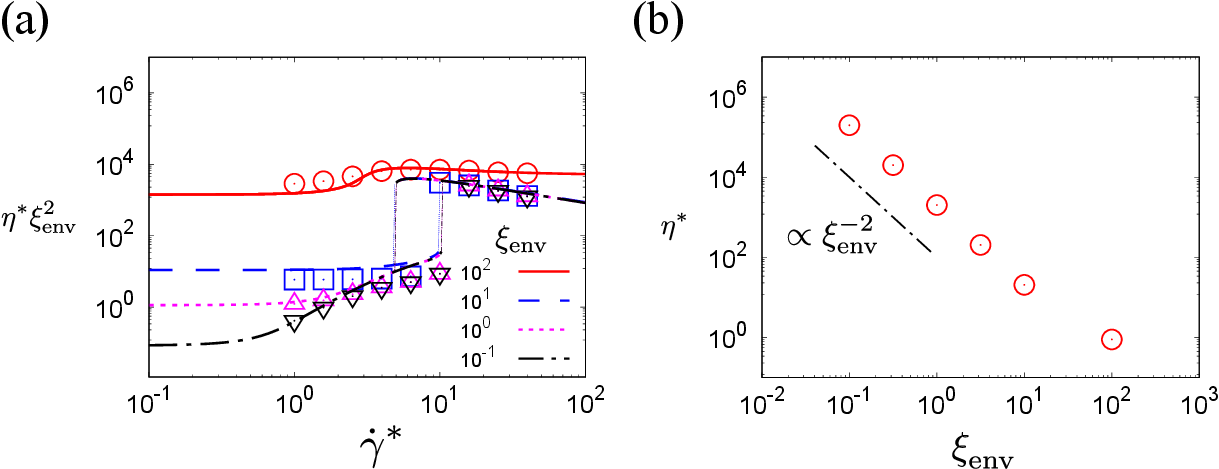}
    \caption{(a) Plot of $\eta^*\xi_\mathrm{env}^2$ against $\dot\gamma^*$ for $\xi_\mathrm{env}=10^{-1}$, $10^0$, $10^1$, and $10^2$ when we fix $\varphi=0.30$ and $\varepsilon^*=10^4$.
    (b) Plot of $\eta^*$ at $\dot\gamma^*=10^{1.6}$ against the strength of the noise $\xi_\mathrm{env}$ when we fix $\varphi=0.30$ and $\varepsilon^*=10^4$.
    The dot-dashed line represents $\eta^*\propto \xi_\mathrm{env}^{-2}$.
    }
    \label{fig:flow_curves_scaled_}
\end{figure}

Figure \ref{fig:flow_curves_scaled_} also shows that the viscosity in the exploded phase is proportional to $\eta^*\propto (\xi_\mathrm{env})^{-2}$ for $\xi_\mathrm{env}\lesssim 10^1$.
This trend is considered to hold even at lower $\xi_\mathrm{env}$.
This means that the viscosity diverges in the zero $\xi_\mathrm{env}$ limit (athermal situation), which is not reachable in experiments.

\begin{figure}[htbp]
    \centering
    \includegraphics[width=\linewidth]{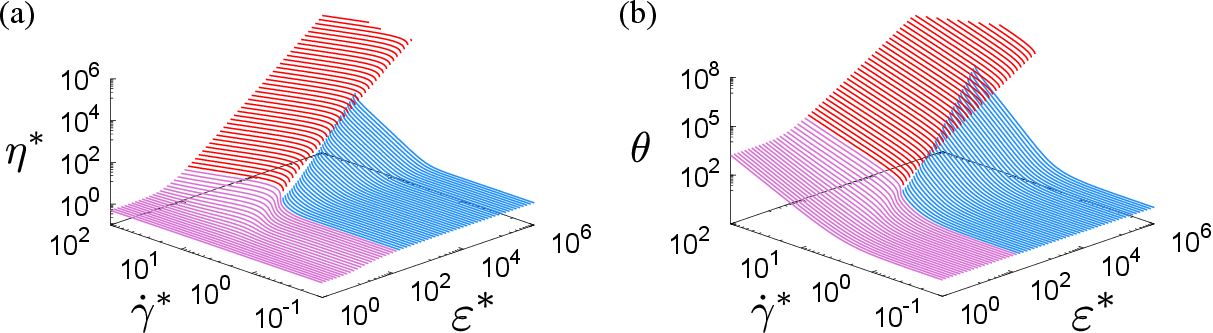}
    \caption{Plots of the theoretical (a) $\eta^*$ and (b) $\theta$ against $\dot\gamma^*$ and $\varepsilon^*$ when we fix $\varphi=0.30$ and $\xi_\mathrm{env}=1.0$, where the theoretical curves are drawn based on Eqs.~\eqref{eq:dynamic_eqs_ver2}--\eqref{eq:nondim_quantities}.
    The purple lines represent the flow curves for the CST.
    The red and the blue lines express the flow curves in the exploded and the CST phases, respectively.
    }
    \label{fig:flow_curves_3D_eps}
\end{figure}

\begin{figure}[htbp]
    \centering
    \includegraphics[width=\linewidth]{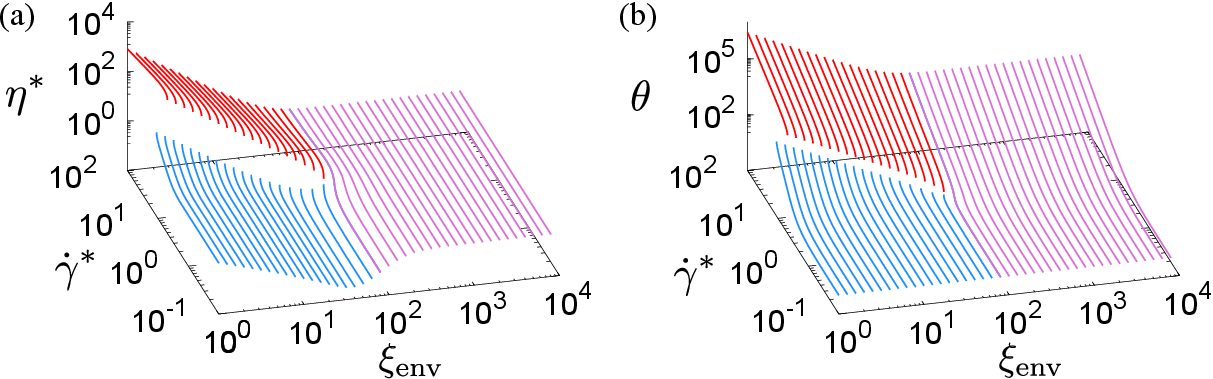}
    \caption{Plots of the theoretical (a) $\eta^*$ and (b) $\theta$ against $\dot\gamma^*$ and $\xi_\mathrm{env}$ when we fix $\varphi=0.30$ and $\varepsilon^*=10^4$, where the theoretical curves are drawn based on Eqs.~\eqref{eq:dynamic_eqs_ver2}--\eqref{eq:nondim_quantities}.
    The purple lines represent the flow curves for the CST.
    The red and the blue lines express the flow curves in the exploded and the CST phases, respectively.
    }
    \label{fig:flow_curves_3D_XI}
\end{figure}

Figures \ref{fig:flow_curves_3D_eps} presents the results of the simulation to show how $\eta^*$ and $\theta$ depend on $\varepsilon^*$ and $\dot\gamma^*$ for $\varphi=0.4$ and $\xi_\mathrm{env}=1.0$.
These figures indicate that both $\eta^*$ and $\theta$ are continuous for small $\varepsilon^*$ but become discontinuous if $\varepsilon^*$ is larger than a critical value.
Similarly, Fig.~\ref{fig:flow_curves_3D_XI} illustrates how $\eta^*$ and $\theta$ depend on $\dot\gamma^*$ and $\xi_\mathrm{env}$ for $\varphi=0.40$ and $\varepsilon^*=10^4$.
These figures indicate that the discontinuous changes of $\eta^*$ and $\theta$ become continuous if $\xi_\mathrm{env}$ is larger than a critical value.


Finally, we comment on the validity of the theoretical assumptions (i)--(vi) listed in the last paragraph of the previous section.
Although we cannot obtain the explicit expression of $\Lambda_{\alpha\beta}$ in moderately dense suspensions, we can write its form explicitly in the dilute limit with the assumptions (i) and (v) (see Appendix~\ref{sec:dilute}). 
This leads to the precise results as shown in Ref.~\cite{Sugimoto20}.
The validity of assumption (vi) is discussed in Appendix \ref{sec:convergence}.
The additional assumptions used in (ii)--(iv) which cannot be justified lead to slight deviations of the theoretical predictions from the simulation results.
Nevertheless, it is unexpected that our theoretical results reasonably agree with the simulation results.

\section{Concluding remarks}\label{sec:conclusion}

This study provides compelling evidence for the occurrence of DST-like discontinuities in both viscosity and kinetic temperature within moderately dense inertial suspensions of frictionless soft-core particles, as demonstrated through Langevin simulations. 
By making several theoretical assumptions, the kinetic theory successfully captures the discontinuous transitions in both viscosity and kinetic temperature across a broad range of parameter space, under the condition that hydrodynamic interactions between particles are neglected. 
Noteworthy, our results substantiate that these discontinuous transitions persist even when hydrodynamic interactions are accounted for, as illustrated in Ref.~\cite{hydro} (see also Appendix \ref{hydro}). 
Furthermore, we highlight that recent molecular dynamics simulations of a mixture support the applicability of the kinetic theory in the absence of hydrodynamic interactions~\cite{Gonzalez23}. 
This work uncovers a novel mechanism for DST-like behavior, driven by the CST-exploded transition in frictionless soft-core particles.

The approximation in Eq.~\eqref{r_min_to_d} is indeed crude. 
Nevertheless, it surprisingly provides reasonable agreement with the simulation results. 
We can offer the following observations:
First, in the dilute limit, the contact contribution becomes negligible. Additionally, the kinetic stress $\sigma^k_{\alpha\beta}$ is not influenced by the approximation in Eq.~\eqref{r_min_to_d}. 
As a result, when the total stress is dominated by the kinetic stress in the exploded phase, the outcome remains unaffected by this simplification.
Second, the approximation may remain valid even when contact stress is considered. 
The collision speed $v^*$ is proportional to the square of the temperature. 
For small $v^*$, $1-r_\mathrm{min}^*$ is small as shown in Fig.~\ref{fig:chi_r_min_vs_v}, which justifies the approximation $r_\mathrm{min}\approx d$.
This indicates that, in this regime, the overlap between particles does not significantly contribute, even though the exact overlap value is not constant.
In contrast, $1-r_\mathrm{min}^*$ is no longer small in the large $v^*$ regime. 
We believe this accounts for the observed discrepancies between the theoretical predictions and simulation results, as shown in Fig.~\ref{fig:flow_curves_dilute}. 
However, in this regime, $1-r_\mathrm{min}^*$ tends to remain nearly constant. 
Consequently, the deviation primarily results in a systematic shift in the predictions by a factor.

It is hard to observe the discontinuous changes of $\eta^*$ and $\theta$ obtained in our model experimentally if the solvent is a liquid.
The significant increase in kinetic temperature during the ignited phase, possibly reaching $10^6$ times that of the quenched phase, might lead to liquid evaporation due to intense stirring effects caused by suspended particles. 
Our model does not encompass the heat-up of $T_\mathrm{env}$ due to stirring effects.
Nevertheless, the indication of the existence of DST-like changes of $\eta^*$ and $\theta$ for frictionless soft-core particles even for relatively dense suspensions is important.

\begin{figure}[htbp]
    \centering
    \includegraphics[width=0.6\linewidth]{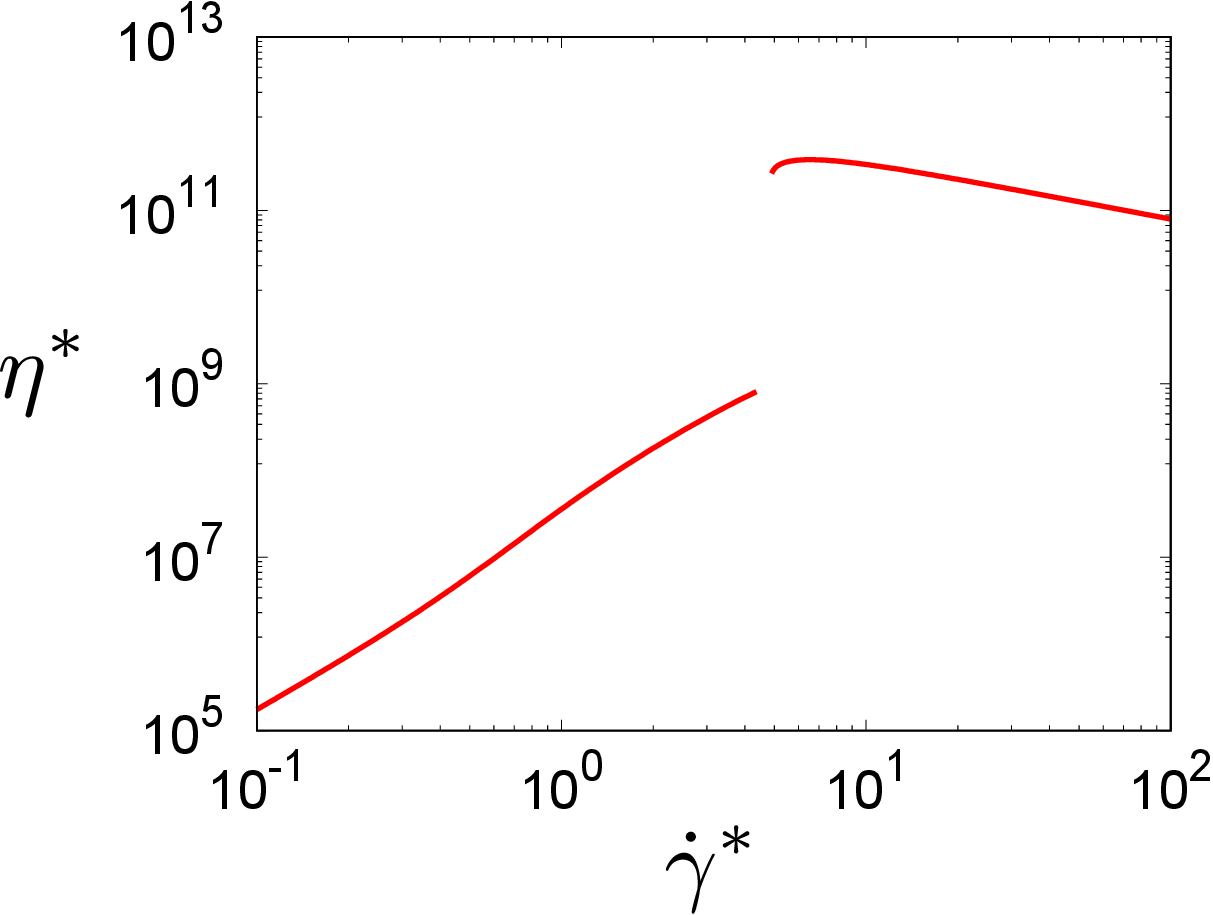}
    \caption{Plot of $\eta^*$ against $\dot\gamma^*$ for $\xi_\mathrm{env}=10^{-4}$ obtained by Eqs.~\eqref{eq:dynamic_eqs_ver2}--\eqref{eq:nondim_quantities} with fixed $\varphi=0.30$ and $\varepsilon^*=10^4$.
    }
    \label{fig:flow_curves_1e-4}
\end{figure}

Let us discuss whether this behavior is observable in experiments of colloidal suspensions. 
Assume the diameter, mass density, and Young's modulus of the colloidal particles are given as $d=2\times 10^{-6}\ \mathrm{m}$, $\rho= 10^3\ \mathrm{kg/m^3}$, and $Y= 10\ \mathrm{GPa}$, respectively. 
From these parameters, the particle mass becomes $m\sim 4\times 10^{-15}\ \mathrm{kg}$.
If the solvent is water at room temperature, with viscosity $\eta_{0}\sim 10^{-3}\ \mathrm{Pa\cdot s}$, the corresponding drag coefficient is $\zeta=3\pi \eta_{0}d/m\sim 5\times 10^6\ \mathrm{s}^{-1}$.
Under the assumption of the roughness parameter $\delta\sim10^{-2}$, we estimate $\varepsilon^*\sim 6\times 10^5\delta\sim 6\times 10^3$.
Notably, this value of $\varepsilon^*$ is close to that used in Fig.~\ref{fig:flow_curves_hydro_0.40_} in Appendix \ref{hydro}. 
It is important to note that the viscosity behavior scales with ${\varepsilon^*}^{4/3}$, indicating that the discontinuous shear thickening (DST) discussed in this study can be observed in the regime of high P\'{e}clet numbers.
A significant challenge arises from the value of $\xi_\mathrm{env}$. 
Assuming $T_\mathrm{env}=4\times 10^{-21}\ \mathrm{J}$, corresponding to room temperature, we find $\xi_\mathrm{env}\sim 2\times 10^{-4}$.
This value is much smaller than the critical range ($10<\xi_\mathrm{env}^\mathrm{c}<10^2$) where DST is absent. 
Therefore, our analysis suggests that DST is likely to occur.
Moreover, we have verified that DST persists for realistic parameter values within the kinetic theory framework, as shown in Fig.~\ref{fig:flow_curves_1e-4}. 
However, observing such DST in frictionless small grains, which are expected to behave as Brownian suspensions, presents a significant experimental challenge.

Our theory is not applicable beyond the Alder transition point ($\varphi \approx 0.49$), where phenomena such as dynamical heterogeneity~\cite{Kob95, Garrahan02, Dauchot05, Berthier05, Charbonneau07, Berthier11}, spatial correlations~\cite{Ernst97, Ernst98, Ernst00, Otsuki09PRE, Otsuki09EPJ}, long-time correlations~\cite{Otsuki09JSM, Otsuki09EPJ}, and many-body correlations~\cite{Suzuki15, Suzuki19} become significant. 
Addressing these correlation effects is essential for understanding the behavior of dense inertial suspensions near the glass or jamming transition.
We will study such correlation effects in dense inertial suspensions in the near future.

Needless to say, it is important to analyze suspensions of frictional grains corresponding to the typical experimental setup for colloidal suspensions.
This subject would be also our future task.

\section*{Acknowledgements}
The authors thank Takeshi Kawasaki, Takashi Uneyama, Michio Otsuki, and Pradipto for their helpful comments.
This research was partially supported by the Grant-in-Aid of MEXT for Scientific Research (Grants No.~\href{https://kaken.nii.ac.jp/grant/KAKENHI-PROJECT-20K14428}{JP20K14428}, No.~\href{https://kaken.nii.ac.jp/grant/KAKENHI-PROJECT-21H01006}{JP21H01006}, No.~\href{https://kaken.nii.ac.jp/grant/KAKENHI-PROJECT-24K06974}{JP24K06974}, and No.~\href{https://kaken.nii.ac.jp/grant/KAKENHI-PROJECT-24K07193/}{JP24K07193}).
H.H. was also supported by the Kyoto University Foundation.
The authors thank YITP activity \href{https://www.yukawa.kyoto-u.ac.jp/seminar/s52935?lang=en-GB}{YITP–X–21-13}.

\appendix
\section{Summary of Ref.~\cite{hydro}}\label{hydro}
In Ref.~\cite{hydro}, we have also investigated the effect of the lubrication force between particles based on simulations.
In this appendix, we briefly explain the main results of Ref.~\cite{hydro}.

\subsection{Details of the lubrication force between particles}
First, let us explain the explicit form of the lubrication force between particles.
We have adopted the framework of the Lubrication-Friction Discrete Element Method (LF-DEM).
The equation of motion of the suspended particle is given by
\begin{equation}\label{eq:Langevin_LFDEM}
    \frac{d\bm{p}_i}{dt} 
    = \sum_{j\neq i}\left(\bm{F}_{ij}+\bm{F}^\mathrm{H}_{ij}\right)
    + \bm{\xi}_i,
\end{equation}
where $\bm{F}^\mathrm{H}_{ij}$ is the lubrication force.
The explicit form of the lubrication forces and the relating stresses act between them are given by
\begin{equation}
    \begin{bmatrix}
    \bm{F}^\mathrm{H}_{ij} \\ \bm{F}^\mathrm{H}_{ji} \\ 
    \overleftrightarrow{\overline{\sigma}^\mathrm{H}_{ij}} \\ \overleftrightarrow{\overline{\sigma}^\mathrm{H}_{ji}}
    \end{bmatrix}
    =-\eta_0
    \begin{bmatrix}
    \overleftrightarrow{A_{ij}^{(11)}} & 
    \overleftrightarrow{A_{ij}^{(12)}} &  
    \overleftrightarrow{\widetilde{G}_{ij}^{(11)}} & 
    \overleftrightarrow{\widetilde{G}_{ij}^{(12)}}\\
    \overleftrightarrow{A_{ij}^{(21)}} & 
    \overleftrightarrow{A_{ij}^{(22)}} &  
    \overleftrightarrow{\widetilde{G}_{ij}^{(21)}} & 
    \overleftrightarrow{\widetilde{G}_{ij}^{(22)}}\\
    \overleftrightarrow{G_{ij}^{(11)}} & 
    \overleftrightarrow{G_{ij}^{(12)}} & 
    \overleftrightarrow{M_{ij}^{(11)}} & 
    \overleftrightarrow{M_{ij}^{(12)}}\\
    \overleftrightarrow{G_{ij}^{(21)}} & 
    \overleftrightarrow{G_{ij}^{(22)}} & 
    \overleftrightarrow{M_{ij}^{(21)}} & 
    \overleftrightarrow{M_{ij}^{(22)}}
    \end{bmatrix}
    \begin{bmatrix}
    \bm{v}_i-\overleftrightarrow{E}\bm{x}_i\\
    \bm{v}_j-\overleftrightarrow{E}\bm{x}_j\\
    -\overleftrightarrow{E} \\ 
    -\overleftrightarrow{E}
    \end{bmatrix}.
    \label{eq:F_hydro}
\end{equation}
Because we are only interested in the simple shear condition, the matrix $\overleftrightarrow{E}$ is reduced to
\begin{equation}
    E_{\alpha\beta}=\dot\gamma \delta_{\alpha x}\delta_{\beta y}.
\end{equation}
Note that $\overleftrightarrow{E}$ is independent of particle indices $i$ and $j$.

It is known that the coefficients $A^{(k\ell)}_{ij,\alpha\beta}$, $G^{(k\ell)}_{ij,\alpha\beta\gamma}$, and $M^{(k\ell)}_{ij,\alpha\beta\gamma\delta}$ in Eq.~\eqref{eq:F_hydro} satisfy the following symmetries~\cite{KimKarrila}:
\begin{subequations}
\begin{align}
    A^{(k\ell)}_{ij,\alpha\beta} &= A^{(\ell k)}_{ij,\beta\alpha},\\
    G^{(k\ell)}_{ij,\alpha\beta\gamma} &= \widetilde{G}^{(\ell k)}_{ij,\gamma\alpha\beta},
    \label{eq:G_sym}\\
    M^{(k\ell)}_{ij,\alpha\beta\gamma\delta} &= M^{(\ell k)}_{ij, \gamma\delta \alpha\beta}.
\end{align}
\end{subequations}
The explicit forms of the coefficients are written as
\begin{subequations}\label{eq:F4}
\begin{align}
    A^{(k\ell)}_{ij,\alpha\beta}
    &= X^{A(k\ell)}_{ij} \hat{k}_{ij,\alpha} \hat{k}_{ij,\beta}
    + Y^{A(k\ell)}_{ij} \left(\delta_{\alpha\beta}-\hat{k}_{ij,\alpha} \hat{k}_{ij,\beta}\right),
    \label{eq:A_X_Y}\\
    G^{(k\ell)}_{ij,\alpha\beta\gamma}
    &= X^{G(k\ell)}_{ij}
    \left(\hat{k}_{ij,\alpha} \hat{k}_{ij,\beta}
    -\frac{1}{3}\delta_{\alpha\beta}\right)\hat{k}_{ij,\gamma}
    + Y^{G(k\ell)}_{ij}
    \left(\hat{k}_{ij,\alpha}\delta_{\beta\gamma} 
    + \hat{k}_{ij,\beta}\delta_{\alpha\gamma} 
    - 2\hat{k}_{ij,\alpha} \hat{k}_{ij,\beta} \hat{k}_{ij,\gamma}\right),
    \label{eq:G_X_Y}\\
    M^{(k\ell)}_{ij,\alpha\beta\gamma\delta}
    &=X^{M(k\ell)}_{ij}\hat{k}^{(0)}_{ij,\alpha\beta\gamma\delta}
    + Y^{M(k\ell)}_{ij}\hat{k}^{(1)}_{ij,\alpha\beta\gamma\delta}
    + Z^{M(k\ell)}_{ij}\hat{k}^{(2)}_{ij,\alpha\beta\gamma\delta},
    \label{eq:M_X_Y}
\end{align}
\end{subequations}
respectively,
where we have introduced $X^{A(k\ell)}_{ij}$, $Y^{A(k\ell)}_{ij}$, $X^{G(k\ell)}_{ij}$, $Y^{G(k\ell)}_{ij}$, $X^{M(k\ell)}_{ij}$, $Y^{M(k\ell)}_{ij}$, and $Z^{M(k\ell)}_{ij}$ which are functions of the separation length $h_{ij}\equiv 2(r_{ij}-d_\mathrm{H})/d_\mathrm{H}=2(r_{ij}/d_\mathrm{H}-1)$, as well as $\hat{{k}}^{(0)}_{ij,\alpha\beta\gamma\delta}$, $\hat{{k}}^{(1)}_{ij,\alpha\beta\gamma\delta}$, and $\hat{k}^{(2)}_{ij,\alpha\beta\gamma\delta}$ as~\cite{KimKarrila}
\begin{subequations}
\begin{align}
    \hat{k}^{(0)}_{ij,\alpha\beta\gamma\delta}
    &\equiv \frac{3}{2}
    \left(\hat{k}_{ij,\alpha} \hat{k}_{ij,\beta}
    -\frac{1}{3}\delta_{\alpha\beta}\right)
    \left(\hat{k}_{ij,\gamma} \hat{k}_{ij,\delta} 
    -\frac{1}{3}\delta_{\gamma\delta}\right),\\
    \hat{k}^{(1)}_{ij,\alpha\beta\gamma\delta}
    &\equiv
    \frac{1}{2}
    \left(\hat{k}_{ij,\alpha}\hat{k}_{ij,\gamma} \delta_{\beta\delta}
    +\hat{k}_{ij,\beta} \hat{k}_{ij,\gamma} \delta_{\alpha\delta}
    +\hat{k}_{ij,\alpha} \hat{k}_{ij,\delta} \delta_{\beta\gamma}
    +\hat{k}_{ij,\beta} \hat{k}_{ij,\delta} \delta_{\alpha\gamma}
    -4\hat{k}_{ij,\alpha} \hat{k}_{ij,\beta} \hat{k}_{ij,\delta} \hat{k}_{ij,\gamma}\right),\\
    \hat{k}^{(2)}_{ij,\alpha\beta\gamma\delta}
    &\equiv
    \frac{1}{2}
    \left(\delta_{\alpha\gamma}\delta_{\beta\delta} + \delta_{\alpha\delta}\delta_{\beta\gamma}
    - \delta_{\alpha\beta} \delta_{\gamma\delta} 
    +\hat{k}_{ij,\alpha} \hat{k}_{ij,\beta} \delta_{\gamma\delta}
    +\hat{k}_{ij,\gamma} \hat{k}_{ij,\delta} \delta_{\alpha\beta}
    -\hat{k}_{ij,\alpha} \hat{k}_{ij,\gamma} \delta_{\beta\delta}\right.\nonumber\\
    &\hspace{3em}\left.
    -\hat{k}_{ij,\beta} \hat{k}_{ij,\gamma} \delta_{\alpha\delta}
    -\hat{k}_{ij,\alpha} \hat{k}_{ij,\delta} \delta_{\beta\gamma}
    -\hat{k}_{ij,\beta} \hat{k}_{ij,\delta} \delta_{\alpha\gamma}
    +\hat{k}_{ij,\alpha} \hat{k}_{ij,\beta} \hat{k}_{ij,\delta} \hat{k}_{ij,\gamma}\right),
\end{align}
\end{subequations}
When we consider their leading terms for $h_{ij}$, these coefficients for the monodisperse particles are reduced to~\cite{Jeffrey84, Jeffrey92}.
Note that these expressions are dimensional. We should be careful when we compare these results with those given in Refs.~\cite{Jeffrey84, Jeffrey92, KimKarrila}.
\begin{subequations}\label{eq:F6}
\begin{align}
    X^{A(11)}_{ij}&=-X^{A(12)}_{ij}= -X^{A(21)}_{ij} = X^{A(22)}_{ij}
    = 3\pi d_\mathrm{H} \frac{1}{4}\frac{1}{h_{ij}},\\
    Y^{A(11)}_{ij}&=-Y^{A(12)}_{ij}=-Y^{A(21)}_{ij}=Y^{A(22)}_{ij}
    = 3\pi d_\mathrm{H} \frac{1}{6}\ln \frac{1}{h_{ij}},\\
    X^{G(11)}_{ij}&=-X^{G(12)}_{ij}=X^{G(21)}_{ij}=-X^{G(22)}_{ij}
    = \frac{3}{8}\pi d_\mathrm{H}^2 \frac{1}{h_{ij}},\\
    Y^{G(11)}_{ij}&=-Y^{G(12)}_{ij}
    = Y^{G(21)}_{ij}=-Y^{G(22)}_{ij}
    = \frac{1}{8}\pi d_\mathrm{H}^2 \ln\frac{1}{h_{ij}},\\
    X^{M(11)}_{ij}&= X^{M(12)}_{ij}= X^{M(21)}_{ij}= X^{M(22)}_{ij}
    =\frac{1}{8}\pi d_\mathrm{H}^3 \frac{1}{h_{ij}},\\
    Y^{M(11)}_{ij}&=Y^{M(22)}_{ij}
    =\frac{1}{10}\pi d_\mathrm{H}^3 \ln \frac{1}{h_{ij}},\quad
    Y^{M(12)}_{ij}=Y^{M(21)}_{ij}
    =\frac{1}{40}\pi d_\mathrm{H}^3 \ln \frac{1}{h_{ij}},\\
    Z^{M(11)}_{ij}&=Z^{M(22)}_{ij}
    \approx \frac{5}{6}\pi d_\mathrm{H}^3,\quad
    Z^{M(12)}_{ij}=Z^{M(21)}_{ij}
    \approx -\frac{3}{16}\pi d_\mathrm{H}^3,
\end{align}
\end{subequations}
respectively.

Although smooth hard-core particles are not allowed to contact each other~\cite{KimKarrila,Jeffrey84,Jeffrey92}, the LF-DEM does allow contact between rough particles, and so we introduce the roughness parameter $\delta \equiv (d-d_\mathrm{H})/d_\mathrm{H}$ with the hydrodynamic diameter $d_\mathrm{H}$~\cite{Seto13,Pradipto20,Mari15}.
This parameter equates to a simplified description of dimples on the surface of particles.

\subsection{Results}
In this subsection, we show the main results obtained by the simulation based on the LF-DEM.

Let us introduce the Peclet number $\mathrm{Pe}$ defined as~\cite{Hunt02}
\begin{equation}
    \mathrm{Pe}
    \equiv \frac{3\pi \eta_0 d^3}{4T_\mathrm{env}}\dot\gamma.
    \label{eq:Pe}
\end{equation}
Since $\xi_{\rm env}$ introduced in Eq.~\eqref{eq:dimless_quantities} is proportional to $\sqrt{m T_\mathrm{env}/\eta_0^2d^4}$, as shown in Eq.~\eqref{eq:dimless_quantities}, $\xi_{\rm env}$ increases as the inertia becomes important. 
Using $\dot\gamma^*$, we introduce the dimensionless viscosity $\eta^*$ as
\begin{equation}
    \eta^*
    \equiv\frac{\sigma_{xy}}{n T_\mathrm{env}\mathrm{Pe}}.
\end{equation}    

\begin{figure}[htbp]
    \centering
    \includegraphics[width=0.6\linewidth]{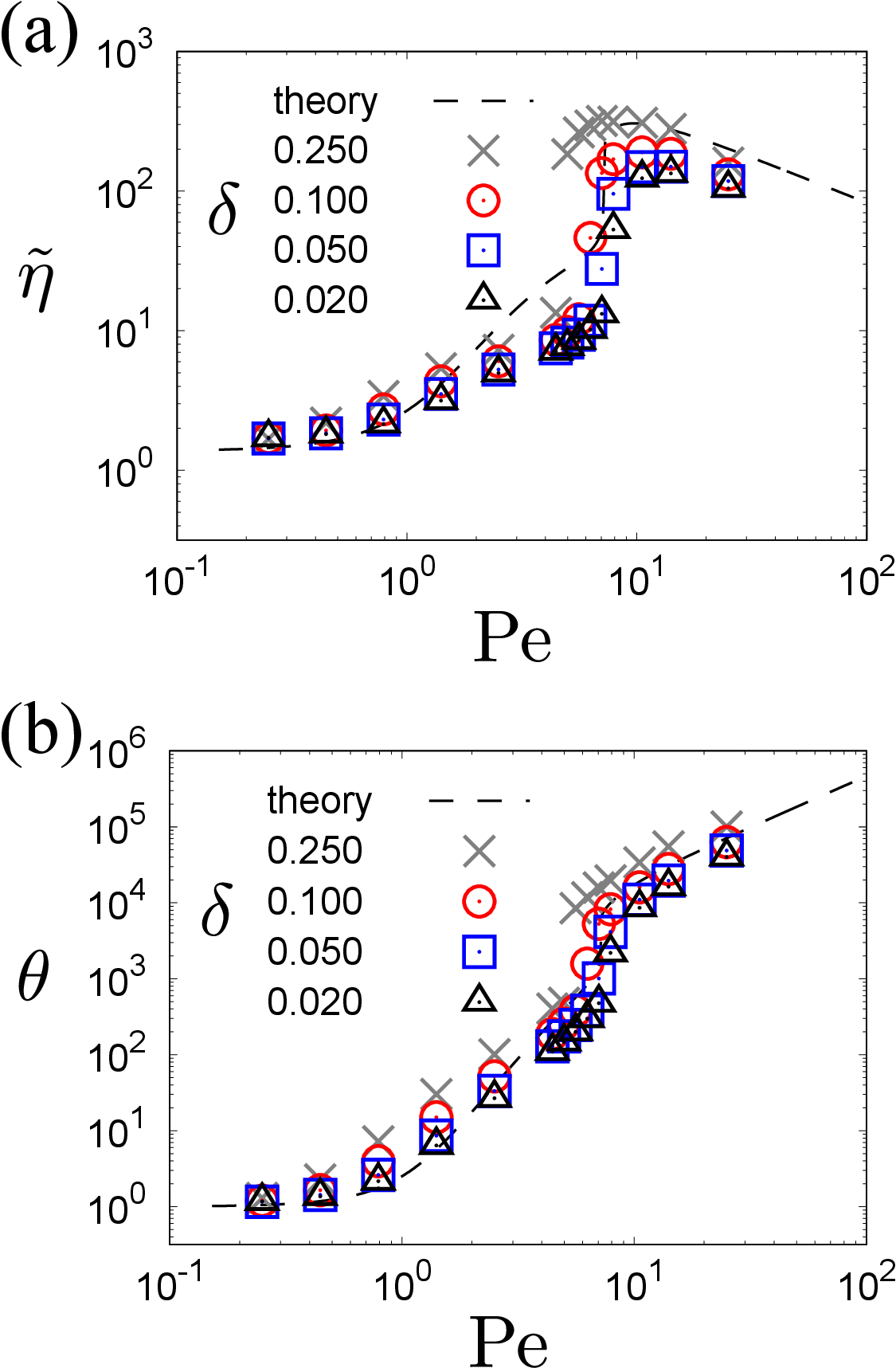}
    \caption{Plots of (a) $\tilde{\eta}$ and (b) $\theta$ against $\mathrm{Pe}$ for $\delta=0.02$ (open circles), $0.05$ (open squares), $0.10$ (open triangles), and $0.25$ (crosses) with fixed values of $\varphi=0.30$, $\varepsilon^*=10^4$, and $\xi_\mathrm{env}=1.0$.
    The dashed line represents the prediction of the kinetic theory in the main text.
    }
    \label{fig:flow_curves_hydro_0.30}
\end{figure}
 
We examine $N=1000$ and $10$ ensemble averages in the simulations.
We control $\mathrm{Pe}$, $\xi_\mathrm{env}$, $\varepsilon^*$, and $\delta$ in the range $0.25\le \mathrm{Pe}\le 25$, $1\le\xi_\mathrm{env}\le10^2$, $10^2\le \varepsilon^*\le 10^8$, and $0.02\le\delta\le0.25$, respectively.

Figure \ref{fig:flow_curves_hydro_0.30} shows how the scaled viscosity $\tilde{\eta}\equiv \eta^*/\eta_\mathrm{a}$ and the dimensionless temperature $\theta$ depend on $\mathrm{Pe}$ with fixed values of $\varphi=0.30$, $\varepsilon^*=10^4$, and $\xi_\mathrm{env}=1.0$ for various $\delta$, where 
the empirical expression of the apparent viscosity $\eta_\mathrm{a}$ in the low shear limit $\eta_\mathrm{a}= 1+(5/2)\varphi + 4\varphi^2+42\varphi^3$~\cite{deKruif85}.
Note that we do not have any theoretical result for $\eta_{\rm a}$ until $O(\varphi^3)$ as far as we have been able to ascertain, though the expression until $O(\varphi^2)$ is well known for hard-core suspensions~\cite{Batchelor72}.

Although the quantitative agreement between the kinetic theory without consideration of hydrodynamic interactions shown in the main text and hydrodynamic simulation tend to be poor for $\varphi>0.3$, the kinetic theory captures the qualitative behavior of the DST.
Moreover, the agreement between the theory and simulation is reasonable for $\varphi=0.30$ (Fig.~\ref{fig:flow_curves_hydro_0.30}(a)).
We note that the results of the simulation based on the LF-DEM approach the theoretical prediction reported in the main text as $\delta$ increases, though we have used the empirical expression for $\eta_{\rm a}$.
Figure \ref{fig:flow_curves_hydro_0.30} indicates that these DSTs are caused by the ignited-quenched transition.

Thus, we have confirmed that DST and the ignited-quenched transition of inertial suspensions for soft and frictionless particles can survive even if hydrodynamic interactions between particles exist. 
We have also confirmed the relevance of the kinetic theory developed in the main text to describe the suspensions with hydrodynamic interactions.

\begin{figure}[htbp]
    \centering
    \includegraphics[width=0.6\linewidth]{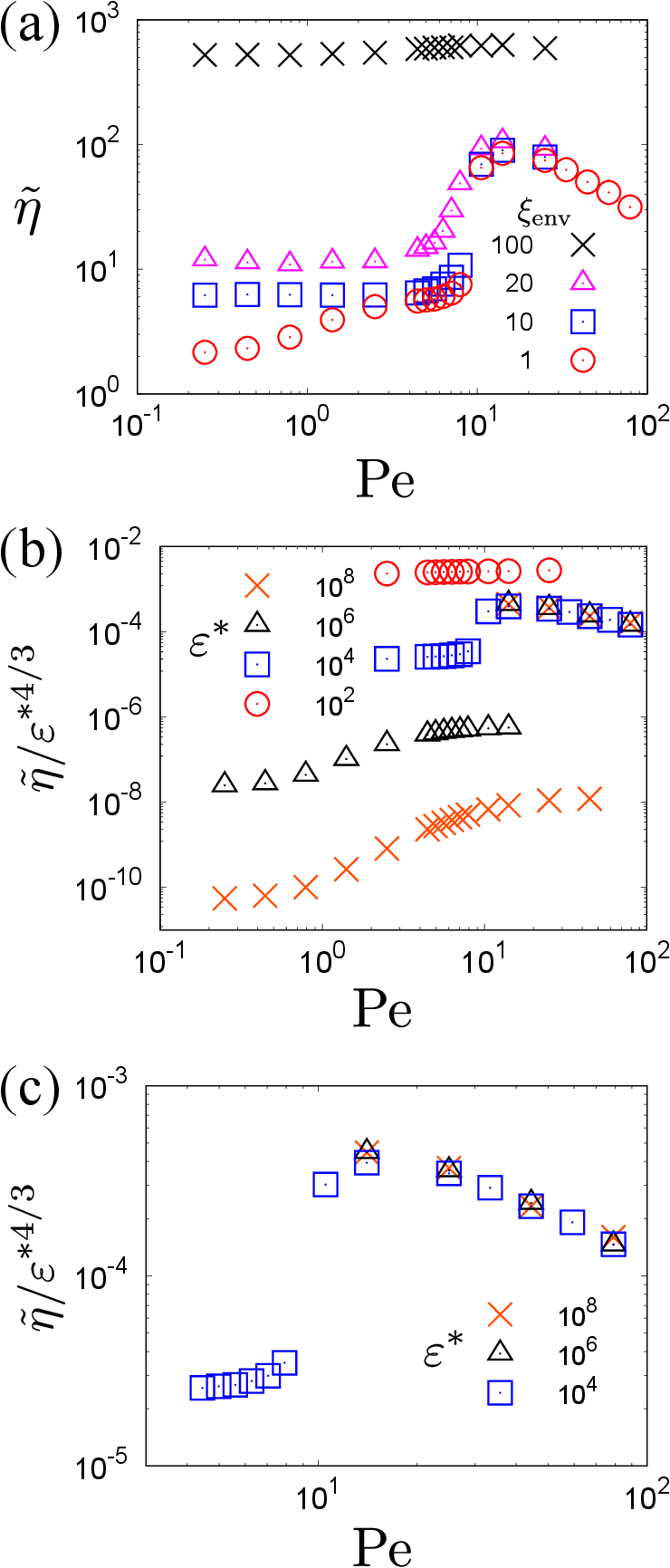}
    \caption{(a) Plot of $\tilde{\eta}$ against $\mathrm{Pe}$ for various $\xi_\mathrm{env}$.
    (b) Plot of $\tilde{\eta}/\varepsilon^{*4/3}$ for various $\varepsilon^*$.
    Here, we fix (a) $\varepsilon^*=10^4$ and (b) $\xi_\mathrm{env}=1$, respectively.
    (c) A detailed view of the high $\mathrm{Pe}$ regime of (b).
    Parameters are fixed as $\varphi=0.40$ and $\delta=0.05$.
    }
    \label{fig:flow_curves_hydro_0.40_}
\end{figure}

Figure \ref{fig:flow_curves_hydro_0.40_}(a) is the plot of $\tilde{\eta}$ against ${\rm Pe}$ for various $\xi_{\rm env}$ with fixed values of $\delta=0.05$ and $\varphi=0.40$.
We verify that DST still persists for $\xi_{\rm env}=10$ but disappears around $10\lesssim \xi_{\rm env}\lesssim 20$.
Figure \ref{fig:flow_curves_hydro_0.40_}(b) is the plot of $\tilde{\eta}/\varepsilon^{*4/3}$ against ${\rm Pe}$ for various $\varepsilon^*$ with fixed values of $\delta=0.05$ and $\varphi=0.40$.
Except for the case of $\varepsilon^*=10^2$, $\tilde{\eta}$ exhibits DST.
As the softness parameter $\varepsilon^*$ increases, the upper branch also increases (see Fig.~\ref{fig:flow_curves_hydro_0.40_}(b)).
As shown in Figs.~\ref{fig:flow_curves_hydro_0.40_}(b) and (c) and Fig.~\ref{fig:flow_curves_scaled} in the main text, the viscosity in the upper branch is scaled as $\tilde{\eta}/\varepsilon^{*4/3}$ except for $\varepsilon^*=10^2$.
Thus, we conclude that DST with frictionless soft particles can be observed for systems with $\xi_\mathrm{env}\le20$ and $\varepsilon^*\ge10^4$.

\section{The basis of the kinetic theory of inertial suspension of soft particles}\label{frame_McLennan}

In this appendix, we explain the basis of the kinetic theory.
The appendix consists of two parts.
In the first subsection, we explain the framework of the kinetic theory. 
In the second subsection, we derive the kinetic equation of the inertial suspension of soft particles.

\subsection{Framework of kinetic model}\label{sec:framework}
Let us consider an $N$-body distribution function $f^{(N)}(\{\bm{r}_i\}, \{\bm{p}_i\}, t)$.
We adopt the abbreviation $f^{(N)}\equiv f^{(N)}(\{\bm{r}_i\}, \{\bm{p}_i\}, t)$ to simplify the notation.
From the conservation of probability, one gets~\cite{Resibois,McLennan,Zwanzig,Evans}
\begin{equation}
    \frac{\partial f^{(N)}}{\partial t}
    + \sum_i \frac{\partial}{\partial \bm{r}_i}\cdot\left(\frac{d\bm{r}_i}{dt}f^{(N)}\right)
    + \sum_i \frac{\partial}{\partial \bm{p}_i}\cdot\left(\frac{d\bm{p}_i}{dt}f^{(N)}\right)=0
    \label{eq:eq_P2}
\end{equation}
for soft particles.
Substituting Eq.~\eqref{eq:Langevin} into Eq.~\eqref{eq:eq_P2}, we obtain the stochastic-Liouville equation
\begin{align}\label{stochastic_Liouville_eq}
    &\frac{\partial f^{(N)}}{\partial t}
    + \sum_i \frac{\partial}{\partial \bm{r}_i}\cdot\left[\left(\frac{\bm{p}_i}{m}+\dot\gamma y_i \hat{\bm{e}}_x \right)f^{(N)}\right]
    + \sum_i \frac{\partial}{\partial \bm{p}_i}\cdot\left[\left(\sum_{j\neq i}\bm{F}_{ij} 
    - \zeta \bm{p}_i + \bm{\xi}_i\right) f^{(N)}\right]=0.
\end{align}
With the aid of the noise average, Eq.~\eqref{stochastic_Liouville_eq} can be rewritten as
\begin{align}
    &\left[\frac{\partial}{\partial t}
    + \sum_i \bm{V}_i \cdot \frac{\partial}{\partial \bm{r}_i}
    -\sum_i\dot\gamma V_{i,y}\frac{\partial}{\partial V_{i,x}}\right]
    \langle f^{(N)} \rangle\nonumber\\
    &=\sum_i
    \frac{\partial}{\partial \bm{V}_i}\cdot\left[
    \zeta \left(\bm{V}_i + \frac{T_\mathrm{env}}{m}\frac{\partial}{\partial \bm{V}_i}\right)\langle f^{(N)}\rangle\right]
    - \sum_{i,j} \frac{\partial}{\partial \bm{V}_i}\cdot\left(\frac{\bm{F}_{ij}}{m} 
    \langle f^{(N)}\rangle\right),
    \label{eq:eq_av_P}
\end{align}
where $\langle f^{(N)}\rangle$ is the noise-averaged distribution function and we have introduced $\bm{V}_i\equiv \bm{p}_i/m$.
Integrating Eq.~\eqref{eq:eq_av_P} over $\bm{r}_i$ and $\bm{V}_i$ for $i=2,3,\cdots, N$, we obtain
\begin{align}
    &\left[\frac{\partial}{\partial t}
    + \bm{V}_1 \cdot \frac{\partial}{\partial \bm{r}_1}
    -\dot\gamma V_{1,y}\frac{\partial}{\partial V_{1,x}}\right]
    f(X_1;t)\nonumber\\
    &=
    \frac{\partial}{\partial \bm{V}_1}\cdot\left[\zeta\left(\bm{V}_1 + \frac{T_\mathrm{env}}{m}\frac{\partial}{\partial \bm{V}_1}\right)f(X_1;t)\right]
    - \frac{1}{N-1}\sum_{j} \frac{\partial}{\partial \bm{V}_1}\cdot\left(\frac{\bm{F}_{1j}}{m} 
    f^{(2)}(X_1,X_j;t)\right)\nonumber\\
    &\approx
    \frac{\partial}{\partial \bm{V}_1}\cdot\left[\zeta\left(\bm{V}_1 + \frac{T_\mathrm{env}}{m}\frac{\partial}{\partial \bm{V}_1}\right)f(X_1;t)\right]
    - \frac{\partial}{\partial \bm{V}_1}\cdot\left(\frac{\bm{F}_{12}}{m} 
    f^{(2)}(X_1,X_2;t)\right),
    \label{eq:eq_f_f2}
\end{align}
where we have introduced one- and two-body distribution functions as
\begin{subequations}
\begin{align}
    f(X_1;t)
    &\equiv \frac{1}{(N-1)!(L^3)^{N-1}}\int \prod_{j=2}^NdX_j \langle f^{(N)} \rangle,\\
    f^{(2)}(X_1,X_2; t)
    &\equiv \frac{1}{(N-2)!(L^3)^{N-2}}\int \prod_{j=3}^N dX_j  \langle f^{(N)} \rangle 
\end{align}
\end{subequations}
with $dX_j\equiv d\bm{r}_j d\bm{p}_j$, and $X_j\equiv (\bm{r}_j,\bm{p}_j)$.
To obtain the last expression in Eq.~\eqref{eq:eq_f_f2}, we put $j=2$ and attach $N-1$ factor to the second and third terms on the right-hand side (RHS) of Eq.~\eqref{eq:eq_f_f2} in the first line.
As shown in Appendix~\ref{McLennan}
the last term on the RHS of Eq.~\eqref{eq:eq_f_f2} can be written as the collision integral
$J[\bm{V}_1|f^{(2)}(1,2)]$, at least, in the low density limit.
Thus, we can reach Eq.~\eqref{eq:kinetic_eq}.

\subsection{Derivation of the equation of one-body distribution function}\label{McLennan}
In this subsection, we derive the collision operator for a dilute gas without solvents based on Ref.~\cite{McLennan}.

The stochastic Liouville equation can be rewritten as
\begin{equation}\label{Liouville_tot}
    \frac{\partial f^{(N)}}{\partial t}=(\mathcal{L}^{(N)}+\mathcal{L}_{\dot\gamma}+\mathcal{L}_\mathrm{hyd})f^{(N)}
\end{equation}
where we have introduced three parts of Liouvillian acting on a variable $A$ as
\begin{align}\label{Liouville_Poisson}
    \mathcal{L}^{(N)} A&\equiv\{A,H^{(N)} \}, \\
    \mathcal{L}_{\dot\gamma} A&\equiv -\dot\gamma \sum_{i=1}^N \left\{\frac{\partial}{\partial x_i} (y A)-\frac{\partial}{\partial p_{i,x}} ({p}_{i,y}A) \right\}\label{L_gamma} \\
    \mathcal{L}_\mathrm{hyd} A&\equiv
    \sum_{i=1}^N \frac{\partial}{\partial \bm{p}_i}\cdot\left[(\zeta\bm{p}_i+\bm{\xi}_i)A\right] 
\end{align}
with Poisson's bracket $\{A,B\}$:
\begin{equation}
    \{A,B\}\equiv \sum_{i=1}^N\left(\frac{\partial A}{\partial \bm{r}_i}\cdot\frac{\partial B}{\partial \bm{p}_i}
    -\frac{\partial B}{\partial \bm{r}_i}\cdot\frac{\partial A}{\partial \bm{p}_i}
    \right),
\end{equation}
and the total Hamiltonian $H_N$:  
\begin{equation}
    H^{(N)}
    \equiv \sum_{i=1}^N \left\{\frac{\bm{p}_i^2}{2m}+\frac{1}{2}\sum_{j=1}^N U(r_{ij}) \right\}.
\end{equation}
Since Eq.~\eqref{Liouville_tot} is a linear equation for $f^{(N)}$, the linear combination of the solution of each Liouville equation
\begin{align}\label{Liouville_0}
    \frac{\partial f_0^{(N)}}{\partial t}&=\mathcal{L}^{(N)}f_0^{(N)} \\
    \label{Liouville_shear}
    \frac{\partial f_{\dot\gamma}^{(N)}}{\partial t}&=\mathcal{L}_{\dot\gamma}f_{\dot\gamma}^{(N)} \\
    \frac{\partial f_\mathrm{hyd}^{(N)}}{\partial t}&=\mathcal{L}_\mathrm{hyd} f_\mathrm{hyd}^{(N)}
    \label{Liouville_hydr}
\end{align}
i.e. $f^{(N)}=c_1f_0^{(N)}+c_2f_{\dot\gamma}^{(N)}+c_3 f_\mathrm{hyd}^{(N)}$ with constants $c_1$, $c_2$, and $c_3$
is a solution of Eq.~\eqref{Liouville_tot}.

Now, let us rewrite Eq.~\eqref{Liouville_0}.
Here, the one-particle distribution $f_0^{(1)}(1)$ with $1\equiv (\bm{r}_1,\bm{p}_1;t)$ satisfies
\begin{align}
    &\frac{\partial f_0^{(1)}(1)}{\partial t}
    +\left\{f_0^{(1)}(1), H^{(1)}(1)\right\}
    =L^3 \int \prod_{j=2}^NdX_j \left\{H^{(N)}-H^{(1)}, f_0^{(N)}\right\},    
\end{align}
where $H^{(1)}(1)$ is the one-particle Hamiltonian and $dX_j\equiv d\bm{r}_jd\bm{p}_j$.
It is obvious that the kinetic Hamiltonian is commutable with $f^{(N)}$, because $\int d\bm{p}_j \bm{p}_j\cdot \partial f^{(N)}/\partial \bm{r}_j$ vanishes at the boundary.
Thus, the Liouville equation for the one-particle can be expressed as \cite{McLennan}
\begin{equation}
    \frac{\partial f_0^{(1)}(1)}{\partial t}
    +\left\{f_0^{(1)}(1),H^{(1)}(1)\right\}
    =n\int dX_2\left\{U^{(2)},f_0^{(2)}(1,2)\right\},
\end{equation}
with
\begin{equation}
    U^{(2)} \equiv U(r_{12}).
\end{equation}

Poisson's bracket with $H^{(1)}(1)$ is $\{f^{(1)}(1),H^{(1)}(1)\}=\bm{v}\cdot\nabla f^{(1)}(1)$ with $\bm{v}=\bm{p}/m$.
Thus, the one-body distribution satisfies
\begin{equation}\label{McL7.10}
    \left(\frac{\partial}{\partial t}+\bm{v}\cdot\nabla\right) f_0^{(1)}(1)=nJ(1),
\end{equation}
where 
\begin{equation}\label{McL7.11}
    J(1)\equiv \int dX_2\left\{U(\bm{r}_1,\bm{r}_2),f_0^{(2)}(1,2)\right\} . 
\end{equation}
It should be noted that Eqs.~\eqref{McL7.10} and \eqref{McL7.11} with the assumption of naive molecular chaos $f_0^{(2)}(1,2)=f_0^{(1)}(1)f_0^{(1)}(2)$ does not lead to the Boltzmann equation but the Vlasov equation
\begin{equation}
   \left(\frac{\partial}{\partial t}+\bm{v}\cdot\nabla\right)    f_0^{(1)}(1)=\nabla_{\bm{p}}\cdot\mathcal{F}(1) f_0^{(1)}(1) , 
\end{equation}
where $\nabla_{\bm{p}}$ denotes the divergence in momentum space and
\begin{equation}
    \mathcal{F}(1)\equiv -n \frac{\partial}{\partial \bm{r}_1}\int dX_2 f_0^{(1)}(2)U(\bm{r}_1,\bm{r}_2) .
\end{equation}
Thus, the derivation of the Boltzmann equation is non-trivial.

If the density is low enough, we may neglect the contribution of the collision integral in the equation for $f^{(2)}(1,2)$.
Under this assumption, we can write the equation
\begin{equation}
    \frac{\partial}{\partial t}f_0^{(2)}(1,2)
    +\left\{f_0^{(2)}(1,2),H^{(2)}(1,2)\right\}=0.    
\end{equation}
Now, we write
\begin{equation}
    f_0^{(2)}(1,2)
    =\Omega f_0^{(1)}(1)f^{(1)}_0(2).
    \label{eq:f0_2_12}
\end{equation}
with an operator
\begin{equation}
    \Omega\equiv \lim_{\tau\to\infty}S^{(2)}(-\tau)S^{(0)}(\tau).
\end{equation}
Here, $S^{(s)}(t)$ is a streaming operator satisfying
\begin{equation}
    \frac{\partial}{\partial t}S^{(s)}(t)A
    = \{S^{(s)}(t)A, H^{(s)}\}.
\end{equation}
This operator also satisfies
\begin{equation}
    S^{(s)}(0)=1,\quad
    S^{(s)}(t)S^{(s)}(u)=S^{(s)}(t+u),\quad
    S^{(s)-1}(t)=S^{(s)}(-t).
\end{equation}
Substituting Eq.~\eqref{eq:f0_2_12} into the collision operator in Eq.~\eqref{McL7.11} we obtain
\begin{align}\label{J_2}
    &J(1)\approx J_2(1);\nonumber\\ 
    &J_2(1) \equiv \int dX_2 \{U(\bm{r}_1,\bm{r}_2), 
    \Omega f_0^{(1)}(1)f_0^{(1)}(2)\}.
\end{align}
Let us introduce a new operator $\vartheta$ defined as
\begin{align}
    \vartheta A(1,2)
    &\equiv \{U(\bm{r}_1,\bm{r}_2),
    A(1,2)\} \nonumber\\
    &=\frac{\partial U(\bm{r}_1,\bm{r}_2)}{\partial \bm{r}_1}\cdot\left(\frac{\partial}{\partial \bm{p}_1}-\frac{\partial}{\partial \bm{p}_2}\right)A(1,2).
\end{align}
Since the total momentum is conserved, $\Omega$ does not affect the mass-center position or velocity, one can write
\begin{equation}
    \vartheta \Omega=\bm{v}_{12}\cdot\nabla_r\Omega-\Omega \bm{v}_{12}\cdot\nabla_r .
\end{equation}
The resulting form of $J_2(1)$ can be written as
\begin{equation}
    J_2(1)=\int dX_2 [\bm{v}_{12}\cdot\nabla_r\Omega-\Omega \bm{v}_{12}\cdot\nabla_r] f_0^{(1)}(1)f_0^{(1)}(2).
\end{equation}

When the system is almost spatial homogeneous, $J_2(1)$ can be approximated as
\begin{equation}
    J_2(1)=\int dX_2\bm{v}_{12}\cdot\nabla_r\Omega f_0^{(1)}(\bm{p}_1;t)f_0^{(1)}(\bm{p}_2;t) .
\end{equation}
To evaluate the integral we use the cylindrical coordinate where $z-$axis is parallel to $\bm{v}_{12}$.
Then, $J_2$ can be rewritten as
\begin{align}\label{J_2_final}
    J_2(1)
    &=\int d^3p_2\int bdb d\phi \int_{-\infty}^\infty dz 
    v_{12}\frac{d}{dz}\Omega f_0^{(1)}(\bm{p}_1;t)f_0^{(1)}(\bm{p}_2;t)\nonumber\\
    &=\int d^3p_2\int v_{12} bdb d\phi
    [\Omega f_0^{(1)}(\bm{p}_1;t)f_0^{(1)}(\bm{p}_2;t)|_{z=\infty}
    -\Omega f_0^{(1)}(\bm{p}_1;t)f_0^{(1)}(\bm{p}_2;t)|_{z=-\infty}] ,
\end{align}
where $b$ is the impact parameter.
In this case $S^{(0)}$ does not play any role, and then, $\Omega$ is equivalent to $\lim_{\tau\to\infty}S^{(2)}(-\tau)$.
Here, there is no interaction between two particles for $|z|\gg 1$ in the region of $z<0$, which means that the second term in the bracket of Eq.~\eqref{J_2_final} is $f_0^{(1)}(\bm{p}_1;t)f_0^{(1)}(\bm{p}_2;t)$.
In the first term, $S^{(2)}(-\tau)$ takes the particles back through a collision and converts $\bm{p}_1,\bm{p}_2$ to the pre-collisional momenta $\bm{p}_1^\prime,\bm{p}_2^\prime$.
Thus, $J_2(1)$ is reduced to Boltzmann's collision operator.

It is straightforward to rewrite Eq.~\eqref{Liouville_shear} as
\begin{equation}
    \frac{\partial f_{\dot\gamma}^{(1)}(1)}{\partial t}-\dot\gamma V_y\frac{\partial}{\partial V_x} f_{\dot\gamma}^{(1)}(1)=0 
\end{equation}
for the one-body distribution function.

Equation \eqref{Liouville_hydr} can be rewritten as
\begin{equation}
    \frac{\partial \langle f_\mathrm{hydr}^{(1)}(1)\rangle }{\partial t}
    =\frac{\partial}{\partial \bm{V}_1}\cdot
    \left[\zeta\left(\bm{V}_1+\frac{T_\mathrm{env}}{m}\frac{\partial}{\partial \bm{V}_1}\right) \right]\langle f_\mathrm{hydr}^{(1)}(1)\rangle.
\end{equation}

Through the argument of this appendix, 
we can use the additive approximation of three contributions for the equation of the one-body distribution function.
As the result of the arguments in this appendix, we obtain Eq.~\eqref{eq:kinetic_eq}.

\section{Detailed expressions of $\chi$ and $r_\mathrm{min}$}\label{sec:scattering}
\begin{figure}[htbp]
    \centering
    \includegraphics[width=0.6\linewidth]{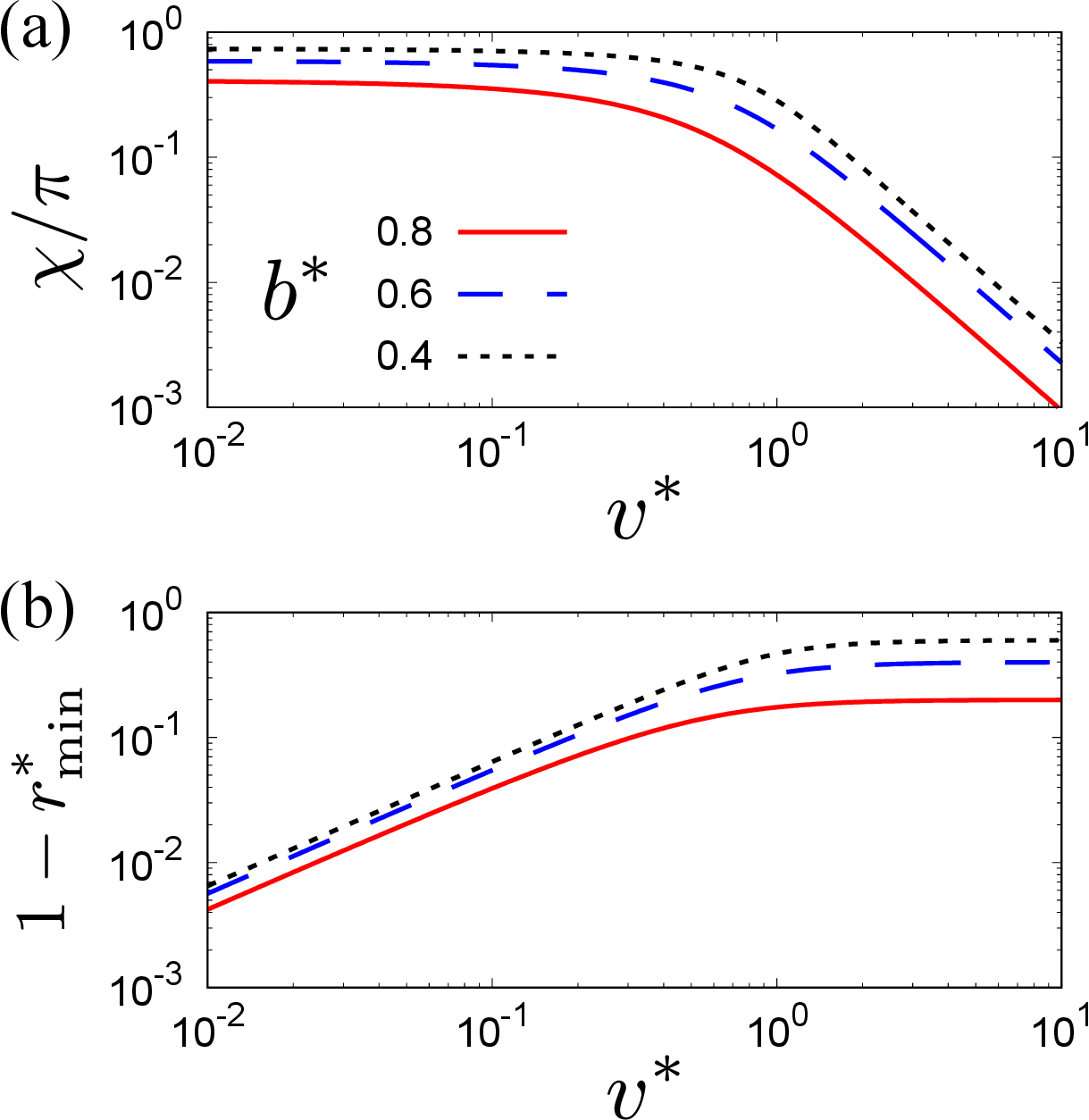}
    \caption{Velocity dependence of (a) the scattering angle $\chi$ and (b) the closest distance $r_\mathrm{min}$ against the dimensionless relative speed $v^*$ for $b^*\equiv b/d=0.4$, $0.6$, and $0.8$.
    Here, we have introduced $r_\mathrm{min}^*\equiv r_\mathrm{min}/d$.}
    \label{fig:chi_r_min_vs_v}
\end{figure}

In this appendix, we present the explicit expressions of the scattering angle $\chi$ and the turning point $r_\mathrm{min}$. 
Because their derivations are given in Ref.~\cite{Sugimoto20}, we only show the final results as functions of $b$ and $v$:
\begin{subequations}\label{eq:chi_r_min}
\begin{align}
    \chi 
    &= \pi-2\sin^{-1}\frac{b}{d}\nonumber\\
    &\hspace{1em}
    -
    \begin{cases}
    \displaystyle 
    \frac{4\sqrt{A_1A_2}}{(D_1D_2)^{1/4}}
	\left\{\frac{\alpha_2}{A_1}\left[F\left(\frac{\pi}{2}, \frac{A_2^2}{A_1^2}\right) - F\left(\sin^{-1}\left(\frac{w_0}{A_2}\right),\frac{A_2^2}{A_1^2}\right)\right]\right.\\
	\displaystyle 
    \hspace{2em}\left.+\frac{\alpha_1-\alpha_2}{A_1}\left[\Pi\left(A_2^2;\frac{\pi}{2}\left| \frac{A_2^2}{A_1^2}\right.\right)
        -\Pi\left(A_2^2; \sin^{-1}\left(\frac{w_0}{A_2}\right) \left| \frac{A_2^2}{A_1^2}\right.\right)\right]\right.\\
	\displaystyle 
    \hspace{2em}\left.+\frac{\alpha_1-\alpha_2}{\sqrt{(A_1^2-1)(1-A_2^2)}}
    \tan^{-1}\left(\sqrt{\frac{(A_1^2-1)(A_2^2-w_0^2)}{(1-A_2^2)(A_1^2-w_0^2)}}\right)\right\} & (D_1\ge 0)\\
	\displaystyle 
    \frac{4\sqrt{A_1A_2}}{(-D_1D_2)^{1/4}}
	\left\{\frac{\alpha_2}{\sqrt{A_1^2+A_2^2}}
	F\left(\cos^{-1}\left(\frac{w_0}{A_2}\right),\frac{A_2^2}{A_1^2+A_2^2}\right)\right.\\
	\displaystyle 
	\hspace{2em}\left.
	+\frac{\alpha_1-\alpha_2}{(1-A_2^2)\sqrt{A_1^2+A_2^2}}
		\Pi\left(-\frac{A_2^2}{1-A_2^2};\cos^{-1}\left(\frac{w_0}{A_2}\right) \left|\frac{A_2^2}{A_1^2+A_2^2}\right.\right) \right.\\
	\displaystyle 
	\hspace{2em}\left.+\frac{\alpha_1-\alpha_2}{\sqrt{(A_1^2+1)(1-A_2^2)}}
	\tan^{-1}\left(\sqrt{\frac{(A_1^2+1)(A_2^2-w_0^2)}{(1-A_2^2)(A_1^2+w_0^2)}}\right)\right\} & (D_1<0)
	\end{cases},\\
    r_\mathrm{min}
    &= \frac{2d}{\sqrt{\beta}+\sqrt{D_2}},
    \label{eq:r_min}
\end{align}
\end{subequations}
where $b^*\equiv b/d$, $v^*\equiv v/(\sqrt{\varepsilon/m})$, $p\equiv (2-v^{*2})/(b^{*2}v^{*2})$, $q\equiv -4/(b^{*2}v^{*2})$, $r\equiv -q/2$, $P\equiv -(p^2/3 + 4r)$, $Q\equiv -(2/27)p^3-q^2+(8/3)pr$, $\Delta\equiv (Q/2)^2+(P/3)^3$,
\begin{equation}
    \beta \equiv 
    \begin{cases}
        \displaystyle -\frac{2p}{3}+\left(-\frac{Q}{2}+\sqrt{\Delta}\right)^{1/3}+\left(-\frac{Q}{2}-\sqrt{\Delta}\right)^{1/3} & (\Delta \ge 0)\\
        \displaystyle -\frac{2p}{3}+3\sqrt{-\frac{P}{3}}\cos\left\{\frac{1}{3}\cos^{-1}\left[-\frac{Q}{2}\left(-\frac{3}{P}\right)^{3/2}\right]\right\} & (\Delta < 0)
    \end{cases},
\end{equation}
and
\begin{subequations}
\begin{align}
    D_1 &\equiv 
    -\beta-2p+ \frac{2q}{\sqrt{\beta}},\quad
    D_2 \equiv 
    -\beta-2p- \frac{2q}{\sqrt{\beta}},\\
    w_0 &\equiv -\frac{\sqrt{q^2+2\beta^2(p+\beta)}+q-2\beta}{\sqrt{q^2+2\beta^2(p+\beta)}-q+2\beta},\\
    \alpha_1 &\equiv 
    \frac{q+ \sqrt{q^2+2\beta^2(p+\beta)}}{2\beta},\quad
    \alpha_2 \equiv 
    \frac{q- \sqrt{q^2+2\beta^2(p+\beta)}}{2\beta},\\
    A_1 &\equiv 
    \sqrt{-\frac{\beta^2 D_1}{\sqrt{q^2+2\beta^2(p+\beta)}-q-\beta^{3/2}}}
    = \sqrt{q+\beta^{3/2}+\sqrt{q^2+2\beta^2(p+\beta)}},\\
    A_2 &\equiv 
    \sqrt{-\frac{\beta^2 D_2}{\sqrt{q^2+2\beta^2(p+\beta)}-q+\beta^{3/2}}}
    = \sqrt{-q+\beta^{3/2}+\sqrt{q^2+2\beta^2(p+\beta)}}.
\end{align}
\end{subequations}
Here, $F(\phi,\mathfrak{m})\equiv \int_0^\phi d\phi^\prime/(1-\mathfrak{m}\sin^2\phi^\prime)^{1/2}$ and $\Pi(a,\phi | \mathfrak{m})\equiv \int_0^\phi d\phi^\prime/[(1-a\sin^2\phi^\prime)(1-\mathfrak{m}\sin^2\phi^\prime)^{1/2}]$ are the elliptic integrals of the first and third kind, respectively~\cite{Abramowitz}.
We note that the sign of the inside of the square root for $A_2$ in Ref.~\cite{Sugimoto20} should be minus.
The velocity dependence of the scattering angle $\chi$ and the closest distance $r_\mathrm{min}$ are shown in Fig.~\ref{fig:chi_r_min_vs_v}.
These have a complicated dependence on the softness of particles.
We note that these results are obtained without considering three-body collisions and the effect of the drag from the background solvent.

\section{Comparison between Cauchy's contact stress and the collisional contribution to the stress in a hard-core system}\label{sec:replacement}
\begin{figure}[htbp]
    \centering
    \includegraphics[width=0.6\linewidth]{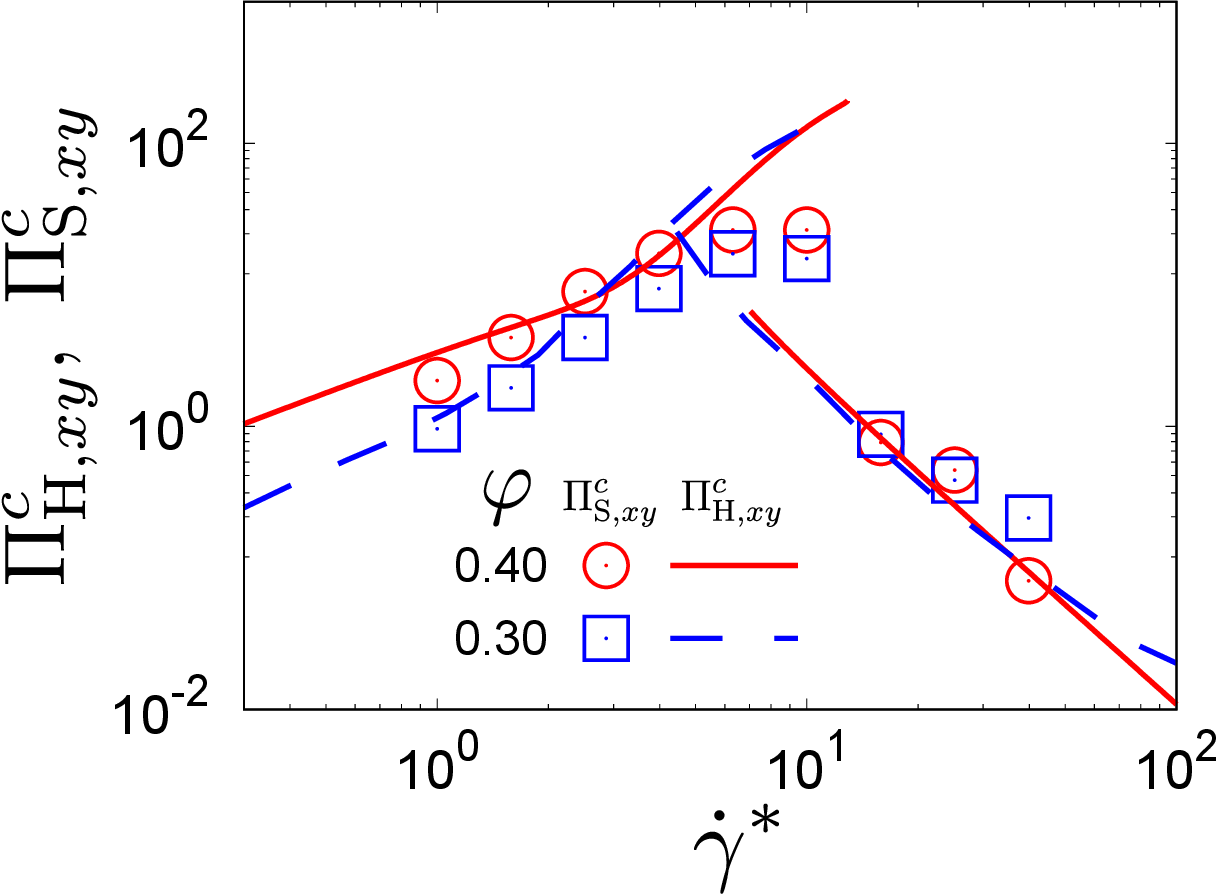}
    \caption{Shear rate dependence of the contact shear stresses $\Pi^c_{\mathrm{H},xy}$ and $\Pi^c_{\mathrm{S},xy}$ for $\varphi=0.30$ and $0.40$ by fixing $\xi_\mathrm{env}=1.0$ and $\varepsilon^*=10^4$.
    The lines represent the theoretical results $\Pi_{\mathrm{H},xy}^c$ obtained from Eq.~\eqref{eq:Pc}.
    The symbols correspond to the simulation results $\Pi_{\mathrm{S},xy}^c$ obtained from Eq.~\eqref{eq:sigma_sim}.}
    \label{fig:stress_comp}
\end{figure}

In this appendix, we examine the validity of $\Pi_{\mathrm{H}, xy}^c$ in Eq.~\eqref{eq:nondim_quantities} with Eq.~\eqref{eq:Pc_2} used for the theoretical analysis to evaluate the contact stress $\Pi_{\mathrm{S},xy}^c$ for the soft-core systems Eq.~\eqref{eq:Pi_c_S} with \eqref{eq:sigma_sim} from the comparison between two expressions.
Figure \ref{fig:stress_comp} illustrates that $\Pi_{\mathrm{H},xy}^c$ is reasonably close to $\Pi_{\mathrm{S},xy}^c$ in the wide range of the parameters' space.
This is the indirect evidence of the validity of the replacement of Eq.~\eqref{eq:sigma_sim} by Eq.~\eqref{eq:Pc}, although we cannot justify this treatment theoretically.

\section{Convergence of $\Lambda_{\alpha\beta}^*$ and $\Pi_{\mathrm{H},xy}^{c}$ in the expansions of $\tilde{\dot\gamma}$ and the linear theory}\label{sec:convergence}

In this appendix, we show how the calculated stress depends on the number of truncation terms $N_\mathrm{c}$ in the set of equations\ \eqref{eq:Lambda_expansion}.
We also present some analytic results of the linear theory with $N_\mathrm{c}=1$.

\subsection{Convergence of expansions}

In this subsection, we present how the expansions of $\Lambda_{\alpha\beta}^*$ and $\Pi_{\mathrm{H},xy}^{c}$ in the set of equations\ \eqref{eq:Lambda_expansion} converge as the number of $N_\mathrm{c}$ increases.

\begin{figure}[htbp]
    \centering
    \includegraphics[width=\linewidth]{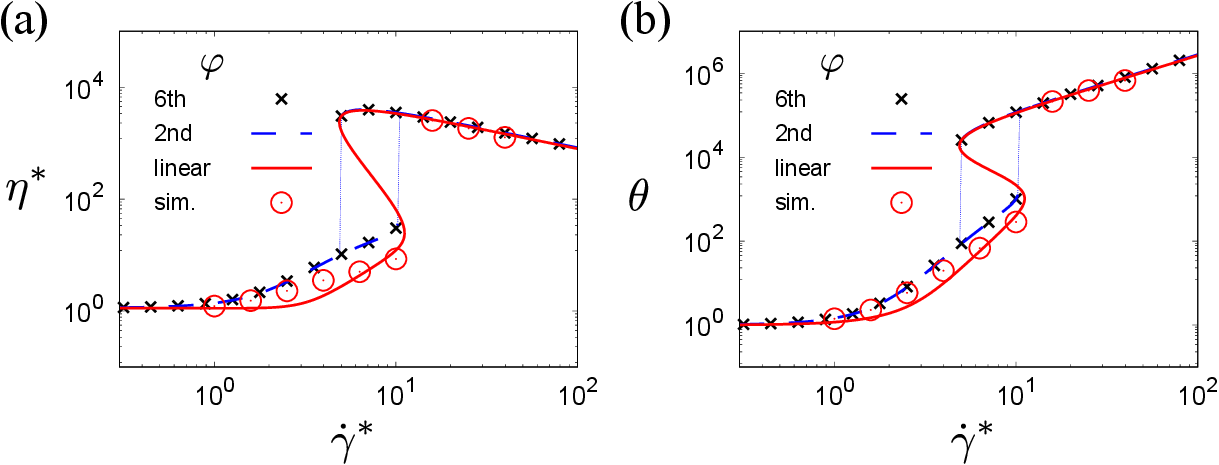}
    \caption{Plots of (a) $\eta^*$ and (b) $\theta$ against $\dot\gamma^*$ for $\varphi=0.30$, $\xi_\mathrm{env}=1.0$, and $\varepsilon^*=10^4$ when we control the truncation $N_\mathrm{c}=1$ (solid line), $2$ (dashed line), and $6$ (crosses).}
    \label{fig:convergence}
\end{figure}

We plot the theoretical viscosity $\eta^*$ against $\dot\gamma^*$ for various $N_\mathrm{c}$ in Fig.~\ref{fig:convergence} with $\varphi=0.30$, $\xi_\mathrm{env}=1.0$, and $\varepsilon^*=10^4$.
We illustrate that the difference between viscosity obtained by $N_\mathrm{c}=2$ and that by $N_\mathrm{c}=6$ is invisible, although the linear theory with $N_\mathrm{c}=1$ deviates from those for $N_\mathrm{c}\ge 2$.
We also obtain the similar results for $\theta$.
Thus, we adopt $N_\mathrm{c}=2$ for the calculations in the main text. 
We note that it is possible to get analytic expressions of $\eta^*$ and $\theta$ in the case of $N_\mathrm{c}=1$ as shown in the next subsection, while it is impossible to obtain the analytic expressions if we adopt $N_\mathrm{c}\ge 2$.

\subsection{Linear theory}\label{app:linear_theory}
In this subsection, let us show the explicit forms when we only consider the terms up to the linear order with respect to the expansion parameter $\tilde{\dot{\gamma}}$ (i.e. $N_\mathrm{c}=1$) in the set of equations~\eqref{eq:Lambda_expansion}.
A set of dynamical equations~\eqref{eq:dynamic_eqs_ver2} in the steady state with $N_\mathrm{c}=1$ is reduced to (see also Ref.~\cite{Hayakawa17}):
\begin{subequations}\label{eq:linear_dynamic_eqs}
\begin{align}
    \frac{2\dot\gamma^*}{3}{\cal C}\Pi_{xy}
    &= 2(\theta-1),\\
    2\dot\gamma^* \Pi_{xy}
    &= (2+\nu^*)\Delta\theta,\\
    2\dot\gamma^* {\cal E}\Pi_{xy}
    &= (2+\nu^*)\delta\theta,\\
    -(2+\nu^*)\Pi_{xy}
    &= \dot\gamma^* \left(\frac{2}{3}{\cal D}\Delta\theta
    -\frac{1}{3}{\cal E}\delta\theta - {\cal C}\theta\right),
    \label{eq:linear_Pi_xy_eq}
\end{align}
\end{subequations}
with
\begin{subequations}
\begin{align}
    {\cal C}&\equiv 1+\frac{8}{5}\varphi g_0 \Omega_{2,2}^*,\\
    {\cal D}&\equiv 1-\frac{4}{35}\varphi g_0 \Omega_{2,2}^*,\\
    {\cal E}&\equiv 1-\frac{16}{35}\varphi g_0 \Omega_{2,2}^*,\\
    \nu^* 
    &\equiv \frac{96}{5\sqrt{\pi}}g_0\varphi
    \xi_\mathrm{env} \sqrt{\theta}\Omega_{2,2}^*.
    \label{eq:nu_HC}
\end{align}
\end{subequations}
We note that these quantities are equivalent to those for hard-core systems \cite{Hayakawa17, Takada20, Hansen, Garzo99, Santos04} for $\Omega_{2,2}^*\to 1$.
The set of solutions of Eqs.~\eqref{eq:linear_dynamic_eqs} is given by
\begin{subequations}\label{eq:sol_linear_dynamic_eqs}
\begin{align}
    \Pi_{xy}
    &= \frac{3}{\dot\gamma^* {\cal C}}(\theta-1),
    \label{eq:Pi_xy_linear}\\
    \Delta \theta
    &= \frac{3}{{\cal C}} \frac{2(\theta-1)}{2+\nu^*},
    \label{eq:Delta_theta_linear}\\
    \delta \theta
    &= \frac{3{\cal E}}{{\cal C}} \frac{2(\theta-1)}{2+\nu^*}.
    \label{eq:delta_theta_linear}
\end{align}
\end{subequations}
Substituting Eqs.~\eqref{eq:sol_linear_dynamic_eqs} into Eq.~\eqref{eq:linear_Pi_xy_eq}, we obtain the expression of $\dot\gamma^*$ as
\begin{equation}
    \dot\gamma^*
    = \sqrt{\frac{-3(1-\theta^{-1})(2+\nu^*)}{{\cal C}{\cal F}}},
    \label{eq:shear_rate_linear}
\end{equation}
where
\begin{align}
    {\cal F}
    &\equiv \frac{2}{3}{\cal D}\frac{\Delta\theta}{\theta} - \frac{1}{3}{\cal E}\frac{\delta\theta}{\theta} - {\cal C}
    = \frac{2{\cal D}-{\cal E}^2}{{\cal C}}\frac{2(1-\theta^{-1})}{2+\nu^*} - {\cal C}.
\end{align}
Equation \eqref{eq:shear_rate_linear} is an equation to determine $\dot\gamma^*$ under a given set of $\varphi$ and $\theta$.
Once we fix $\theta$, we can determine the other rheological quantities $\Pi_{xy}$, $\Delta\theta$, and $\delta\theta$ from Eqs.~\eqref{eq:Pi_xy_linear}--\eqref{eq:delta_theta_linear}, respectively.
Correspondingly, the collisional contribution to the shear stress is given by
\begin{equation}
    \Pi_{\mathrm{H},xy}^{c}
    = \frac{8}{5}\varphi g_0 \Omega_{2,2}^*\left(\Pi_{xy}
    +\frac{1}{\sqrt{\pi}}\frac{\dot\gamma^*}{\xi_\mathrm{env}\sqrt{\theta}}\right).
    \label{eq:Pi_xy_c_linear}
\end{equation}
This quantity is also equivalent to that for hard-core systems for $\Omega_{2,2}^*\to 1$ \cite{Hayakawa17, Takada20}.


\section{Evaluation of the collision moment in dilute soft-core systems}\label{sec:dilute}
In this appendix, we present the evaluation  $\overleftrightarrow{\Lambda}$ in dilute soft-core systems \cite{Sugimoto20, Takada18, Sugimoto} to clarify the difference between the dilute system and the moderately dense system analyzed in this paper.
Note that the results presented in this appendix implicitly used Ref.~\cite{Sugimoto20} but it does not contain any explicit expressions.

Because the two-body distribution function does not include the information of positions, the collision integral $J[\bm{V}|f^{(2)}(1,2)]$ is given by
\begin{align}
    J[\bm{V}|f^{(2)}(1,2)]
    &\approx \int d\bm{V}_2 \int d\bm{\Omega} \mathcal{S}(\chi, V_{12})V_{12}
    \left[f^{(2)}(\bm{V}_1^{\prime\prime}, \bm{V}_2^{\prime\prime};t)
    - f^{(2)}(\bm{V}_1, \bm{V}_2;t)\right].
    \label{eq:J_Boltzmann}
\end{align}
We also assume the decoupling approximation
\begin{align}
    &f^{(2)}(\bm{V}_1, \bm{V}_2;t)
    \approx f(\bm{V}_1, t) f(\bm{V}_2, t).
    \label{eq:decoupling_dilute}
\end{align}
Using this approximation, the collision moment becomes
\begin{align}
    \overleftrightarrow{\Lambda}
    &\equiv 4\times \frac{m}{2} \int d\bm{V}_1 \int d\bm{V}_2
    \int d\hat{\bm{k}}_{12} \mathcal{S}(\chi,V_{12})
    (\bm{V}_{12}\cdot \hat{\bm{k}}_{12})^2\nonumber\\
    &\hspace{1em}
    \times \left[\bm{V}_{12}\hat{\bm{k}}_{12} 
    + \hat{\bm{k}}_{12}\bm{V}_{12}
    - 2(\bm{V}_{12}\cdot \hat{\bm{k}}_{12})
    \hat{\bm{k}}_{12}\hat{\bm{k}}_{12})\right]
    f\left(\bm{V}_1,t\right)
    f\left(\bm{V}_2,t\right),
    \label{eq:Lambda_Boltzmann}
\end{align}
where we have used the tensor representation for later convenience.
After introducing the velocity of center-of-gravity and the relative velocity $\bm{V}_\mathrm{G}$ and $\bm{V}_{12}=\bm{V}_1-\bm{V}_2$ with
\begin{equation}
    \bm{V}_\mathrm{G}\equiv \frac{\bm{V}_1+\bm{V}_2}{2},
\end{equation}
and $\mathcal{S}(\chi, V)\sin\chi = b|\partial b/\partial \chi|$ into Eq.~\eqref{eq:Lambda_Boltzmann}, one obtains
\begin{align}
    \overleftrightarrow{\Lambda}
    &\equiv \frac{m}{2} \int d\bm{V}_\mathrm{G} \int d\bm{V}_{12}
    \int_0^d db \int_0^{2\pi} d\phi\, b V_{12}
    (\bm{V}_{12}\cdot \hat{\bm{k}}_{12})\nonumber\\
    &\hspace{1em}
    \times \left[\bm{V}_{12}\hat{\bm{k}}_{12} 
    + \hat{\bm{k}}_{12}\bm{V}_{12}
    - 2(\bm{V}_{12}\cdot \hat{\bm{k}}_{12})
    \hat{\bm{k}}_{12}\hat{\bm{k}}_{12})\right]
    f\left(\bm{V}_1,t\right)
    f\left(\bm{V}_2,t\right).
    \label{eq:Lambda_Boltzmann_2}
\end{align}
Here, the angle $\phi$ is the same as that appears in Fig.~\ref{fig:scattering}.
Using the Grad approximation, $\Lambda_{\alpha\beta}$ is rewritten as
\begin{align}
    \overleftrightarrow{\Lambda}
    &= \frac{1}{2}\pi^{-3}m n^2 v_\mathrm{T}^3 
    \int d\bm{C}_\mathrm{G} \int d\bm{c}_{12}
    \int_0^d db \int_0^{2\pi} d\phi \, b c_{12}
    \left(\bm{c}_{12}\cdot \hat{\bm{k}}_{12}\right)
    \exp\left(-2C_\mathrm{G}^2-\frac{1}{2}c_{12}^2\right)\nonumber\\
    &\hspace{1em}
    \times \left[\bm{c}_{12}\hat{\bm{k}}_{12} 
    + \hat{\bm{k}}_{12}\bm{c}_{12}
    - 2\left(\bm{c}_{12}\cdot \hat{\bm{k}}_{12}\right)
    \hat{\bm{k}}_{12}\hat{\bm{k}}_{12})\right]
    \left[1-\frac{\Pi_{\alpha\beta}}{\theta}
    \left(2C_{\mathrm{G},\alpha} C_{\mathrm{G},\beta}
    +\frac{1}{2}c_{12,\alpha} c_{12,\beta}\right)\right]\nonumber\\
    &= 2(2\pi)^{-3/2}m n^2 v_\mathrm{T}^3 
    \int d\bm{c}_{12}
    \int_0^d db \int_0^{2\pi} d\phi \, bc_{12}
    \left(\bm{c}_{12}\cdot \hat{\bm{k}}_{12}\right)
    \exp\left(-\frac{1}{2}c_{12}^2\right)\nonumber\\
    &\hspace{1em}
    \times \left[\bm{c}_{12}\hat{\bm{k}}_{12} 
    + \hat{\bm{k}}_{12}\bm{c}_{12}
    - 2\left(\bm{c}_{12}\cdot \hat{\bm{k}}_{12}\right)
    \hat{\bm{k}}_{12}\hat{\bm{k}}_{12}\right]
    \left(1-\frac{\Pi_{\alpha\beta}}{2\theta}
    c_{12,\alpha} c_{12,\beta}\right),
    \label{eq:Lambda_Boltzmann_3}
\end{align}
where we have introduced the dimensional velocities
\begin{equation}
    \bm{C}_\mathrm{G}\equiv \frac{\bm{V}_\mathrm{G}}{v_\mathrm{T}},\quad
    \bm{c}_{12}\equiv \frac{\bm{V}_{12}}{v_\mathrm{T}},
\end{equation}
and used $\Pi_{\alpha\alpha}=0$ in the second equality.

Let us introduce the following four quantities:
\begin{subequations}
\begin{align}
    \overleftrightarrow{\Lambda_1^*}
    &\equiv \int d\bm{c}_{12}
    \int_0^d db \int_0^{2\pi} d\phi\, bc_{12}
    \left(\bm{c}_{12}\cdot \hat{\bm{k}}_{12}\right)
    \exp\left(-\frac{1}{2}c_{12}^2\right)
    \left(\bm{c}_{12}\hat{\bm{k}}_{12} 
    + \hat{\bm{k}}_{12}\bm{c}_{12}\right),\\
    \overleftrightarrow{\Lambda_2^*}
    &\equiv -2\int d\bm{c}_{12}
    \int_0^d db \int_0^{2\pi} d\phi \, bc_{12}
    \left(\bm{c}_{12}\cdot \hat{\bm{k}}_{12}\right)^2
    \exp\left(-\frac{1}{2}c_{12}^2\right)
    \hat{\bm{k}}_{12}\hat{\bm{k}}_{12},\\
    \overleftrightarrow{\Lambda_3^*}
    &\equiv \frac{\Pi_{\alpha\beta}}{2\theta}\int d\bm{c}_{12}
    \int_0^d db \int_0^{2\pi} d\phi \, bc_{12}
    c_{12,\alpha} c_{12,\beta}
    \left(\bm{c}_{12}\cdot \hat{\bm{k}}_{12}\right)
    \exp\left(-\frac{1}{2}c_{12}^2\right)
    \left(\bm{c}_{12}\hat{\bm{k}}_{12} 
    + \hat{\bm{k}}_{12}\bm{c}_{12}\right),\\
    \overleftrightarrow{\Lambda_4^*}
    &\equiv 
    -\frac{\Pi_{\alpha\beta}}{\theta}\int d\bm{c}_{12}
    \int_0^d db \int_0^{2\pi} d\phi \, bc_{12}
    c_{12,\alpha} c_{12,\beta}
    \left(\bm{c}_{12}\cdot \hat{\bm{k}}_{12}\right)^2
    \exp\left(-\frac{1}{2}c_{12}^2\right)
    \hat{\bm{k}}_{12}\hat{\bm{k}}_{12}.
\end{align}
\end{subequations}
Using these quantities, we can rewrite Eq.~\eqref{eq:Lambda_Boltzmann_3} as
\begin{equation}
    \overleftrightarrow{\Lambda}
    = \frac{1}{2}(2\pi)^{-3/2}m n^2 v_\mathrm{T}^3 
    \left(\overleftrightarrow{\Lambda_1^*}
    +\overleftrightarrow{\Lambda_2^*}
    +\overleftrightarrow{\Lambda_3^*}
    +\overleftrightarrow{\Lambda_4^*}\right).
    \label{eq:Lambda_dilute}
\end{equation}

First, we calculate $\overleftrightarrow{\Lambda_1^*}$.
It is convenient to to consider $\bm{c}_{12}$ to be the $z$-axis, and express $\hat{\bm{k}}_{12}$ in the the polar coordinates $(\theta, \phi)$ from $\bm{c}_{12}$.
We note that these $\theta$ and $\phi$ are the same as those in Fig.~\ref{fig:scattering}.
Using these coordinates, it shows the following:
\begin{align}
    \int_0^{2\pi} d\phi
    \left(\bm{c}_{12}\cdot \hat{\bm{k}}_{12}\right)\hat{\bm{k}}_{12}
    &= \int_0^{2\pi} d\phi c_{12}\cos\theta
    \begin{pmatrix}
        \sin\theta \cos\phi\\
        \sin\theta \sin\phi\\
        \cos\theta
    \end{pmatrix}\nonumber\\
    &= 2\pi c_{12}\cos^2\theta
    \begin{pmatrix}
        0 \\ 0 \\ 1
    \end{pmatrix}
    = 2\pi \cos^2\theta
    \bm{c}_{12}.
\end{align}
Then, the quantity $\overleftrightarrow{\Lambda_1^*}$ becomes
\begin{equation}
    \overleftrightarrow{\Lambda_1^*}
    \equiv 4\pi \int d\bm{c}_{12}
    \int_0^d db\, bc_{12}
    \cos^2\theta
    \exp\left(-\frac{1}{2}c_{12}^2\right)
    \bm{c}_{12}\bm{c}_{12}.
    \label{eq:Lambda_1}
\end{equation}
To calculate further, we express $\bm{c}_{12}$ in the polar coordinates $(c_{12}, \theta_\mathrm{c}, \phi_\mathrm{c})$.
Equation \eqref{eq:Lambda_1} becomes
\begin{align}
    \overleftrightarrow{\Lambda_1^*}
    &= 4\pi \int_0^\infty dc_{12}
    \int_0^d db\, bc_{12}^5
    \cos^2\theta
    \exp\left(-\frac{1}{2}c_{12}^2\right)\nonumber\\
    &\hspace{1em}\times 
    \int_0^\pi d\theta_\mathrm{c} \int_0^{2\pi}d\phi_\mathrm{c}
    \sin\theta_\mathrm{c}
    \begin{pmatrix}
        \sin^2\theta_\mathrm{c} \cos^2\phi_\mathrm{c} & \sin^2\theta_\mathrm{c}\sin\phi_\mathrm{c}\cos\phi_\mathrm{c} & \sin\theta_\mathrm{c}\cos\theta_\mathrm{c}\cos\phi_\mathrm{c}\\
        \sin^2\theta_\mathrm{c}\sin\phi_\mathrm{c}\cos\phi_\mathrm{c} & \sin^2\theta_\mathrm{c}\sin^2\phi_\mathrm{c} & \sin\theta_\mathrm{c}\cos\theta_\mathrm{c}\sin\phi_\mathrm{c}\\
        \sin\theta_\mathrm{c}\cos\theta_\mathrm{c}\cos\phi_\mathrm{c} & \sin\theta_\mathrm{c}\cos\theta_\mathrm{c}\sin\phi_\mathrm{c} & \cos^2\theta_\mathrm{c}
    \end{pmatrix}\nonumber\\
    &= \frac{16}{3}\pi^2 \int_0^\infty dc_{12}
    \int_0^d db\, bc_{12}^5
    \cos^2\theta
    \exp\left(-\frac{1}{2}c_{12}^2\right)
    \overleftrightarrow{1},
    \label{eq:Lambda_1_}
\end{align}
where $\overleftrightarrow{1}$ is the unit tensor whose size is $3\times 3$.

Similarly, let us consider the following:
\begin{align}
    \int_0^{2\pi} d\phi 
    \left(\bm{c}_{12}\cdot \hat{\bm{k}}_{12}\right)^2
    \hat{\bm{k}}_{12}\hat{\bm{k}}_{12}
    &= \int_0^{2\pi} d\phi c_{12}^2\cos^2\theta 
    \begin{pmatrix}
        \sin^2\theta \cos^2\phi & \sin^2\theta \sin\phi \cos\phi & \sin\theta \cos\theta \cos\phi\\
        \sin^2\theta \sin\phi \cos\phi & \sin^2\theta \sin^2\phi & \sin\theta\cos\theta \sin\phi\\
        \sin\theta \cos\theta \cos\phi & \sin\theta\cos\theta \sin\phi & \cos^2\theta
    \end{pmatrix}\nonumber\\
    &= \pi c_{12}^2\cos^2\theta
    \begin{pmatrix}
        \sin^2\theta & 0 & 0\\
        0 & \sin^2\theta & 0\\
        0 & 0 & 2\cos^2\theta
    \end{pmatrix}\nonumber\\
    &= \pi \cos^2\theta
    \left[c_{12}^2\sin^2\theta\overleftrightarrow{1} + \left(2\cos^2\theta-\sin^2\theta\right)\bm{c}_{12}\bm{c}_{12}\right].
\end{align}
With the aid of this, the quantity $\overleftrightarrow{\Lambda_2^*}$ is rewritten as
\begin{align}
    \overleftrightarrow{\Lambda_2^*}
    &= -2\pi \int d\bm{c}_{12}
    \int_0^d db \, bc_{12}\cos^2\theta
    \exp\left(-\frac{1}{2}c_{12}^2\right)
    \left[c_{12}^2\sin^2\theta\overleftrightarrow{1} + \left(2\cos^2\theta-\sin^2\theta\right)\bm{c}_{12}\bm{c}_{12}\right]\nonumber\\
    &= -2\pi \int_0^\infty dc_{12}
    \int_0^d db \, bc_{12}^3\cos^2\theta
    \exp\left(-\frac{1}{2}c_{12}^2\right)
    \left[4\pi c_{12}^2\sin^2\theta 
    + \frac{4}{3}\pi\left(2\cos^2\theta-\sin^2\theta\right)\right]\overleftrightarrow{1}\nonumber\\
    &= -\frac{16}{3}\pi^2 \int_0^\infty dc_{12}
    \int_0^d db \, bc_{12}^5\cos^2\theta
    \exp\left(-\frac{1}{2}c_{12}^2\right)\overleftrightarrow{1}.
    \label{eq:Lambda_2}
\end{align}

Third, the quantity $\overleftrightarrow{\Lambda_3^*}$ becomes
\begin{equation}
    \overleftrightarrow{\Lambda_3^*}
    \equiv -\frac{\Pi_{\alpha\beta}}{\theta}2\pi \int d\bm{c}_{12}
    \int_0^d db \, bc_{12}\cos^2\theta
    \exp\left(-\frac{1}{2}c_{12}^2\right)
    c_{12,\alpha} c_{12,\beta}\bm{c}_{12}\bm{c}_{12}.
\end{equation}
Let us consider the diagonal component $\Lambda_{3,xx}^*$.
\begin{align}
    \Lambda_{3,xx}^*
    &= \frac{1}{\theta}2\pi \int d\bm{c}_{12}
    \int_0^d db \, bc_{12}\cos^2\theta
    \exp\left(-\frac{1}{2}c_{12}^2\right)
    \left(\Pi_{xx}c_{12,x} c_{12,x}
    +\Pi_{yy}c_{12,y} c_{12,y}
    +\Pi_{zz}c_{12,z} c_{12,z}\right)c_{12,x}^2\nonumber\\
    &= \frac{\Pi_{xx}}{\theta}\frac{16}{15}\pi^2 
    \int_0^\infty dc_{12}
    \int_0^d db \, bc_{12}^7\cos^2\theta
    \exp\left(-\frac{1}{2}c_{12}^2\right).
\end{align}
Similarly, the nondiagonal component $\Lambda_{3,xy}^*$ is obtained as
\begin{align}
    \Lambda_{3,xy}^*
    &= \frac{\Pi_{xy}}{\theta}2\pi \int d\bm{c}_{12}
    \int_0^d db \, bc_{12}\cos^2\theta
    \exp\left(-\frac{1}{2}c_{12}^2\right)
    2\Pi_{xy}c_{12,x}^2 c_{12,y}^2\nonumber\\
    &= \frac{\Pi_{xy}}{\theta}\frac{16}{15}\pi^2 
    \int_0^\infty dc_{12}
    \int_0^d db \, bc_{12}^7\cos^2\theta
    \exp\left(-\frac{1}{2}c_{12}^2\right).
\end{align}
We can follow the same procedures for other components.
As a result, $\overleftrightarrow{\Lambda_3^*}$ is evaluated as
\begin{equation}
    \overleftrightarrow{\Lambda_3^*}
    = \frac{16}{15}\pi^2 
    \int_0^\infty dc_{12}
    \int_0^d db \, bc_{12}^7\cos^2\theta
    \exp\left(-\frac{1}{2}c_{12}^2\right)\frac{\overleftrightarrow{\Pi}}{\theta}.
    \label{eq:Lambda_3}
\end{equation}

Fourth, we calculate $\overleftrightarrow{\Lambda_4^*}$.
After some calculation, one reaches
\begin{align}
    \overleftrightarrow{\Lambda_4^*}
    &=
    -\frac{16}{15}\pi^2 
    \int_0^\infty dc_{12}
    \int_0^d db \, bc_{12}^7\cos^2\theta
    \left(1-\frac{3}{2}\sin^2\theta\right)
    \exp\left(-\frac{1}{2}c_{12}^2\right)\frac{\overleftrightarrow{\Pi}}{\theta}.
    \label{eq:Lambda_4}
\end{align}

Substituting Eqs.~\eqref{eq:Lambda_1_}, \eqref{eq:Lambda_2}, \eqref{eq:Lambda_3}, and \eqref{eq:Lambda_4} into Eq.~\eqref{eq:Lambda_dilute}, one gets
\begin{align}
    \overleftrightarrow{\Lambda}
    &= -\frac{1}{5}n^2 T 
    \sqrt{\frac{\pi T}{m}}
    \int_0^\infty dc_{12}
    \int_0^d db \, bc_{12}^7\sin^2(2\theta)
    \exp\left(-\frac{1}{2}c_{12}^2\right)\frac{\overleftrightarrow{\Pi}}{\theta}\nonumber\\
    &= -\nu^* \zeta
    \left(\overleftrightarrow{\sigma^k}+p\overleftrightarrow{1}\right),
    \label{eq:Lambda_dilute_sigma}
\end{align}
with the static pressure $p \equiv -\sigma_{\alpha\alpha}^k$ and the collision frequency for hard-core system $\nu_\mathrm{HC}$ as given in Eq.~\eqref{eq:nu_HC}.


\end{document}